\documentclass[twocolumn]{aastex61}
\pdfoutput=1 
\usepackage{amsmath,amstext}
\usepackage[T1]{fontenc}
\usepackage{apjfonts} 
\usepackage{comment}
\usepackage[figure,figure*]{hypcap}
\usepackage{multirow}
\usepackage{amsmath}
\usepackage{nccmath}
\usepackage{rotating}
\usepackage{soul}

\setstcolor{red}

\shorttitle{Distances to  GB's Star-Forming Regions with Gaia}
\shortauthors{Dzib et al.}

\begin{document}

\title{Distances and Kinematics of Gould Belt Star-Forming Regions with Gaia~DR2 results}

\correspondingauthor{Sergio A. Dzib}
\email{sdzib@mpifr-bonn.mpg.de}

\author[0000-0001-6010-6200]{Sergio A. Dzib}
\affiliation{Max-Planck-Institut f\"ur Radioastronomie, Auf dem H\"ugel 69,
 D-53121 Bonn, Germany}
 
\author[0000-0002-5635-3345]{Laurent Loinard}
\affiliation{Instituto de Radioastronom\'{\i}a y Astrof\'{\i}sica, Universidad Nacional Aut\'onoma de M\'exico, Morelia 58089, Mexico}
\affiliation{ Instituto de  Astronom\'{\i}a, Universidad Nacional Aut\'onoma de M\'exico, Apartado Postal 70-264, CdMx C.P. 04510, Mexico}
 \author[0000-0002-2863-676X]{Gisela N.~Ortiz-Le\'on}
\affiliation{Max-Planck-Institut f\"ur Radioastronomie, Auf dem H\"ugel 69,
 D-53121 Bonn, Germany}
 \affiliation{Humboldt Fellow}
\author[0000-0003-2737-5681]{Luis F. Rodr\'{\i}guez}
\affiliation{Instituto de Radioastronom\'{\i}a y Astrof\'{\i}sica, Universidad Nacional Aut\'onoma de M\'exico, Morelia 58089, Mexico}
 \author[0000-0003-2271-9297]{Phillip A. B. Galli}
 \affiliation{Laboratoire d'astrophysique de Bordeaux, Univ. Bordeaux, CNRS, B18N, all\'ee 
 Geoffroy Saint-Hillaire, 33615 Pessac, France.}

\begin{abstract}
We present an analysis of the astrometric results from Gaia second data release (DR2) 
to Young Stellar 
Objects (YSOs) in star-forming regions related to the Gould Belt. These regions are
Barnard 59, Lupus 1 to 4, Chamaeleon I and II, $\epsilon$-Chamaeleontis,
the Cepheus flare, IC~5146 and Corona Australis. The mean distance
to the YSOs in each region are consistent with earlier estimations, 
though a significant improvement to the final errors was obtained.
The mean distances to the star-forming regions were used to fit an ellipsoid 
of size $(358\pm7)\times(316\pm13)\times(70\pm4)$~pc,
and centered at $(X_0,Y_0,Z_0)=(-82\pm15, 39\pm7, -25\pm4)$~pc, consistent with 
recently determined parameter of the Gould Belt. 
The mean proper motions were combined with radial velocities from the
literature to obtain the three dimensional motion of the star-forming regions,
which are consistent with a general expansion of the Gould Belt.
We estimate that this expansion is occurring at a velocity of $2.5\pm0.1$~km~s$^{-1}$.
This is the first time that YSOs motions are used to investigate the kinematic of 
the Gould Belt. As an interesting side result, we also identified stars with large peculiar velocities.
\end{abstract}


\nopagebreak
\section{Introduction}\label{intro}

In the 19$^{th}$ century, Sir John Herschel and, a few decades later, Professor 
Benjamin Gould noticed that the brightest stars in the sky defined a partial ring that 
crossed the Milky Way with an inclination angle of $\sim20^\circ$.
Nowadays, we know that this structure is composed of gas and young stars and, in fact,
most of the nearby star-forming regions are part of this  structure, 
now called the Gould Belt (GB).   The classical assumed structure for the GB is an elliptical ring
traced mainly by clouds and OB stars. However analysis of X-ray active stars have shown that a 
filled disk structure may be more appropriate \citep{guillout1998}.
The origin of the GB is still uncertain, but
there are suggestions that its origin is due to the collision of a high-velocity cloud
(or dark matter dominated clump) with the Galactic plane \citep{comeron1993,bekki2009}.

 The GB semi-major axes are between 200 and 400 pc and its center
lies around 100 pc away from the Sun in the direction of the Galactic anti-center 
\citep[e.g.,][]{perrot2003}. Then, even when some star forming regions lie in projection 
to the so-called GB-plane their heliocentric distances, $>500\,$pc, do not match with the main structure, 
and are not considered part of the GB\footnote{ For example, the Monoceros R2 
region lies in the direction of the Orion Nebula Cluster, a star forming region that
belongs to the GB,
it is however at a distance of 900 pc \citep{dzib2016}, thus far away from the 
GB.}. On the other hand, there are star forming regions in the solar neighborhood 
whose positions are projected in directions far apart from the GB-plane 
(e.g., Chamaeleon clouds and Corona Australis) making their 
membership in the GB debatable. These deviations may simple reflect the thickness 
of the GB disk.

Distances to GB star-forming regions have been controversial and historically have been affected by uncertainties of  
around 30\% (e.g. see next subsection). In recent years, 
Very Long Baseline Interferometry (VLBI)
has helped to estimate the distance with a few percents of accuracy to Young
Stellar Objects (YSOs) with strong magnetic activity  that produces 
non-thermal radio emission \cite[e.g.,][and references 
therein]{ortiz2017b, ortiz2017a,kounkel2017,dzib2018,galli2018}. In all cases,
these estimations allowed to obtain a mean distance, the depth, and the kinematics
of the regions where they belong. 

A major milestone in modern astrometry is the Gaia space mission \citep{GC2016}. Gaia has measured 
the positions ($\alpha$, $\delta$), proper motions ($\mu_\alpha$, $\mu_\delta$), and parallaxes ($\pi$)
of billions of stars in its second data release \citep{GC2018}.  The dust obscures the optical emission of many of the YSOs in star-forming regions so 
that they cannot be detected by the Gaia satellite. However, Gaia does provide 
the distance to YSOs that are not highly embedded and to the brightest members, making it possible to determine 
the distance and kinematics to these star-forming regions.
\\

\subsection{Star-forming regions in Gould Belt}

Around a dozen star-forming regions have been associated with the GB.
Our analysis is focused on the regions whose distance, 
position, or kinematic would suggest
an association with the GB \citep{perrot2003,ward2007}.
We introduce these regions below. Other star-forming regions will be discussed 
in parallel papers:  Perseus by \cite{ortiz2018}, Ophiuchus  and Serpens by Ortiz-Le\'on et al. (in prep.), and Taurus by Galli et al. (in prep.). 

{\it Barnard 59} (B59) is the only cloud in the Pipe Nebula complex with star formation activity 
\citep{alves2008}. There are around 20 embedded YSOs in the B59 cloud \citep{brooke2007} which 
will be used as our target sample for this region. Earlier studies suggested a distance
of 160~pc \citep{reipurth1993}. However, recent studies performed by \cite{lombardi2006} and \cite{alves2007} 
yield distances of $130^{+24}_{-58}$ pc and $145\pm16$~pc, respectively. These authors used Hipparcos data 
of field stars. Recently, \cite{dzib2016} obtained radio observations to search for YSOs with non-thermal radio emission 
to later observe them with the VLBI technique and determine a distance, however, even when nine stars were detected
in the radio no clear candidates  to observe with VLBI were found.
Thus, a direct estimation of the distance to this region is necessary.

The {\it  Cepheus Flare}  is a coherent giant molecular cloud complex \citep[size $\sim100$ pc;][]{olano2006} 
with dense gas distributed in a wide range of radial velocities \citep[from -15 to 15 km s$^{-1}$;][]{hu1981,lebrun1986,kun2008}. The wide range of motion 
of the gas may reflect multiple supernova events in the past. These supernovas
may have also sustained the star formation in this region \citep{kun2008}.
Distance determinations to the different components of the complex agree with the conclusion that these
are distributed between 300 pc and 500 pc \citep[][ and references therein]{kun2008}.
 \citet{kun2008} compiled a list of YSOs in the Cepheus region that we will use
as input for our analysis.
Towards the Cepheus flare region, there are other more distant star-forming regions,
of which the most active is NGC 7129. Many of the pre-main sequence 
stars found in the Cepheus flare are part of this region \citep{kun2008}. NGC 7129, however, 
seems to be at a larger distance of 1~kpc \citep{racine1968,straizys2014}.

{\it Chamaeleon} is divided into three different main clouds,
each in a different evolutionary stage.
While Chamaeleon I (Cha I) is actively forming stars, Chamaeleon III
(Cha III) has no evidence of star formation; and Chamaeleon II (Cha II)
is in an intermediate phase of forming stars.  The input list of YSOs
was taken mainly from the stars listed in \citet[][see also the references therein]{luhman2008}
for the two star forming regions in Chamaeleon.
\cite{luhman2008} discussed the distances 
measured in the last few decades. For Cha I,  photometric works point 
to distances between 135 pc and 165 pc, while the Hipparcos parallax 
to star members are more consistent with a distance around 175 pc.
On the other hand, the extinction measurements towards Cha II point
to a distance of $178\pm18$~pc. Recently, \cite{voirin2018}
using data from Gaia DR1, estimated distances of 179$^{+11,+11}_{-10,-10}$~pc,
 181$^{+6,+11}_{-5,-10}$~pc, and 199$^{+8,+12}_{-7,-11}$~pc, for Cha I, Cha II,
and Cha III, respectively. In the area between these clouds, there is a nearby 
(d$\sim$114 pc) and relatively old ($\tau\sim$6 Myr) group of YSOs, known as the 
$\epsilon$ Chamaeleontis ($\epsilon$ Cha) group \citep{feigelson2003,luhman2004}.

{\it Corona Australis} (CrA) is one of the nearest clouds with ongoing low- 
and intermediate-mass star formation with most of its YSO population in the 
Coronet protostar cluster \citep[see][ for a review of this region]{neuhauser2008}. 
 In their review \citet{neuhauser2008} compile the young stellar members on this
region, which we use as the input list.
Distance estimates toward CrA vary from 120 to 170 pc, with the best estimation 
being around 130 pc \citep{neuhauser2008}. It was discarded as part of the GB by 
\citet{olano1982} since it is  located at a Galactic latitude too low compared to 
that of the GB plane. However, it has been suggested that the 3D 
motion of its T Tauri stars is consistent with the cloud being formed by the impact of a 
high-velocity cloud onto the Galactic plane,  and thus consistent with the formation of 
the GB. Nevertheless, alternative scenarios cannot be ruled out \citep{neuhauser2000}.


{\it IC 5146} is a nebula that  is in the transition between a reflection nebula 
and an HII region \citep{herbig2008}. It has been associated with a large population 
of YSOs, whose most massive member, the B0 V star BD+46$^\circ$3474, illuminates 
and ionizes the gas.  The population of YSOs was listed by \citet{herbig2008}
in their review of this region. The distance to IC 5146
 is still quite uncertain. Early works suggested a distance
around 1~kpc \citep{elias1978,crampton1974}, but later \cite{lada1999} suggested a shorter
distance of 460~pc, based on counts of foreground stars. More recently, different works suggest
that the distance to this massive star-forming region is between 900 and 1400 pc
\citep[see discussion by][and references therein]{herbig2008}.  The position of IC~5146
on the plane of the sky is consistent with the GB-plane, thus its membership in the GB
strongly depends on an accurate measurement of its distance.

The {\it Lupus} complex is formed of nine clouds (named as Lupus 1 to 9) and it has been recognized 
as one of the main low-mass star-forming region within 200~pc of the Sun \citep{comeron2008}. 
Most of the star-forming activity is concentrated in Lupus~3, and it has the densest population
of YSOs in the complex. Other clouds that are forming stars are Lupus 1, 2, and 4. Lupus 4
 contains H$^{13}$CO$^{+}$ cores that are signs of future star formation \citep{comeron2008}. The other clouds have no star formation
activity.  The input list of YSOs for this complex was taken from the work by \citet{comeron2008},
who compiled the YSO population at the different evolutionary stages.
In his review of this complex, \cite{comeron2008} argued that a single distance to the whole
complex seems inadequate, and suggested values of 200~pc for Lupus~3, and 150~pc for the rest of the clouds.
More recently, kinematic studies performed by \cite{galli2013} show that the distances to 
Lupus~1 and Lupus~3 are $182^{+7}_{-6}$ and $185^{+11}_{-10}$~pc, respectively, i.e., they are consistent 
with each other within their errors bars.

\begin{table*}
\centering
\renewcommand{\arraystretch}{1.1}
\caption{Astrometric parameters of YSOs used in our analysis from the Gaia-DR2 catalog \citep{GC2018}.
}
\begin{tabular}{lcccccc}
\hline\hline
       & R.A.     &   Dec.    &$\pi$ &$\mu_{\alpha_{*}}$& $\mu_{\delta}$& \\
YSO Name & ($^\circ$) &($^\circ$) & (mas)&   (mas yr$^{-1}$)&(mas yr$^{-1}$) &Region\\
\hline
EM* LkHA 346&257.8220029&$-27.4190194$&$7.21\pm1.11$&$-5.54\pm2.38$&$-10.88\pm1.76$&Barnard 59\\ 
2MASS J17113036-2726292&257.8765466&$-27.4415534$&$6.39\pm0.19$&$2.37\pm0.34$&$-19.94\pm0.23$&Barnard 59\\ 
2MASS J17112701-2723485&257.8625319&$-27.3969175$&$6.02\pm0.53$&$-0.50\pm1.13$&$-16.58\pm0.78$&Barnard 59\\ 
2MASS J17112942-2725367&257.8725897&$-27.4270103$&$6.18\pm0.63$&$-0.83\pm1.17$&$-19.01\pm0.84$&Barnard 59\\ 
2MASS J17112729-2725283&257.8637175&$-27.4246713$&$7.10\pm0.70$&$1.21\pm1.40$&$-18.33\pm0.95$&Barnard 59\\ 
2MASS J17114182-2725477&257.9243060&$-27.4299610$&$6.35\pm0.06$&$-0.91\pm0.10$&$-17.91\pm0.07$&Barnard 59\\ 
2MASS J17111182-2726547&257.7992678&$-27.4486738$&$6.20\pm0.54$&$-1.23\pm1.08$&$-18.92\pm0.76$&Barnard 59\\ 
2MASS J17111445-2726543&257.8102231&$-27.4485307$&$6.26\pm0.21$&$-0.10\pm0.40$&$-19.05\pm0.26$&Barnard 59\\ 
$[$BHB2007$]$ 18NE&257.9238374&$-27.4307035$&$6.18\pm0.09$&$-0.50\pm0.15$&$-18.20\pm0.10$&Barnard 59\\ 
2MASS J17110411-2722593&257.7671511&$-27.3832858$&$5.92\pm0.30$&$-1.41\pm0.59$&$-21.63\pm0.34$&Barnard 59\\ 
\hline\hline
\label{tab:GP}
\end{tabular}
\tablecomments{Table 1 is published in its entirety in the machine-readable format.
      A portion is shown here for guidance regarding its form and content.}
\end{table*}

\section{Data and Methods}

We compiled an initial list of known young stars in the regions mentioned above 
from the literature and we complemented with sources from the Simbad database. Then we searched in the Gaia 
archive database which stars from this list have an astrometric solution for the five parameters
($\alpha$, $\delta$, $\mu_{\alpha,*}=\mu_{\alpha}\cos{\delta}$, $\mu_\delta$, and $\pi$). The maximum allowed separation
between the position in the literature of the YSOs and the objects in the Gaia catalog was $1''$.
We also discarded those YSOs whose trigonometric parallax  is negative or indicative of a
distance well beyond the value estimated for the regions where they belong. These odd results 
indicate that their Gaia solution needs to be revisited or that the stars in question actually do not belong to the region.  The YSOs used for the analysis in the following sections are listed in Table~\ref{tab:GP}.

To obtain a mean parallax and the dispersion measure to the individual regions, a histogram of
the parallaxes was constructed for each region. Then these histograms were fitted with a Gaussian 
distribution model. To obtain the distance to the regions we apply the well known
relation $d[pc]=1/\bar{\pi}['']$.  
In a similar fashion, and following \cite{dzib2017}, we also 
constructed histograms of the proper motions and fitted Gaussian distribution models to them. 
This procedure is also used to recognize stars moving with peculiar proper motions 
\citep[i.e., stars with $|\mu-\bar{\mu}|>3\sigma_\mu$, where $\bar{\mu}$ is the mean proper motion of each cloud, see also][]{dzib2017}. In the regions where only few YSOs are in the Gaia DR2 catalog, we obtain the main values in parallax and proper motions by averaging the individual values.

\begin{table*}
\small
\centering
\renewcommand{\arraystretch}{1.1}
\caption{{ Studied  star forming regions in the Gould Belt and their position in Galactic coordinates.} The number of YSOs used in the studied regions and the mean and dispersion measurement of trigonometric parallaxes 
to each region are presented. The median of individual trigonometrical parallax errors is also presented for 
comparison with the dispersion measurement. The corresponding distances and final errors to each region are also
listed.}
\begin{tabular}{lcccccc|cc}
\hline\hline
        & ${l}$ & ${b}$& \#   & $\bar{\pi}\pm\varepsilon_{\bar{\pi}}$ &$\sigma_{\pi}$&$\bar{\varepsilon_\pi}$& $\bar{D}$ &$\varepsilon_{\bar{D}}$\\
Region &($^{\circ}$)&($^{\circ}$)&YSOs &    (mas)    & (mas)&   (mas)     &   (pc)    &(pc)\\
\hline
Barnard 59        &357.0 &+07.1 &11&$6.12\pm0.03$ &  0.16& 0.40     & $163$ & $5$   \\ 
Cepheus Flare     &110.0 &+15.0 &  47 &$2.79\pm0.04$ & 0.22&  0.10   & $358$ & $32$  \\
Cepheus - NGC 7129&105.4 &+09.9 & 40 & $1.08\pm0.04$ & 0.16& 0.12   & $926$ & $163$ \\
Chamaeleon I      &297.2 &$-15.4$ &  71 &$5.21\pm0.01$  & 0.13& 0.09   & $192$ & $6$   \\
Chamaeleon II     &303.6 &$-14.4$ & 12 &$5.05\pm0.02$ & 0.09& 0.07    & $198$ & $6$   \\
$\epsilon$ Chamaeleontis&300.3 &$-14.0$&19& $9.93\pm0.03$ & 0.11& 0.05& $101$ & $2$   \\
Corona Australis  &359.9 &$-17.8$ & 21 &$6.48\pm0.02$ & 0.13& 0.09    & $154$ & $4$   \\
IC 5146           &094.4 &$-05.5$ & 62 &$1.23\pm0.03$ & 0.12& 0.12    & $813$ & $106$    \\
Lupus 1           &338.8 &+15.7 & 9  & $6.41\pm0.01$ & 0.06& 0.06   & $156$ & $3$   \\
Lupus 2           &338.9 &+12.1 & 4 & $6.31\pm0.01$  & ...& 0.06     & $159$ & $3$     \\
Lupus 3           &339.6 &+09.4 & 49 &$6.16\pm0.01$ & 0.13& 0.08      & $162$ & $3$   \\
Lupus 4           &336.3 &+08.2 & 11 &$6.13\pm0.02$ & 0.10& 0.13       &$163$ & $4$   \\
\hline\hline
\label{tab:DP}
\end{tabular}
\end{table*}

\begin{table*}
\footnotesize
\small
\centering
\renewcommand{\arraystretch}{1.1}
\caption{List of mean proper motions of the studied regions and the corresponding linear velocities.
   The proper motion and velocity dispersions are also listed.  The reflex solar proper motions are also listed.}
\begin{tabular}{lcc|cc|cc|cc|cc}\hline\hline
    & \multicolumn{2}{c|}{$\mathbf{\mu_\odot}$}& \multicolumn{2}{c|}{$\mu$}&\multicolumn{2}{c|}{Velocity$_{\rm sky}$}&\multicolumn{2}{c|}{$\sigma_\mu$}&\multicolumn{2}{c}{$\sigma_{\rm Vel.}$}         \\
   & $\alpha_*$&$\delta$ & $\alpha_*$&$\delta$&$\alpha_*$&$\delta$&$\alpha_*$&$\delta$&$\alpha_*$&$\delta$\\
Region & \multicolumn{2}{c|}{(mas yr$^{-1}$)}& \multicolumn{2}{c|}{(mas yr$^{-1}$)}&\multicolumn{2}{c|}{(km s$^{-1}$)}&\multicolumn{2}{c|}{(mas yr$^{-1}$)}&\multicolumn{2}{c}{(km s$^{-1}$)}\\
\hline
Barnard 59         &$-3.5$&$-17.9$&$-1.2\pm0.2$&$-19.2\pm0.1$& $-0.9\pm0.2$&$-14.9\pm0.1$&$0.23$ & $0.37$ & $0.2$&$0.3$ \\
{ Cepheus Flare}&$8.3$&$3.7$&  $5.8\pm0.3$ &$0.4\pm0.7$ & $9.9\pm0.5$&$0.7\pm1.2$&$1.7$ & $3.1$ & $2.8$&$5.3$  \\
Cepheus - NGC 7129 &$3.2$&$1.1$&$-1.89\pm0.02$&$-3.51\pm0.05$& $-7.8\pm0.1$&$-14.6\pm0.2$&$0.26$ & $0.43$ & $1.1$&$1.8$ \\
Chamaeleon I       &$-18.0$&$1.3$& $-22.8\pm0.1$&$0.3\pm0.1$& $-20.6\pm0.1$&$0.3\pm0.1$&$0.90$ & $1.25$ & $0.8$&$1.1$ \\
{ Chamaeleon II}&$-16.7$&$-7.2$& $-20.7\pm0.1$&$-7.8\pm0.1$& $-19.1\pm0.1$&$-7.3\pm0.1$&$0.72$ & $0.42$ & $0.7$&$0.4$ \\
$\epsilon$ Chamaeleontis&$-16.7$&$-7.2$&$-40.3\pm0.2$&$-7.3\pm0.7$& $-19.2\pm0.1$&$-3.5\pm0.3$&$1.24$ & $1.60$ & $0.5$&$1.1$ \\
Corona Australis  &$7.2$&$-20.7$ &   $4.4\pm0.2$&$-27.3\pm0.2$& $3.2\pm0.2$&$-19.8\pm0.2$&$1.46$ & $1.09$ & $1.1$&$0.8$ \\
IC 5146           &$3.8$&$0.3$&$-2.96\pm0.01$&$-2.96\pm0.03$& $-11.0\pm0.04$&$-11.0\pm0.1$&$0.39$ & $0.41$ & $1.5$&$1.5$ \\
Lupus 1           &$-11.5$&$-18.9$& $-15.0\pm0.1$&$-23.3\pm0.1$& $-11.1\pm0.1$&$-17.7\pm0.1$&$0.74$ & $0.67$ & $0.6$&$0.5$ \\
Lupus 2           &$-10.3$&$-19.3$& $-10.7\pm0.1$&$-22.0\pm0.1$& $-8.0\pm0.1$&$-16.5\pm0.1$&... & ... & ...&... \\
Lupus 3           &$-9.1$&$-19.5$&$-10.0\pm0.1$&$-23.7\pm0.1$& $-7.5\pm0.1$&$-17.9\pm0.1$&$0.78$ & $0.63$ & $0.6$&$0.5$ \\
Lupus 4           &$-8.1$&$-19.4$&$-11.9\pm0.2$&$-23.9\pm0.2$& $-9.2\pm0.2$&$-18.5\pm0.2$&$1.0$ & $0.5$ & $0.8$&$0.4$ \\
\hline\hline
\label{tab:kin}
\end{tabular}
\end{table*}

\section{Results and Discussion of Individual Regions}

Using the methods described in the previous section, we determined the mean trigonometric parallax to 
the  star-forming regions introduced in Section \ref{intro}.  The mean trigonometric parallaxes and 
the corresponding distances are presented in Table \ref{tab:DP}.
The trigonometric parallax dispersion ($\sigma_\pi$) is dominated by the errors of the individual 
parallaxes ($\varepsilon_\pi$), since the median of these errors is similar to the parallax dispersion 
(see Table~\ref{tab:DP}), thus they represent 
an uncertainty in the mean distance to the region rather than a sign of their depth. Our final error,
then, will be obtained by adding in quadrature the error of the mean trigonometric parallax ($\varepsilon_{\bar{\pi}}$),
the parallax dispersion ($\sigma_\pi$) and the systematic errors of 0.1~mas of the Gaia DR2 results 
\citep[][]{luri2018}, being the last two the dominant sources of errors. 
The estimated distances to the star-forming regions are in good agreement with previous estimations 
but we found a significant improvement on the final errors. 

The proper motion measured by the Gaia mission contain the reflex solar motion.
This motion is, in some cases, the dominant contribution to the proper motion 
vector. We have calculated the resulting proper motions by using a Solar motion of  
($U_\odot,V_\odot,W_\odot$)= ($11.1\pm0.7,12.2\pm0.47,7.25\pm0.37$)~km~s$^{-1}$ 
\citep{schonrich2010} and the Galacto-centric distance of 8.4 
kpc \citep{reid2009}. The values for each region are listed in Table~\ref{tab:kin}
to compare them with their estimated mean values.

The estimated global proper motions to each star-forming region are shown in Table~\ref{tab:kin} together with 
the proper motion dispersions. Tangential velocities and velocity dispersions were 
calculated according to the mean distance to each region and are also presented in Table~\ref{tab:kin}.
In the following, we will describe  the individual results for each region.

\subsection{Barnard 59}

\begin{figure*}[!ht]
   \centering
   \begin{tabular}{cc}
  \includegraphics[height=0.33\textwidth, trim=10 0 0 0, clip]{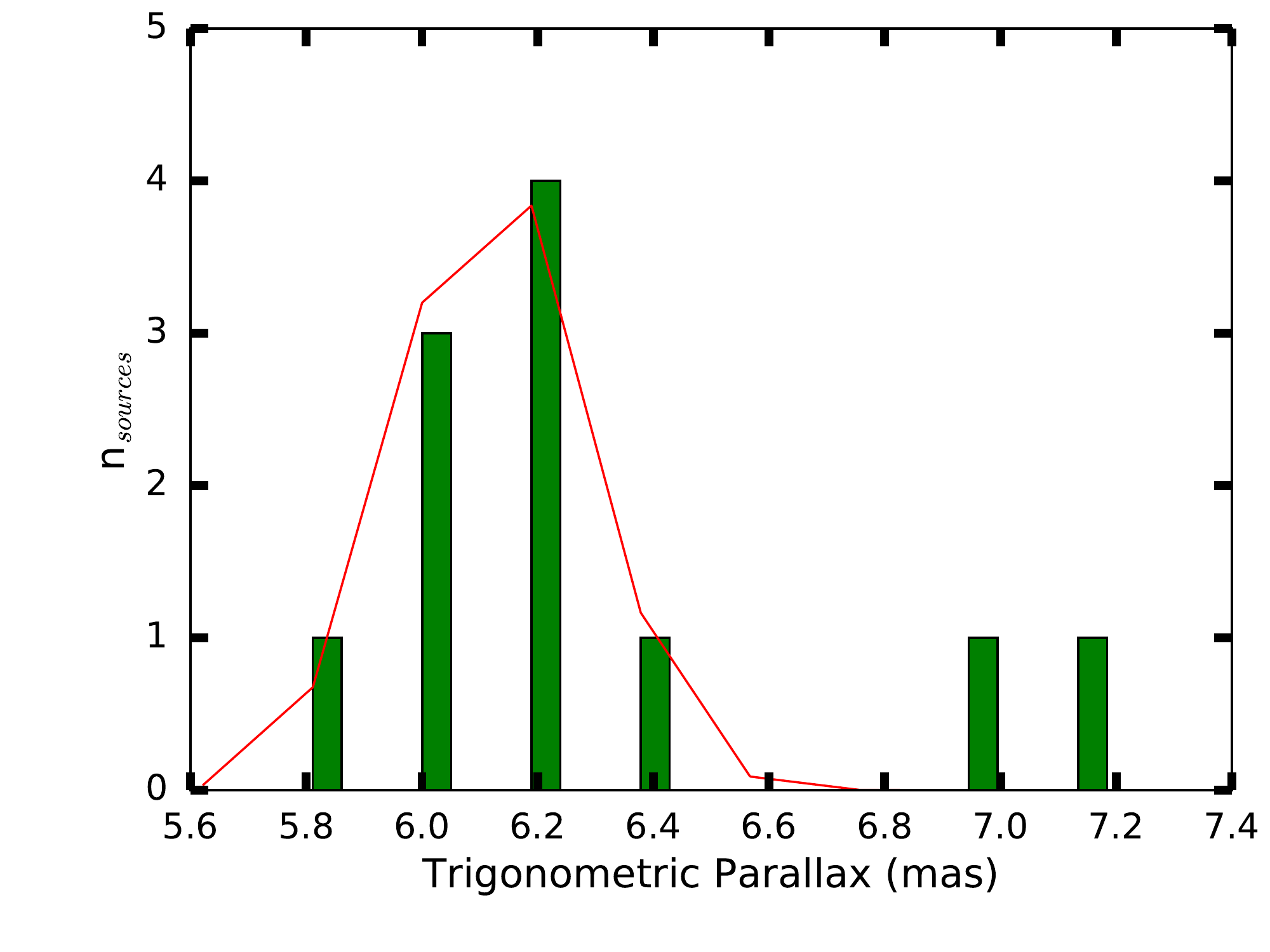} &
  \includegraphics[height=0.33\textwidth, trim=0 0 0 0, clip]{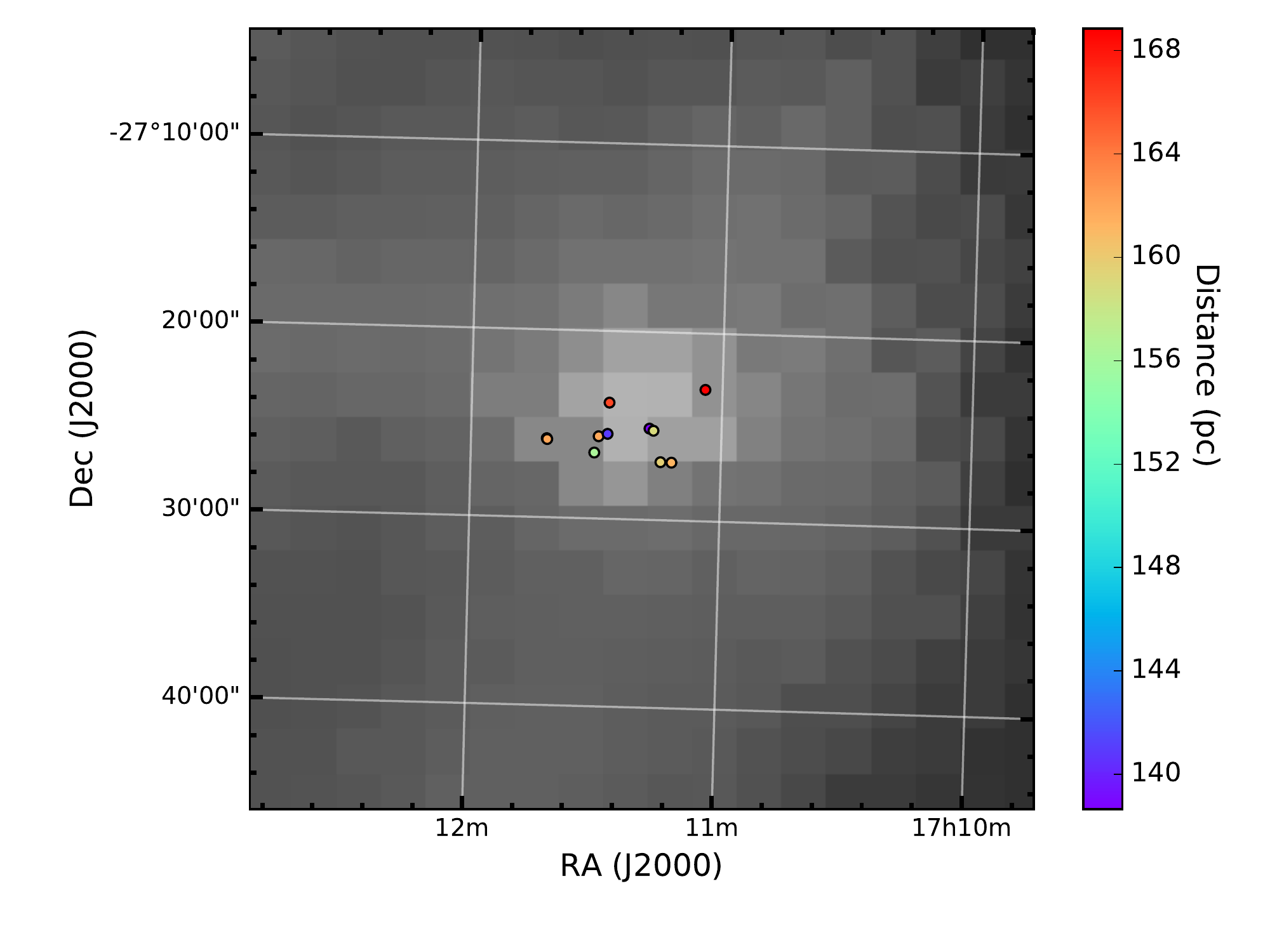} \\
  \includegraphics[height=0.33\textwidth, trim=0 0 40 0, clip]{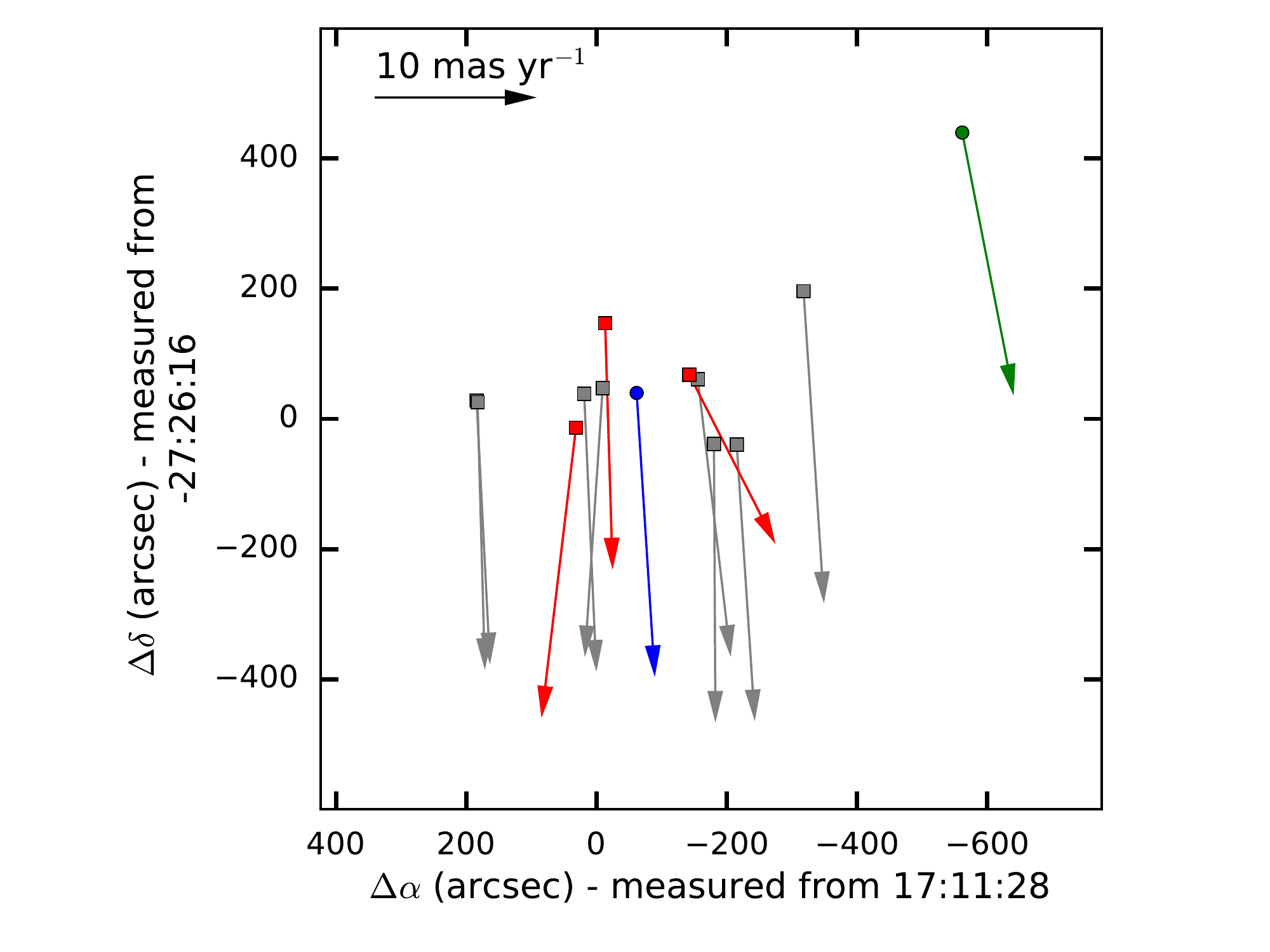} &
  \includegraphics[height=0.40\textwidth, trim=20 25 0 0, clip]{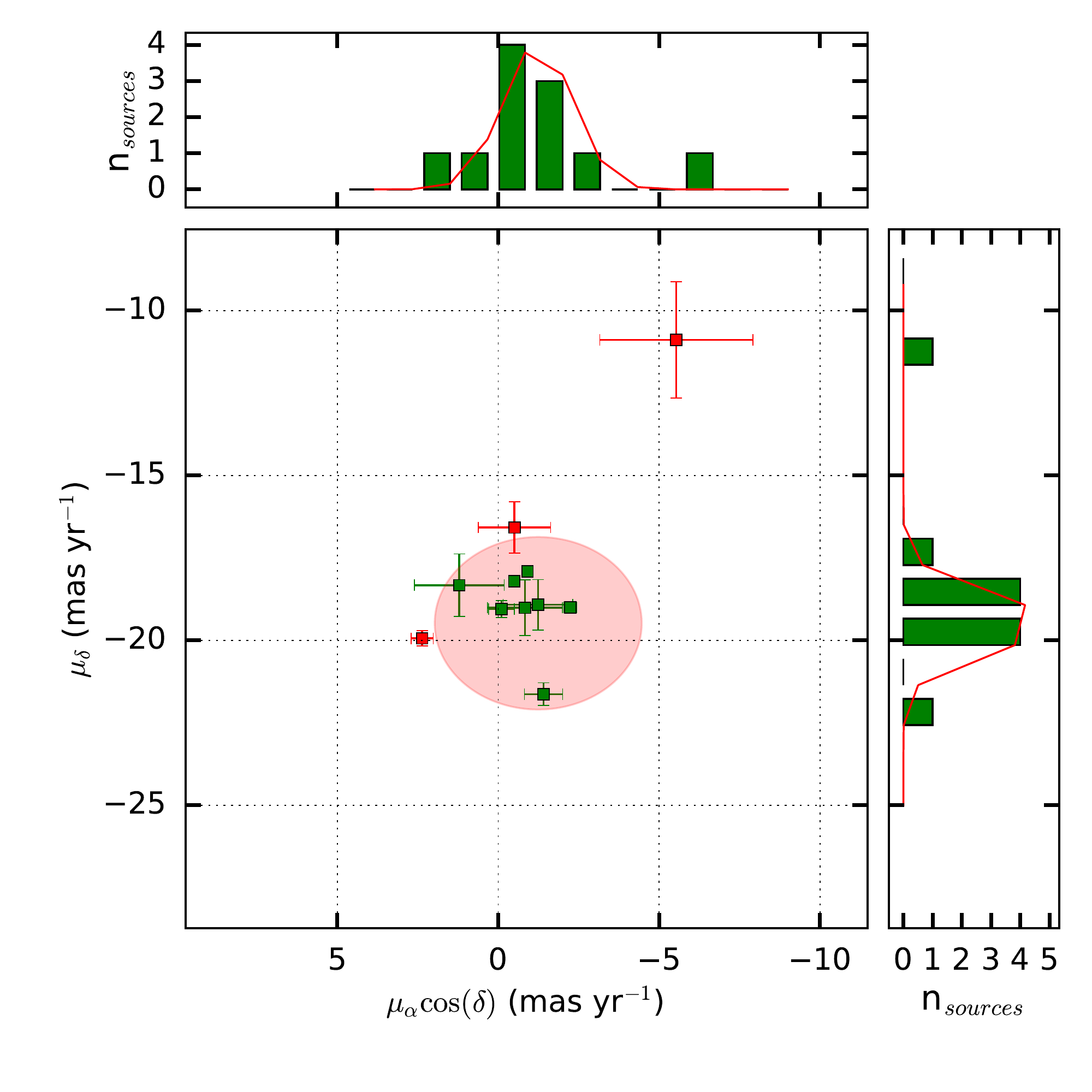}
  \end{tabular}
   \caption{{\it Top-left:} Histogram of individual trigonometric parallaxes to YSOs in the B59 region. 
   The red line is the Gaussian fit to the histogram.
    {\it Top-right:} Distribution of YSOs in the B59 region, colored by their corresponding
    distance from Gaia DR2 trigonometric parallaxes. {\it Bottom-left:} Proper motion vectors of YSOs in B59.
    In grey arrows we plot those proper motions that are within three times the dispersion, while the red 
    arrows are the outliers of the proper motion distribution. The blue arrow indicates the vector of the 
    mean proper motion of the stars in the B59 region (as given in Table~\ref{tab:DP}).  The green arrow
    indicate the reflex proper motion as due to the Solar motion at the Galactic position of B59.
    {\it Bottom-right:} Proper motion distribution of YSOs in B59. Red circle represents 
    three times the dispersion measurement constrain to identify YSOs with peculiar motions. Histograms and 
    gaussian fits (red lines) of the proper motions in R.A. (top) and Dec. (right), are also shown.}
   \label{fig:b59}
\end{figure*}

From the 22 YSOs and YSO candidates in this region, we find that 11 are in the Gaia DR2 catalog.
The mean trigonometric parallax of $6.12\pm0.19$~mas to the stars corresponds to a distance of 
$163\pm5$~pc for this region (see Table~\ref{tab:DP} and top panels of Figure~\ref{fig:b59}).
All the stars agree within  errors with this value, even the two with the larger trigonometric 
parallaxes of $7.2\pm1.1$ mas, and  7.1$\pm0.7$~mas. 

The mean distance to B59 seems to be larger than the earlier estimations, but they agree within the errors.
As the mean distance obtained by us is based on known YSOs in the region, we are more confident 
with it. This new distance also shows an improvement in the accuracy, that is now 3\%.

The proper motion of one of the stars, named [BHB2007]~7 or LkH$\alpha$~346 \citep{brooke2007}, indicates that 
it is also moving differently than the remaining YSOs, as it can be seen in the bottom panels
of Figure~\ref{fig:b59}. The proper motion of this star differs by about 10~mas~yr$^{-1}$ 
from the mean of the rest of the stars. This difference corresponds to about 8~km~s$^{-1}$ 
at a distance of 163 pc. This suggests that LkH$\alpha$~346 may be escaping from the region.

\subsection{Cepheus Flare}

\begin{figure*}[!htb]
   \centering
   \begin{tabular}{cc}
  \includegraphics[height=0.33\textwidth, trim=0 10 0 0, clip]{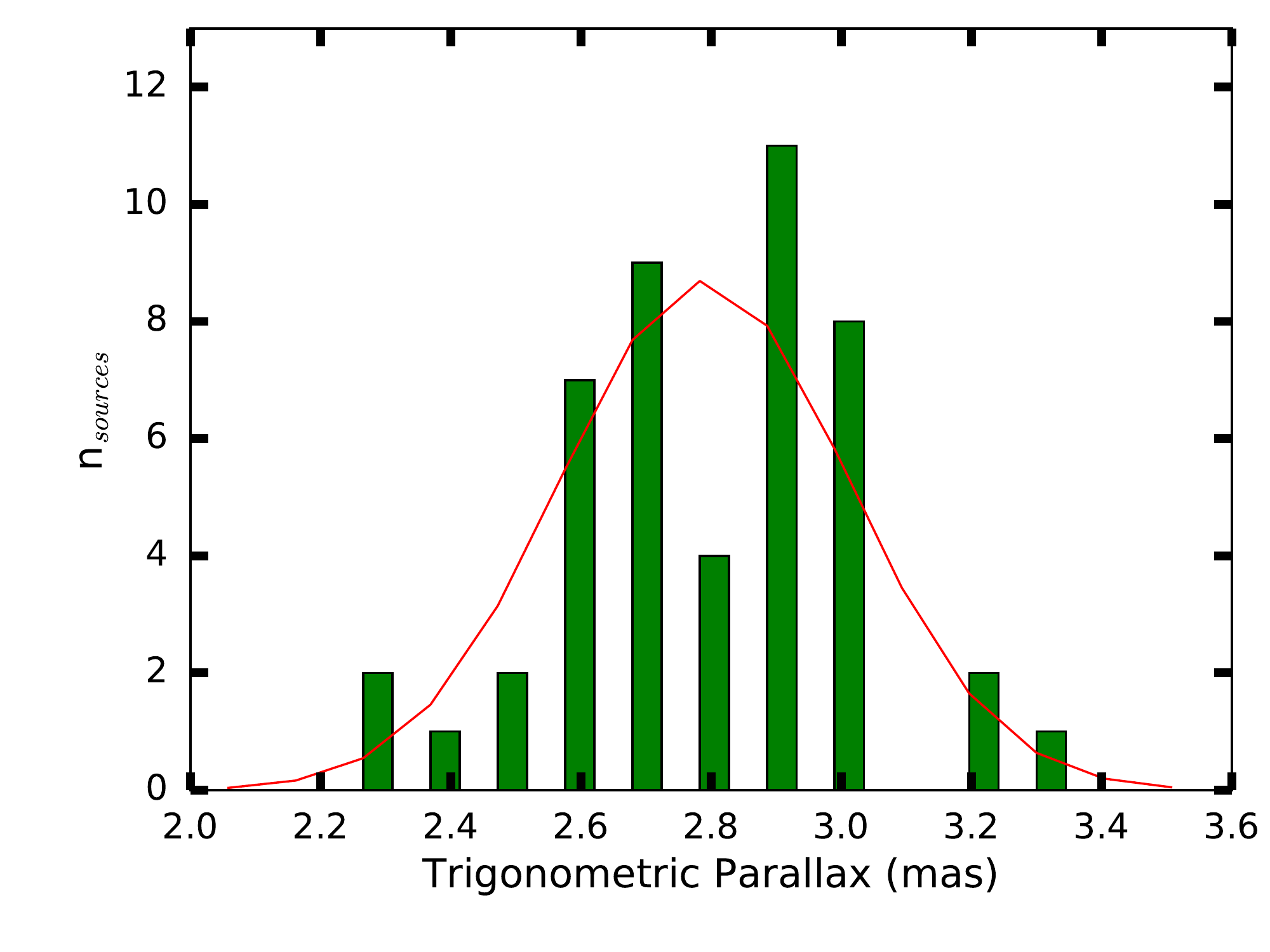} &
  \includegraphics[height=0.33\textwidth, trim=0 10 0 0, clip]{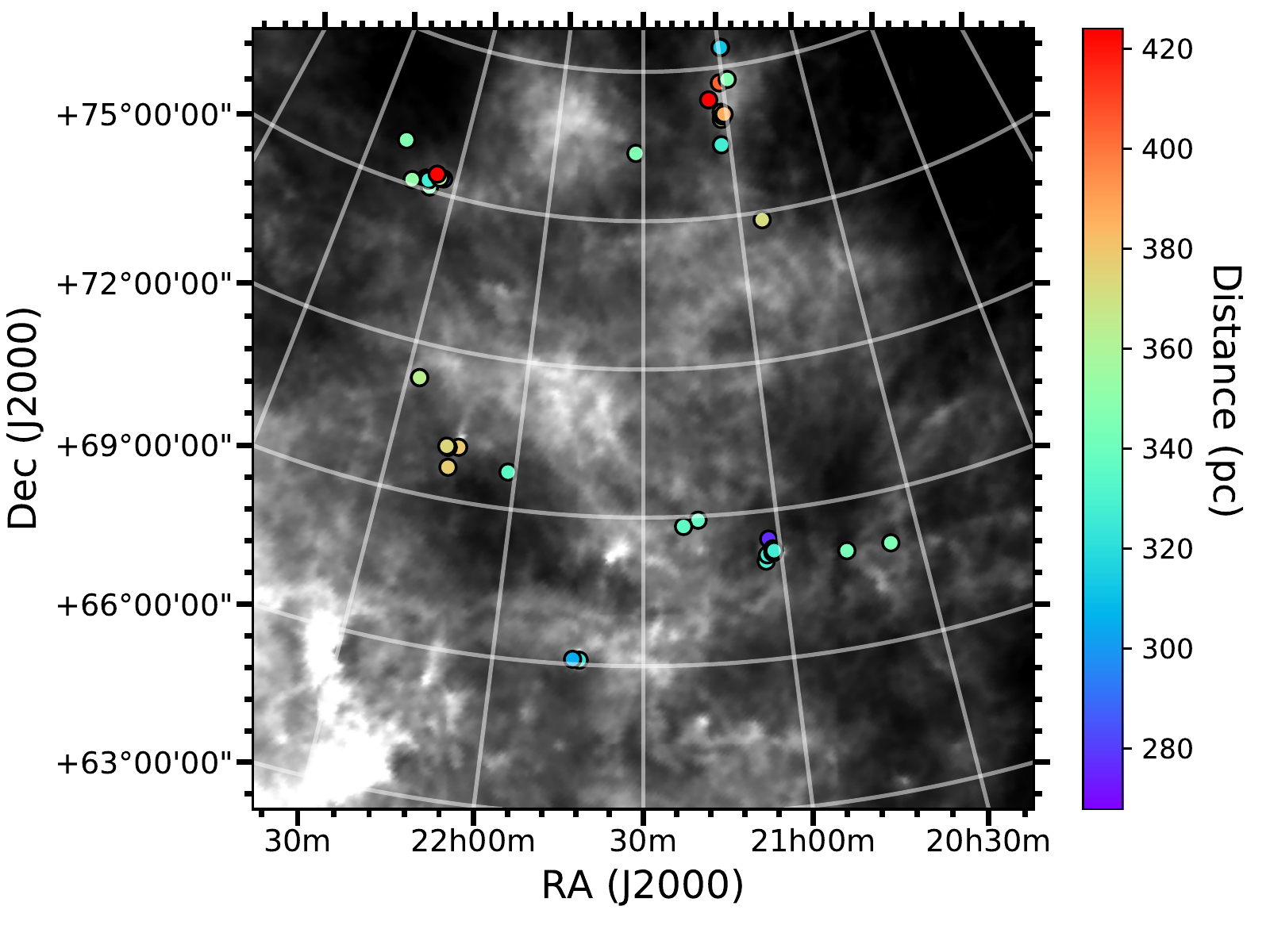}\\
  \includegraphics[height=0.33\textwidth, trim=0 0 40 0, clip]{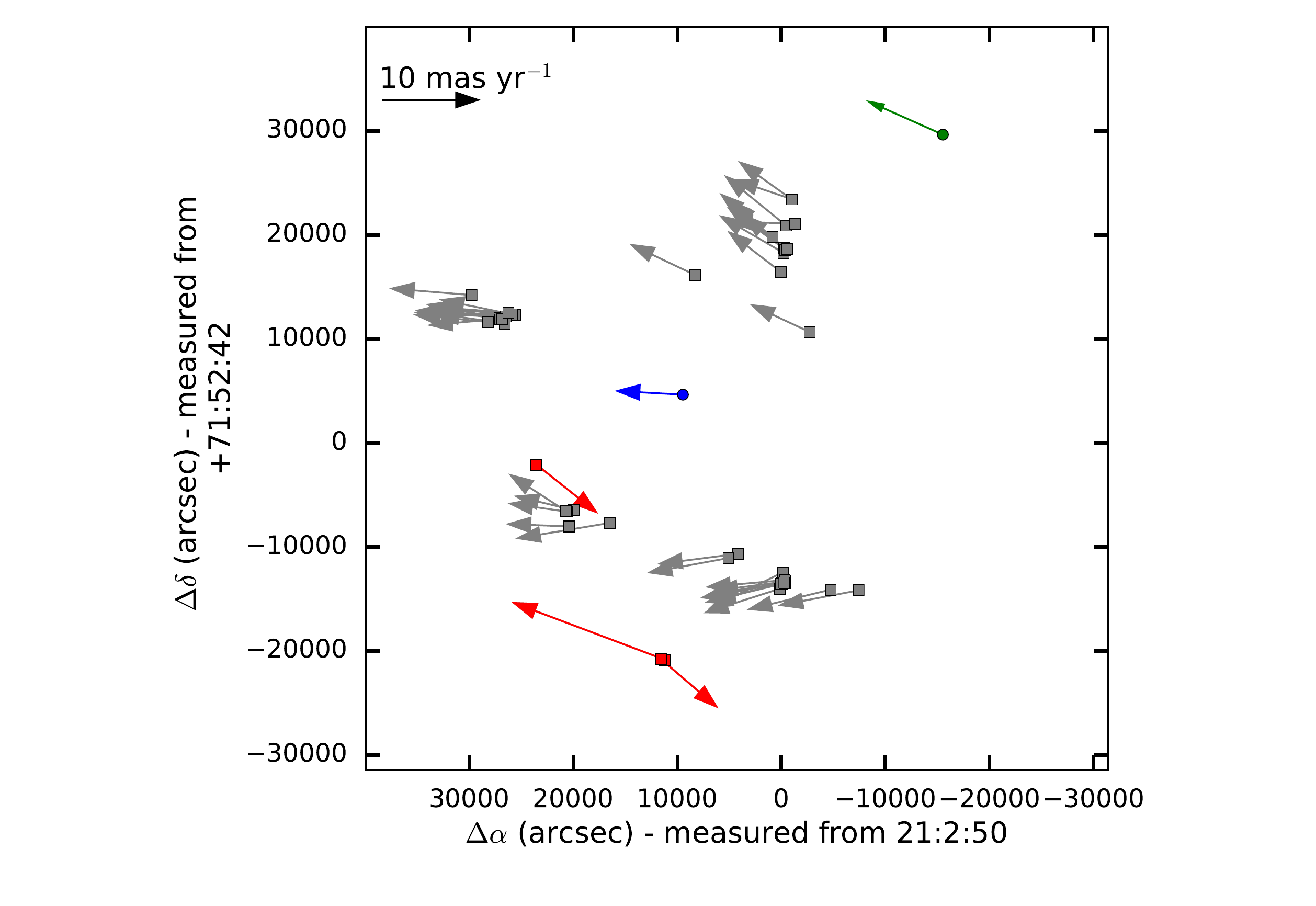} &
  \includegraphics[height=0.40\textwidth, trim=20 35 0 0, clip]{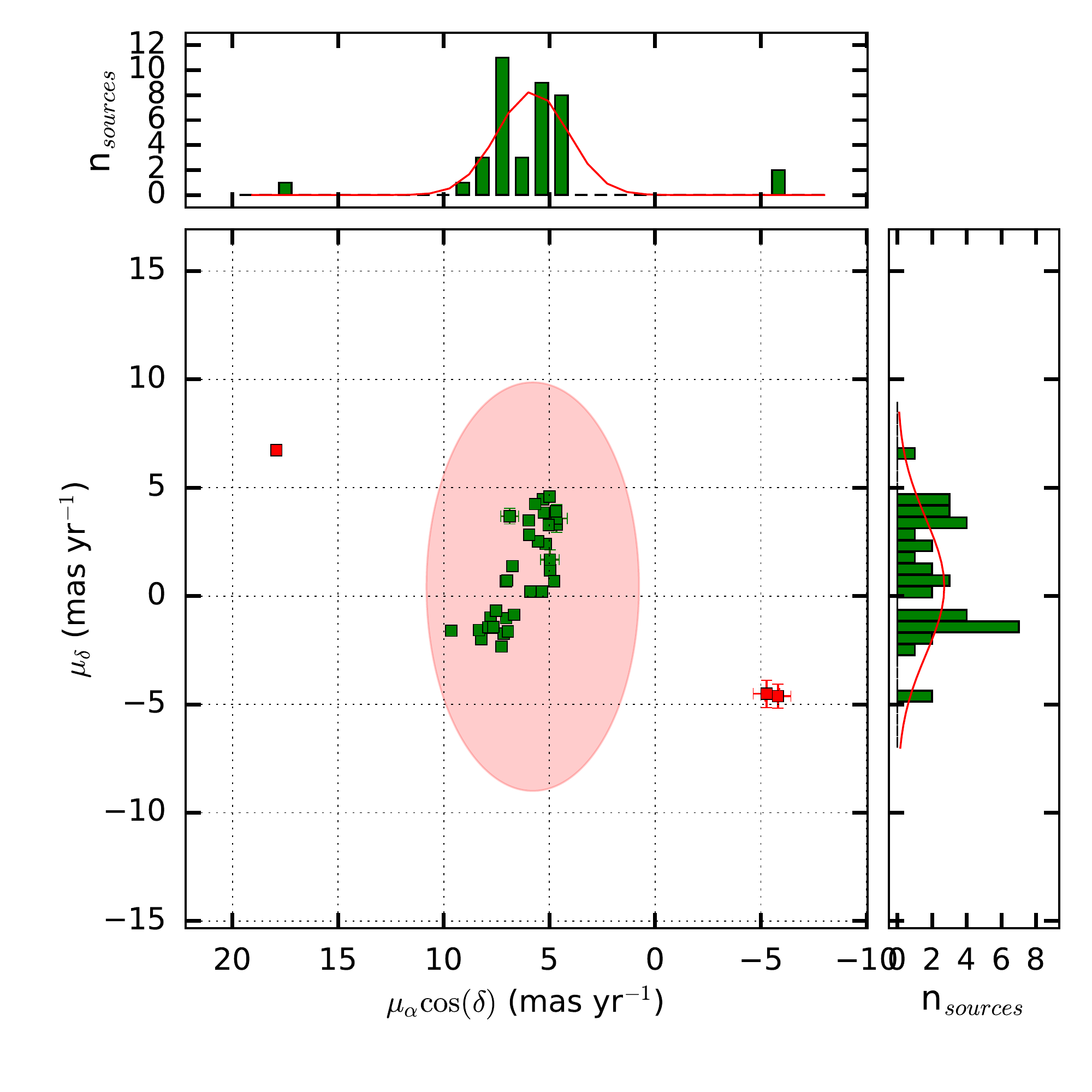}
  \end{tabular}
   \caption{Same as Figure~\ref{fig:b59}, but for the Cepheus flare region.}
   \label{fig:cep}
\end{figure*}

Despite being widely spread on the plane of the sky, the YSOs in the Cepheus flare show a good agreement
between parallax measurements and give a mean value of $2.78\pm0.25$~mas (see  top panels of Figure~\ref{fig:cep}). 
The corresponding distance is $360\pm32$~pc. This value is in the range of the distances, 300 to 500 pc, 
estimated for this complex based on previous determinations on the literature \citep[see][and references 
therein]{kun2008}. In the top-right panel of Figure~\ref{fig:cep}, we show
the spatial distribution of these YSOs colored by their corresponding distances. From this image,
we can see that the YSOs to the north of the complex seem to be  farther than those
in the south. 

The proper motion dispersions are large compared to the other regions (see Table~\ref{tab:kin} 
and Figure~\ref{fig:cep}). This could reflect the fact that the YSOs were formed in different 
clouds within the complex, and thus with slightly different kinematics. 

There are only three stars that  are moving well beyond the mean proper motion of the other stars,
especially in the R.A. direction. These stars are [KBK2009b]~119, 2MASS~J21440537+6605531, and 
BD+65$^{\circ}$1636. The last two objects are projected in the direction of NGC~7129 (see next section).
However, their determined trigonometric parallaxes ($3.3\pm0.4$~mas and $3.09\pm0.04$~mas for 2MASS~J21440537+6605531 and 
BD+65$^{\circ}$1636, respectively) show that they are at a much shorter distance and do not belong to NGC~7129.

\subsection{Cepheus - NGC 7129}

\begin{figure*}[!th]
   \centering
   \begin{tabular}{cc}
  \includegraphics[height=0.33\textwidth, trim=10 0 0 0, clip]{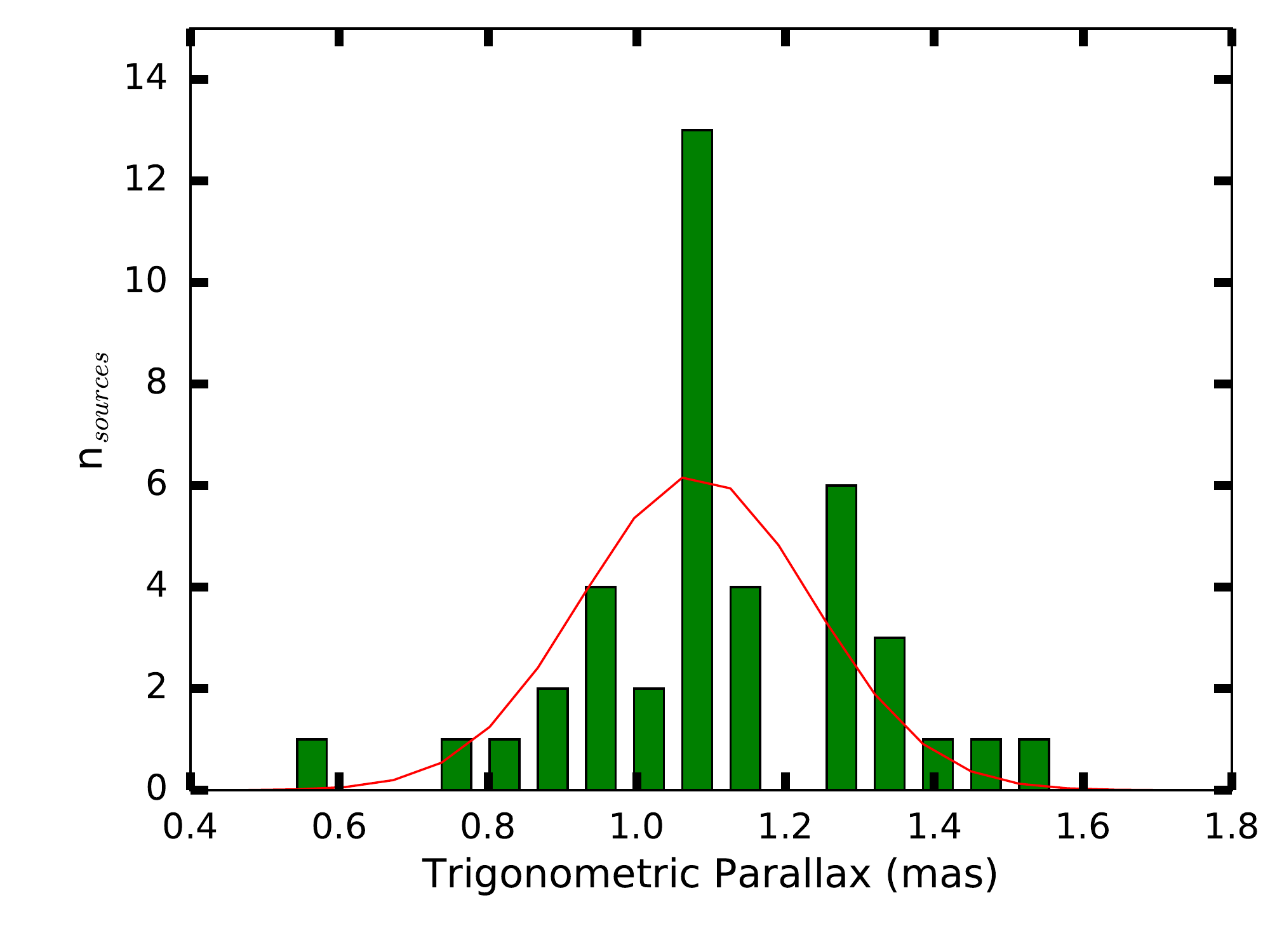} &
  \includegraphics[height=0.33\textwidth, trim=0 10 0 0, clip]{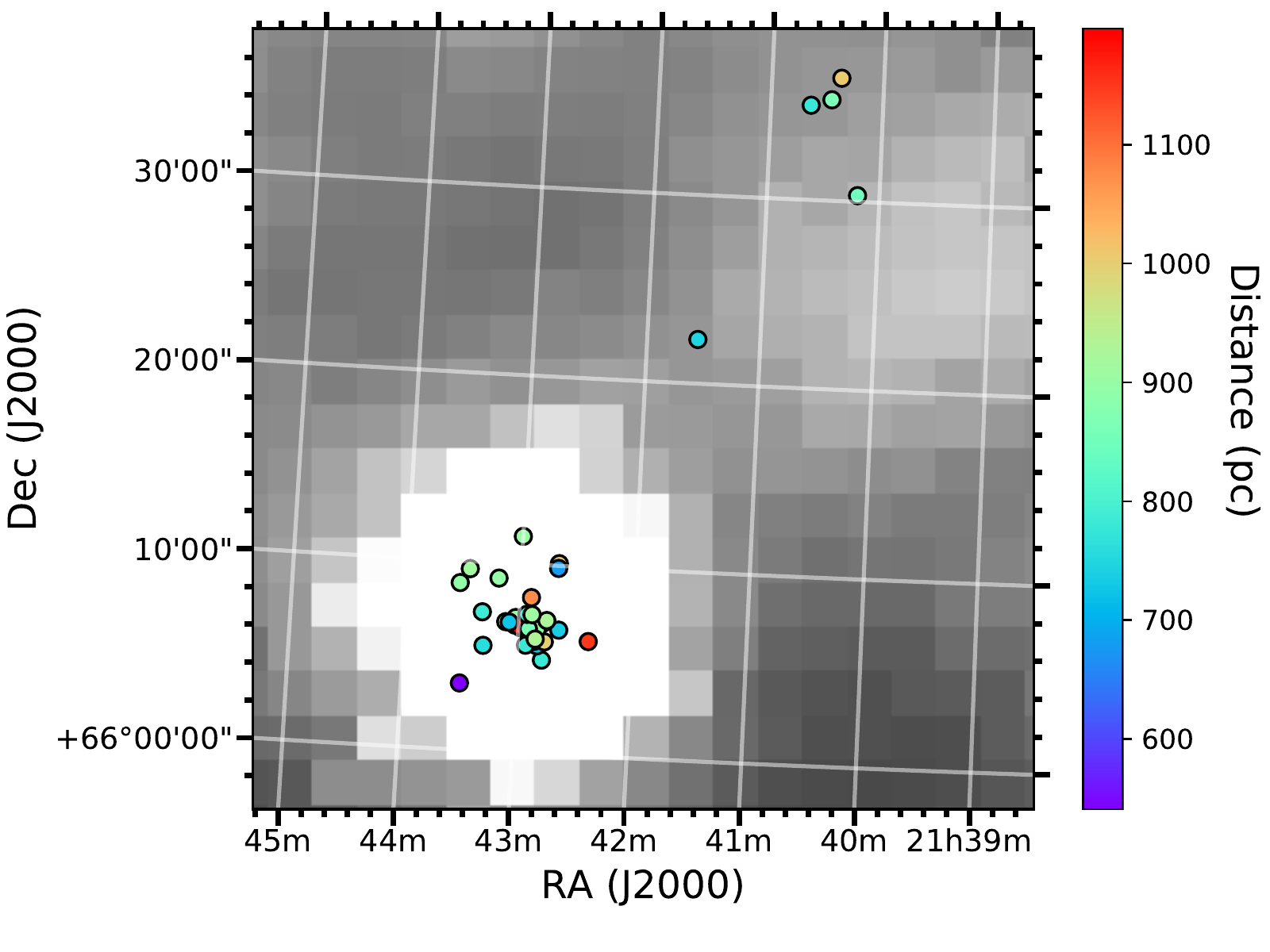}\\
  \includegraphics[height=0.33\textwidth, trim=0 0 40 0, clip]{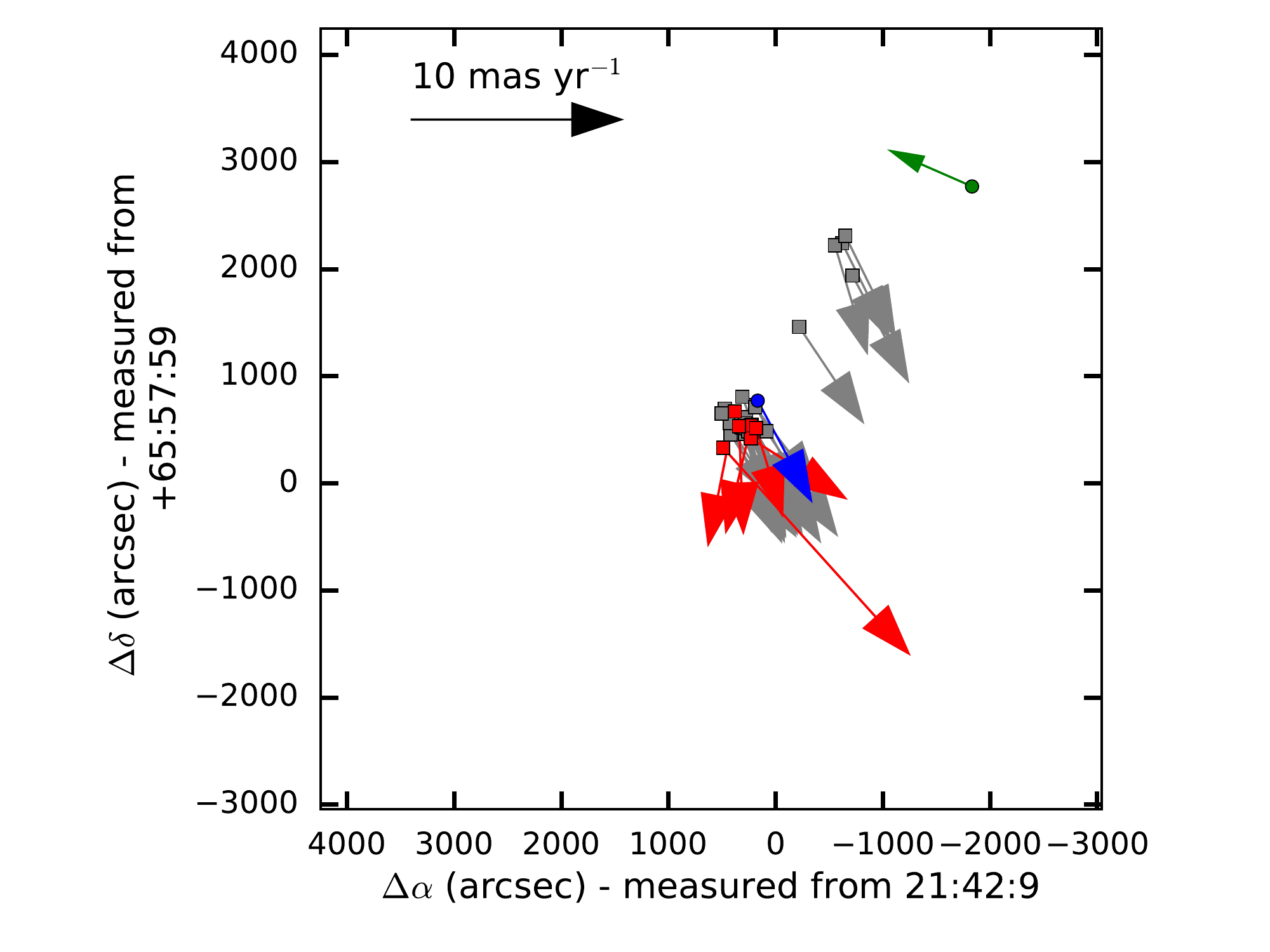} &
  \includegraphics[height=0.40\textwidth, trim=20 35 0 0, clip]{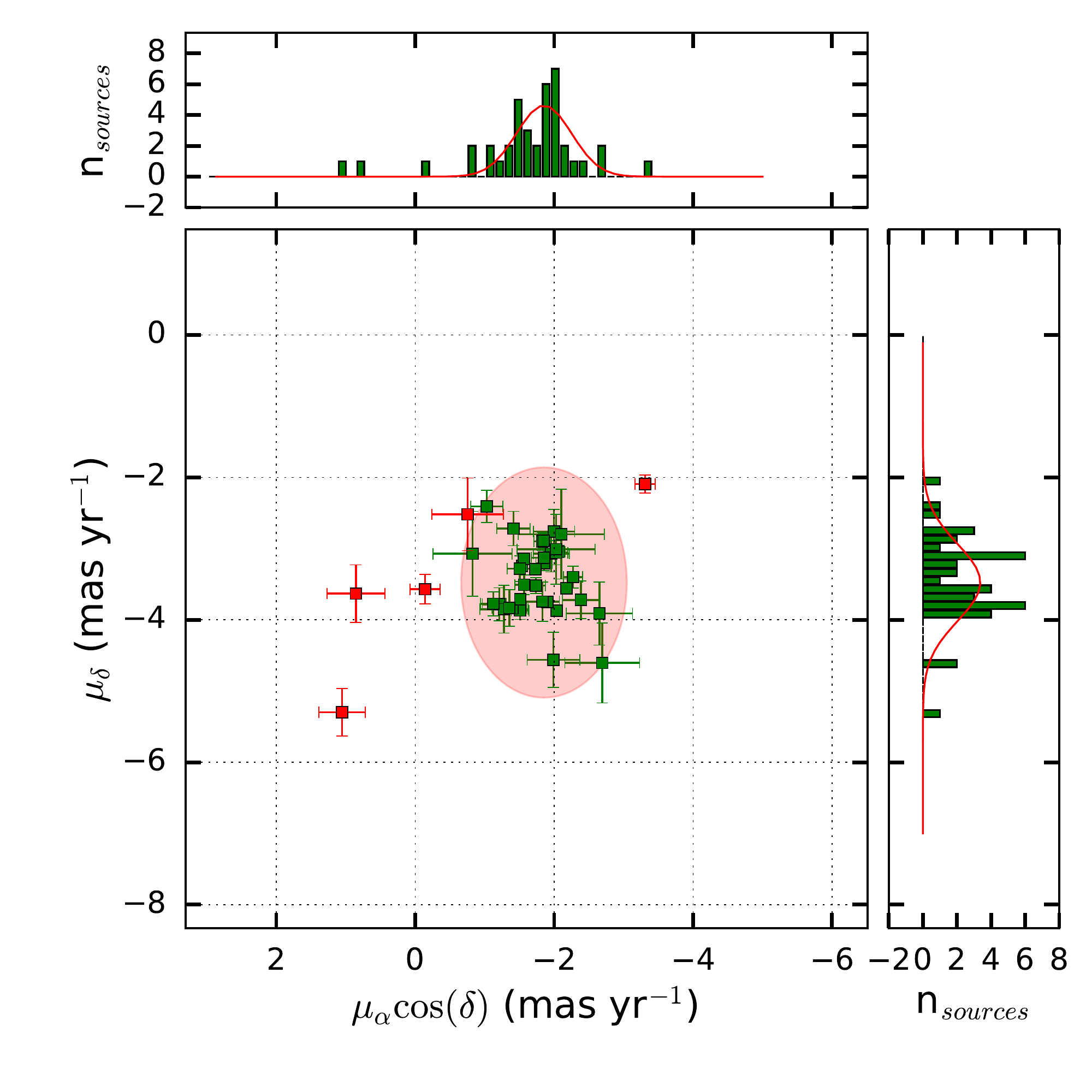}
  \end{tabular}
   \caption{Same as Figure~\ref{fig:b59}, but for the Cepheus - NGC 7129 region.}
   \label{fig:cepf}
\end{figure*}

Almost half of the YSO population in the Cepheus flare with astrometric results
in the Gaia DR2 catalog are concentrated in the NGC~7129 region.  The mean trigonometric 
parallax of the region of $1.08\pm0.19$~mas confirms that the distance to this region,
$926\pm163$~pc (see top panels of Figure~\ref{fig:cepf}), is larger than that of the Cepheus 
flare located at $360\pm32$~pc. Thus, it does not belong to this complex. 

The proper motions of the stars in NGC 7129 are well concentrated around $\mu_\alpha\,=\,-1.89\pm0.02$~mas~yr$^{-1}$, 
and $\mu_\delta\,=\,-3.51\pm0.05$~mas~yr$^{-1}$. As can be seen in the bottom panels
of Figure~\ref{fig:cepf}, four cases are outside the $\pm3\sigma_{\mu}$ from the mean and 
one is consistent with within errors. These objects are 2MASS~J21424705+6604578, 2MASS~J21424687+6606574, 
2MASS J21431161 +6609114, and 2MASS J21430502+6606533. 
In compact clusters such as NGC~7129, fast moving stars may be produced by $n$-body interactions 
\citep{poveda1967}, as has been observed in Orion \citep[e.g.,][]{dzib2017}.
In NGC~7129 this may be the case as well.

\subsection{Chamaeleon}

\begin{figure*}[!th]
   \centering
   \begin{tabular}{cc}
  \includegraphics[height=0.33\textwidth, trim=0 0 0 0, clip]{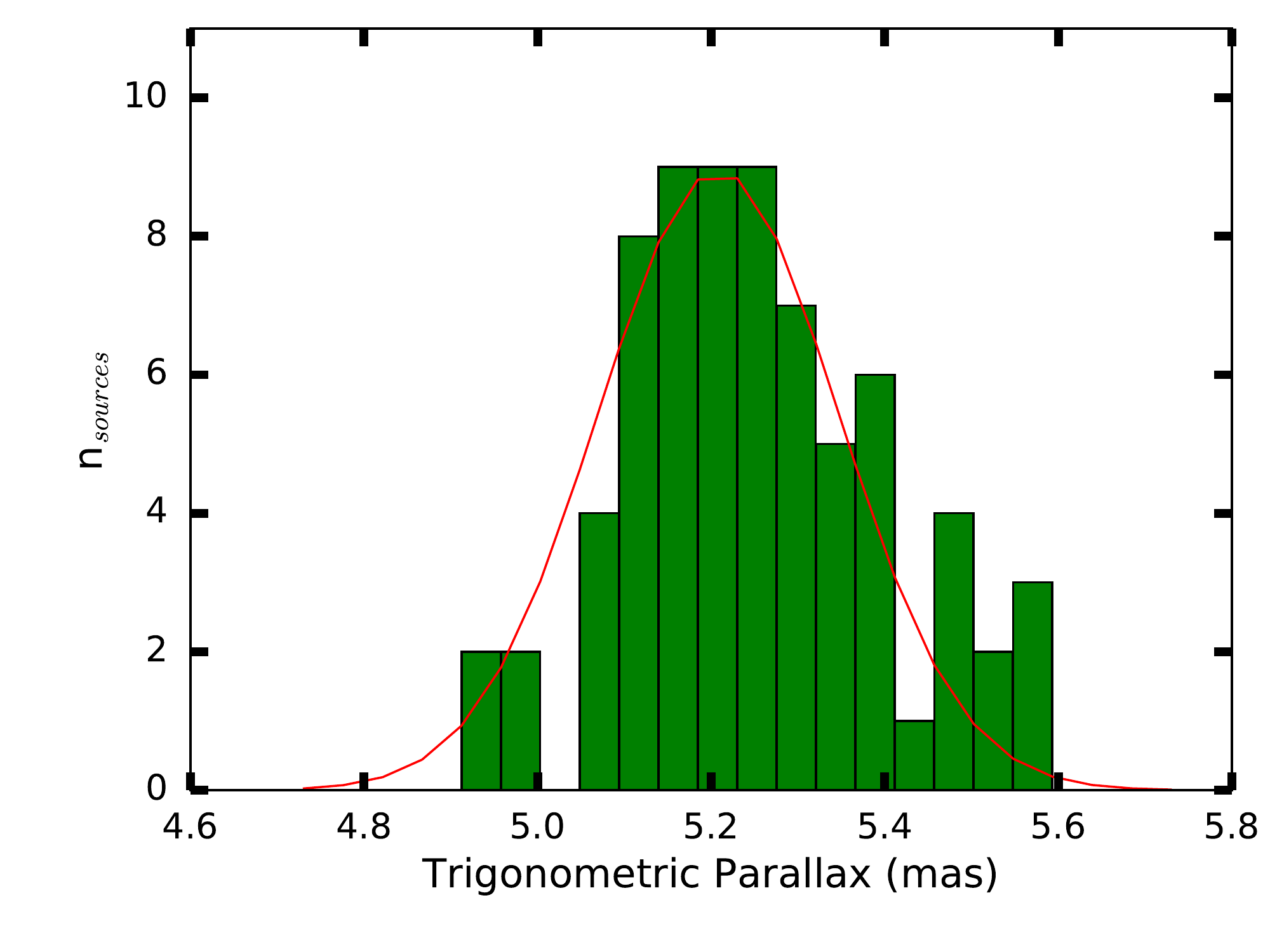} &
  \includegraphics[height=0.33\textwidth, trim=0 10 0 0, clip]{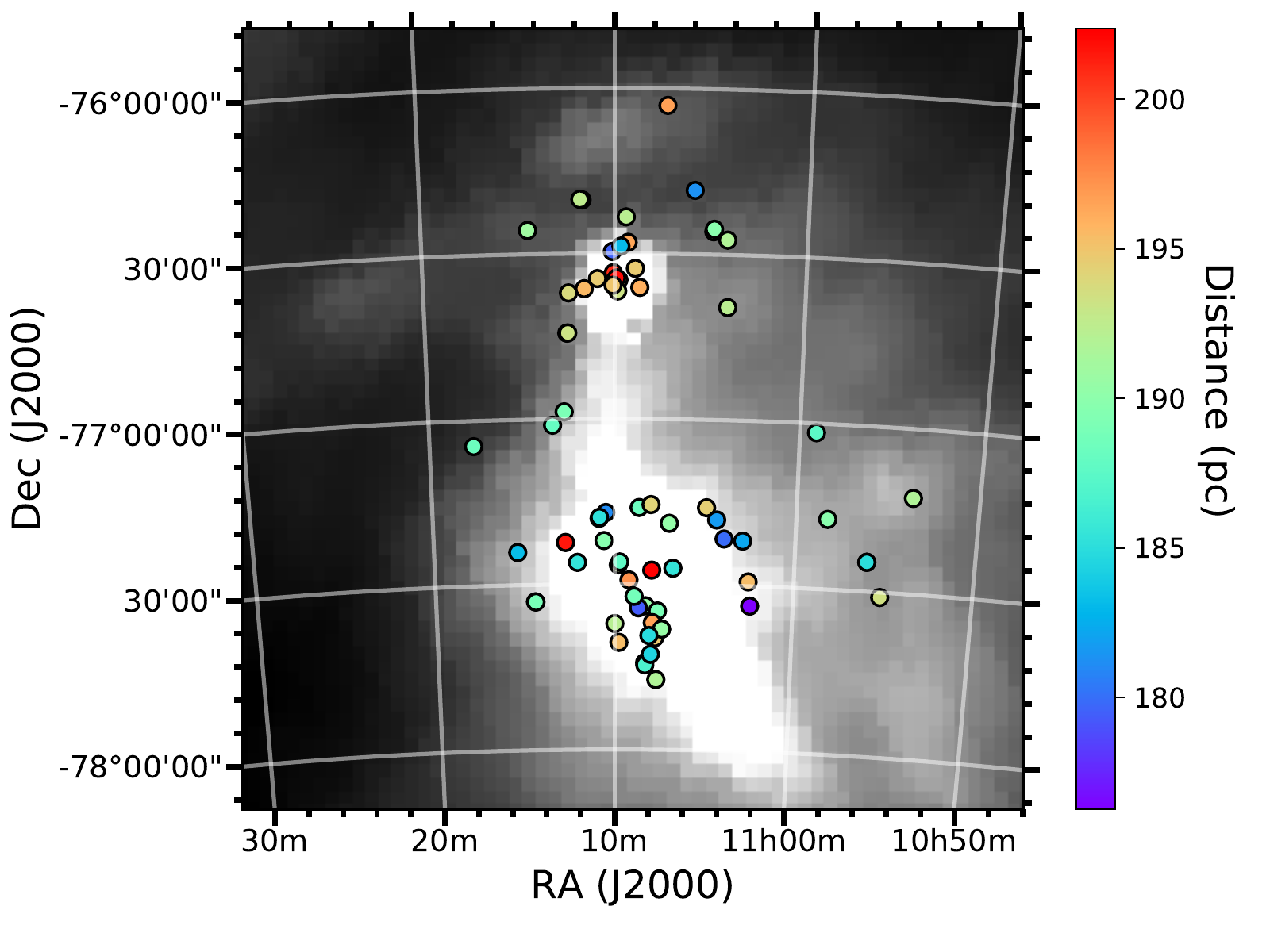}\\
  \includegraphics[height=0.33\textwidth, trim=0 0 40 0, clip]{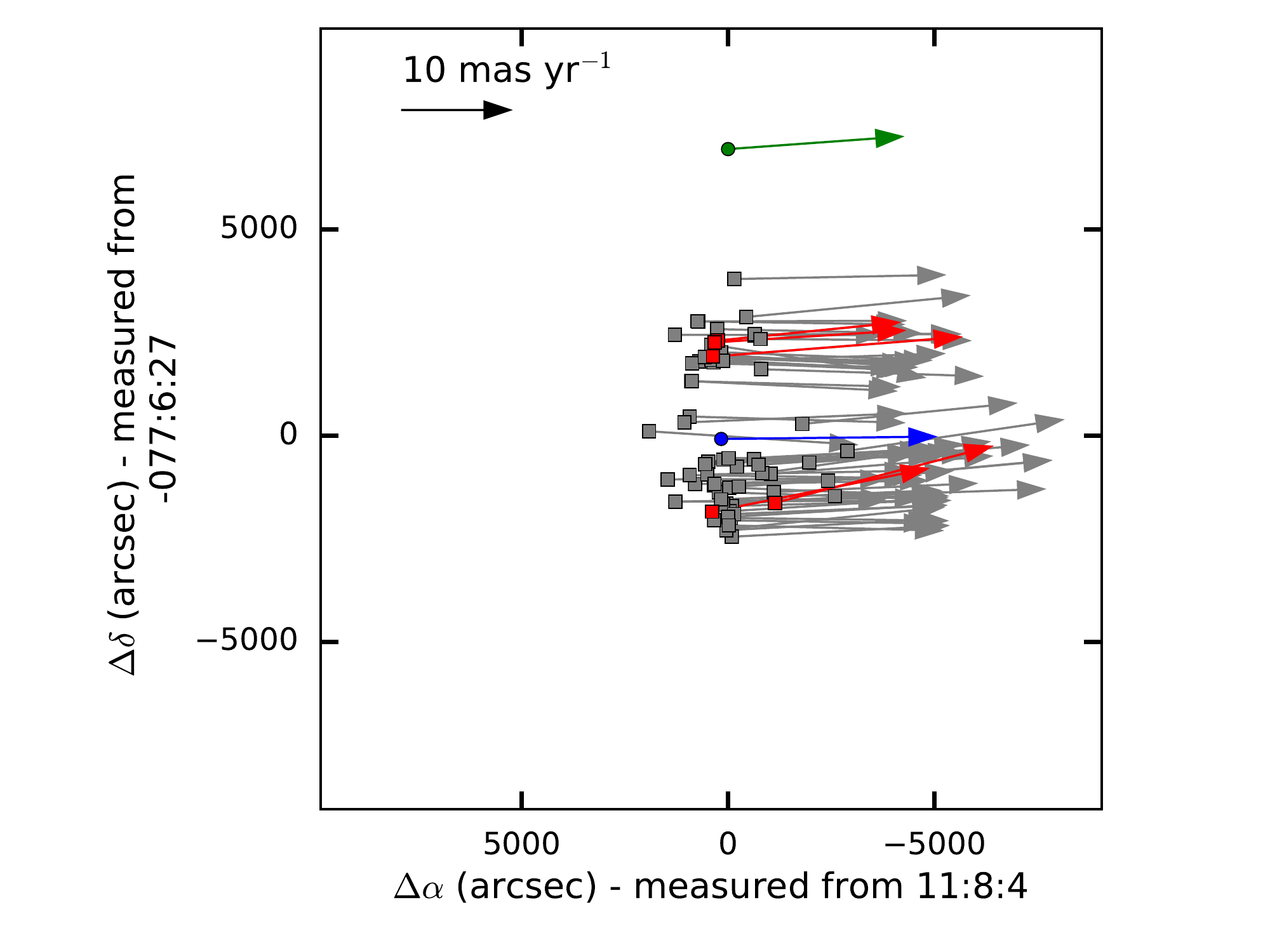} &
  \includegraphics[height=0.40\textwidth, trim=0 35 0 0, clip]{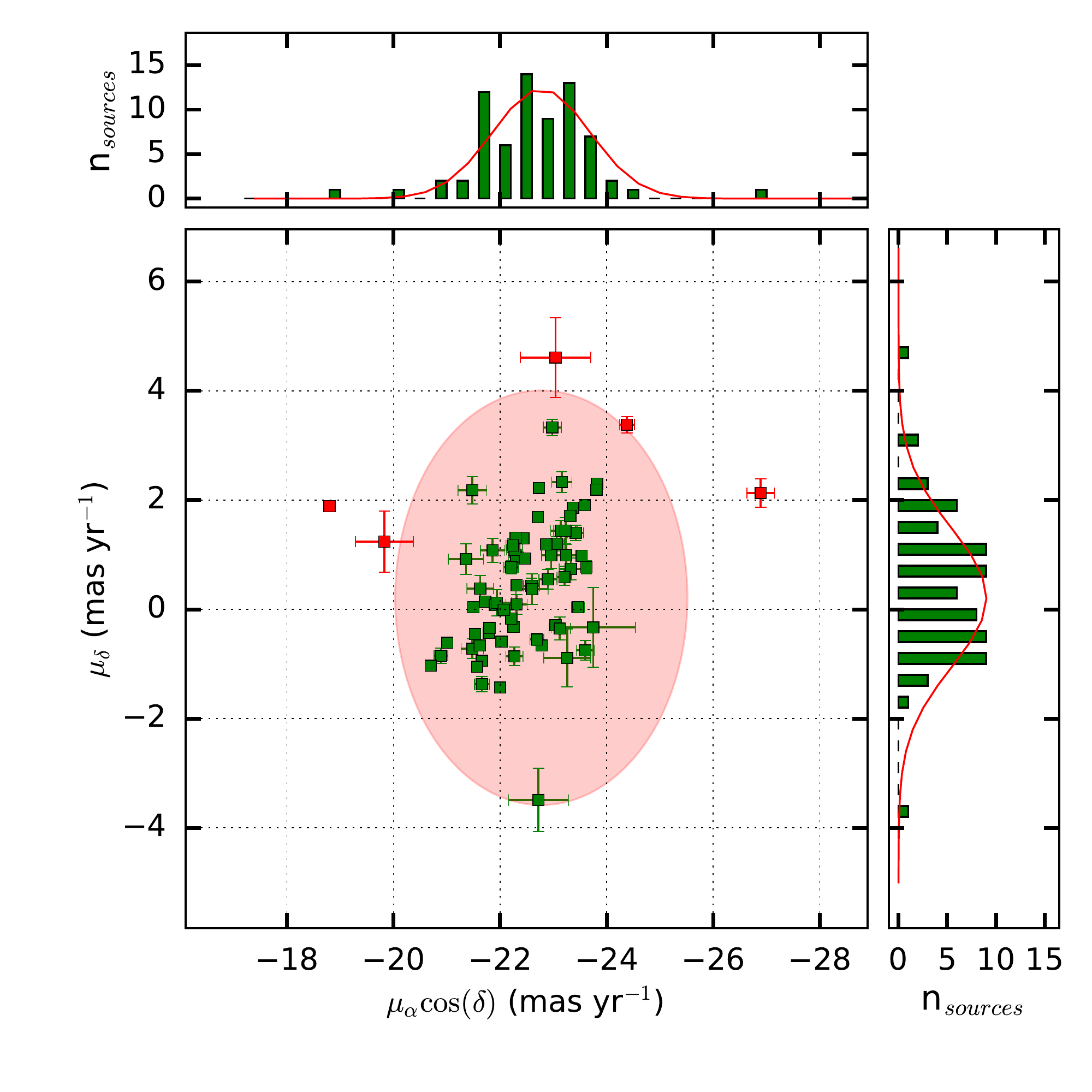}
  \end{tabular}
   \caption{Same as Figure~\ref{fig:b59}, but for Chamaeleon I.}
   \label{fig:chai}
\end{figure*}

\begin{figure*}[!th]
   \centering
   \begin{tabular}{cc}
  \includegraphics[height=0.33\textwidth, trim=0 0 0 0, clip]{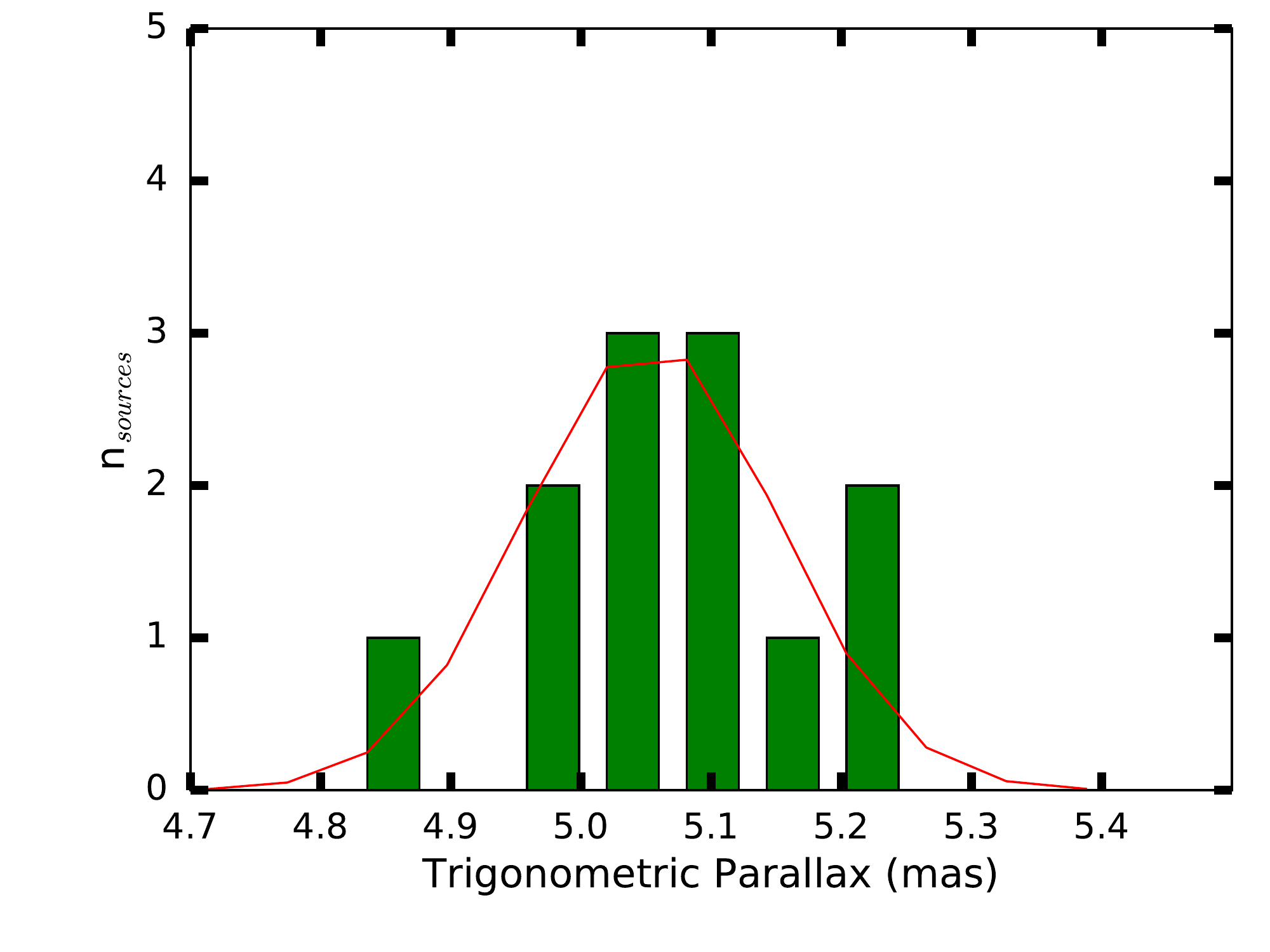} &
  \includegraphics[height=0.33\textwidth, trim=0 10 0 0, clip]{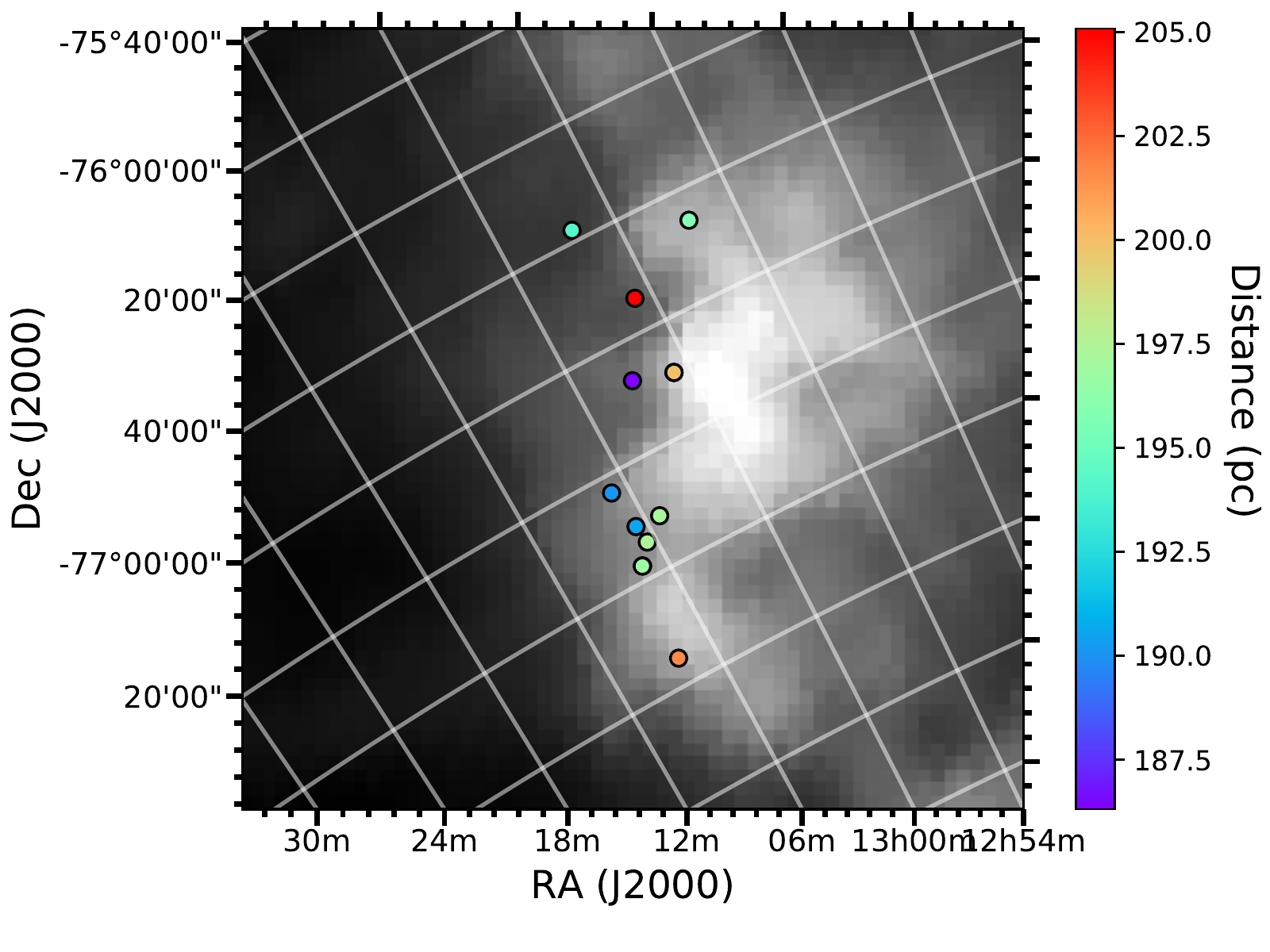}\\
  \includegraphics[height=0.33\textwidth, trim=0 0 40 0, clip]{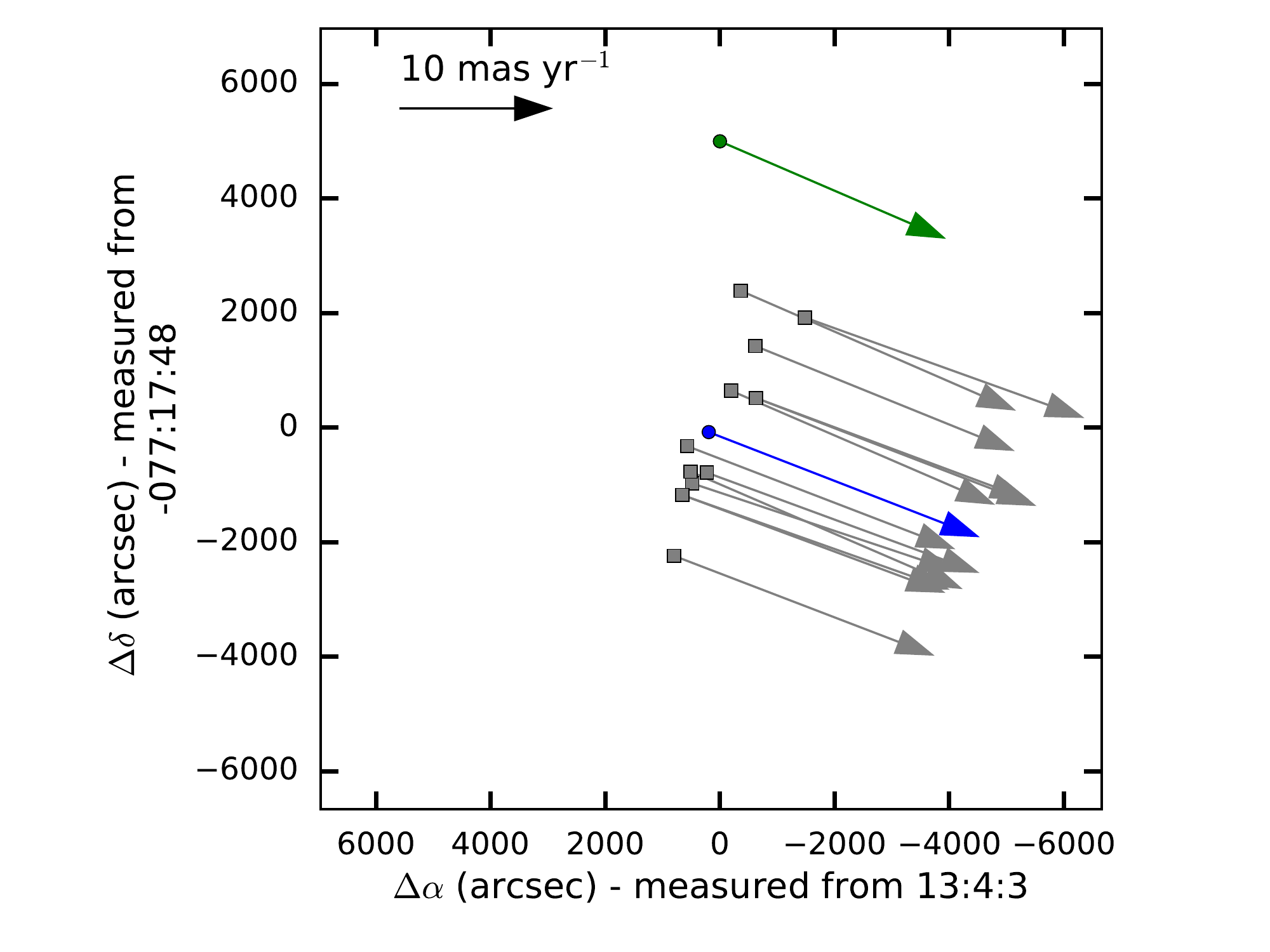} &
  \includegraphics[height=0.40\textwidth, trim=0 35 0 0, clip]{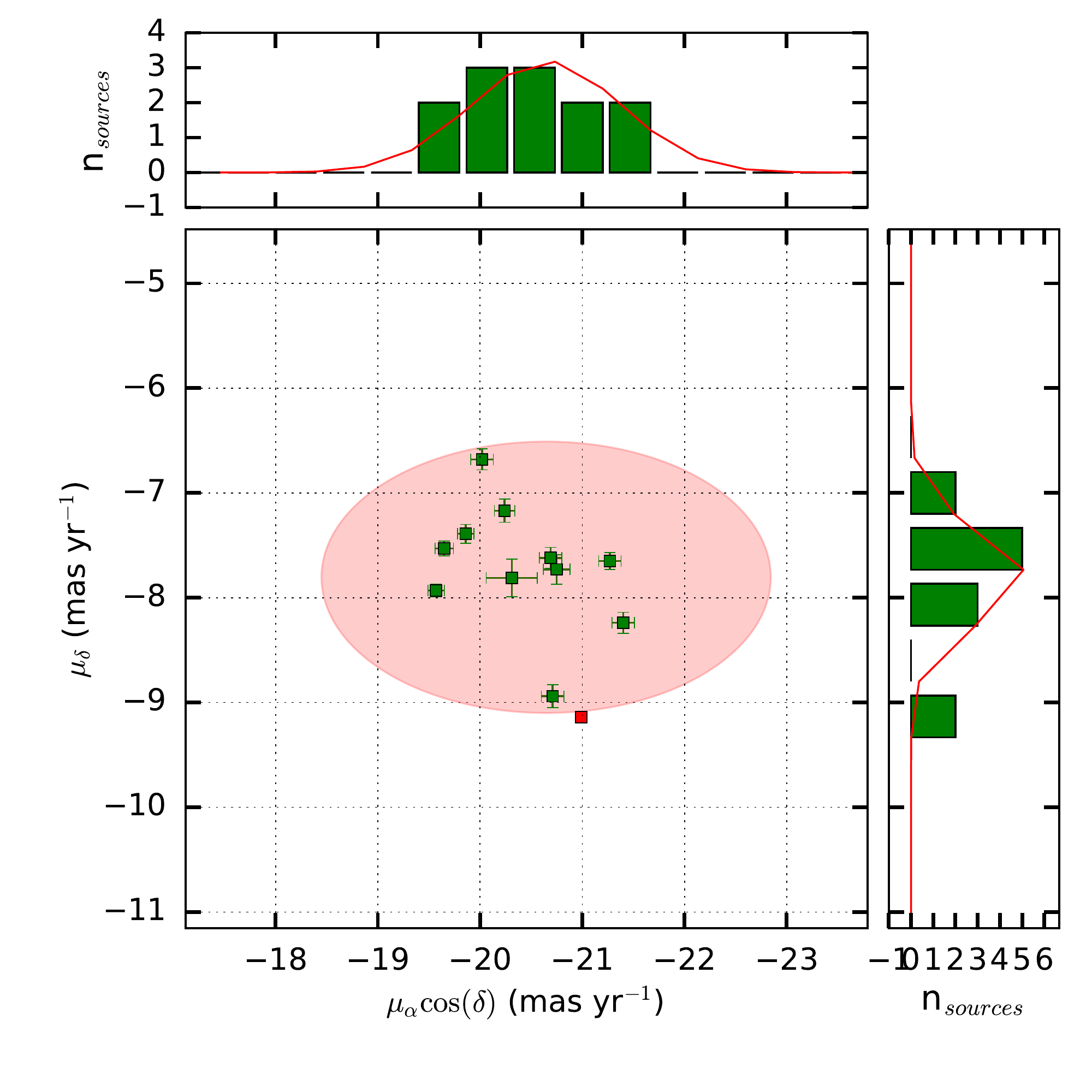}
  \end{tabular}
   \caption{Same as Figure~\ref{fig:b59}, but for Chamaeleon II.}
   \label{fig:chaii}
\end{figure*}

\begin{figure*}[!th]
   \centering
   \begin{tabular}{cc}
  \includegraphics[height=0.33\textwidth, trim=0 0 0 0, clip]{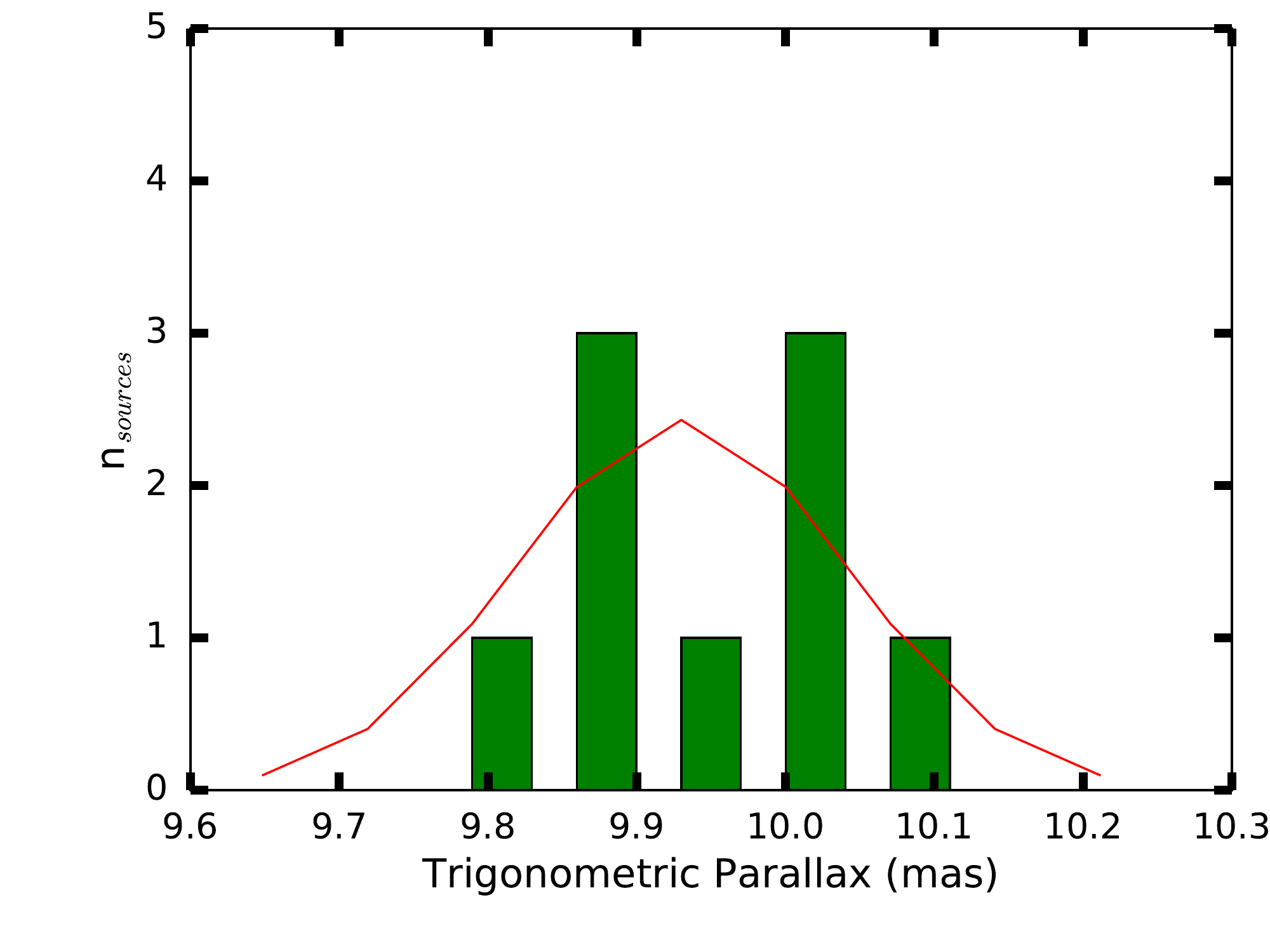} &
  \includegraphics[height=0.33\textwidth, trim=0 10 0 0, clip]{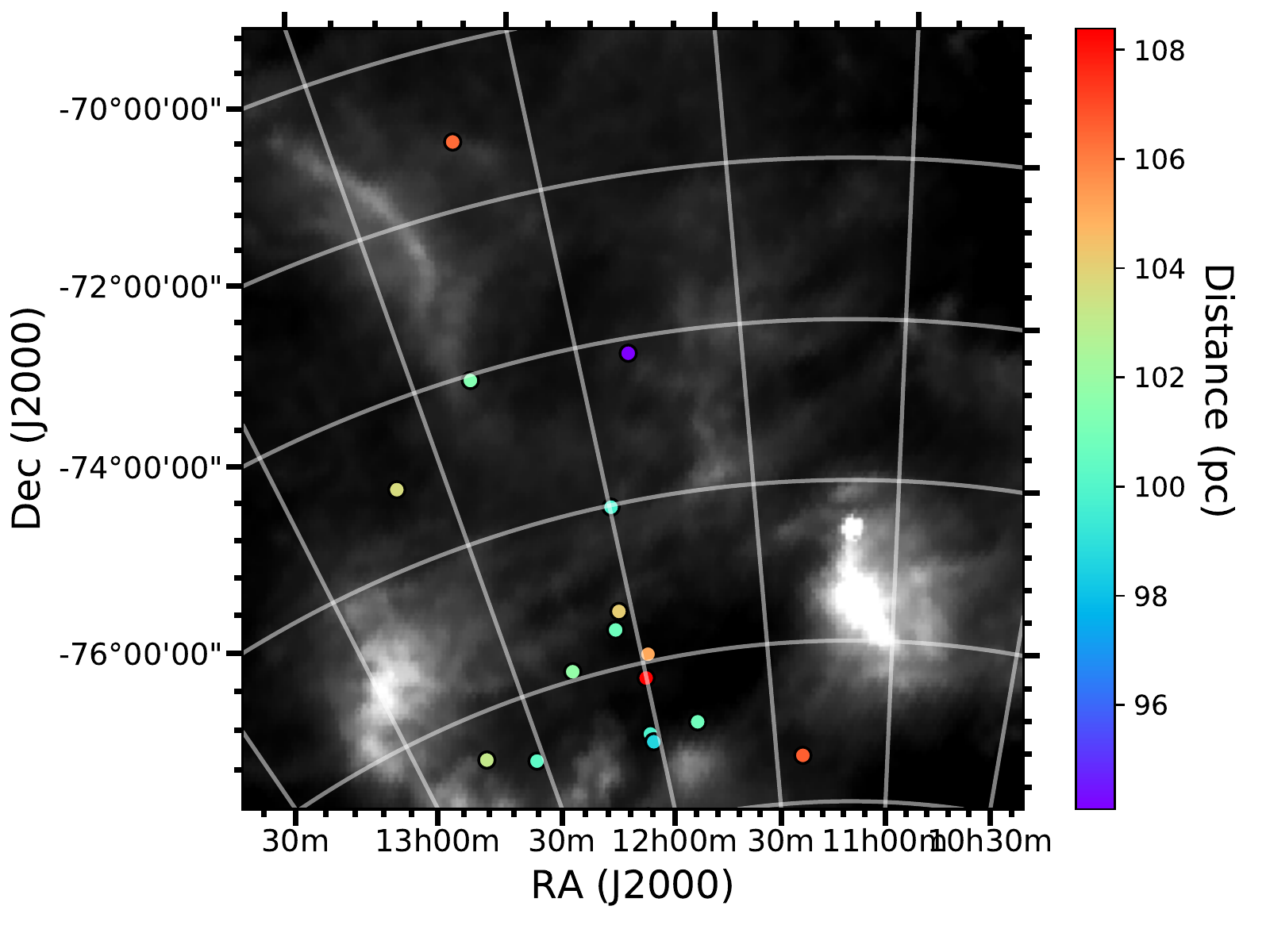}\\
  \includegraphics[height=0.33\textwidth, trim=0 0 40 0, clip]{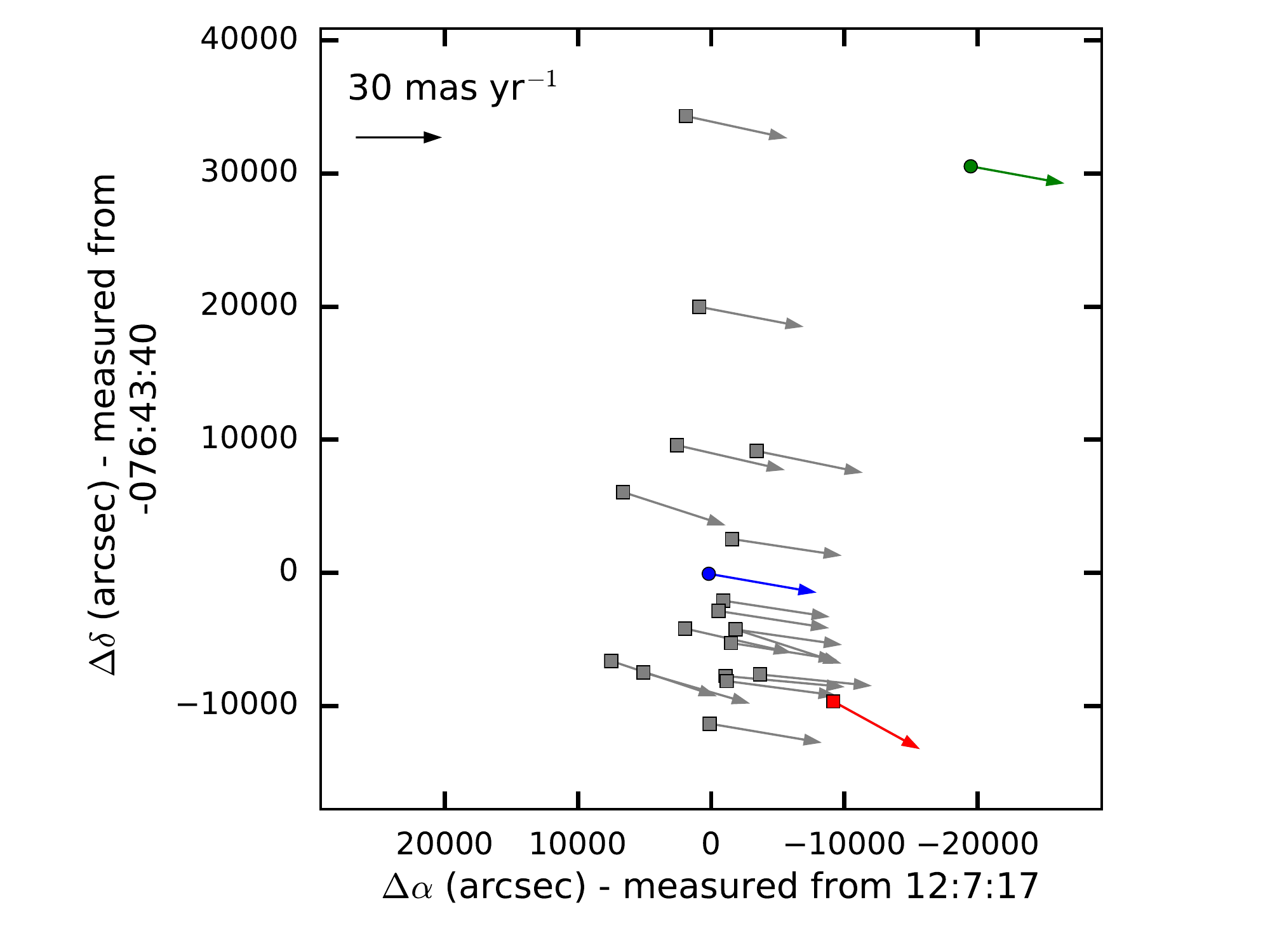} &
  \includegraphics[height=0.40\textwidth, trim=0 35 0 0, clip]{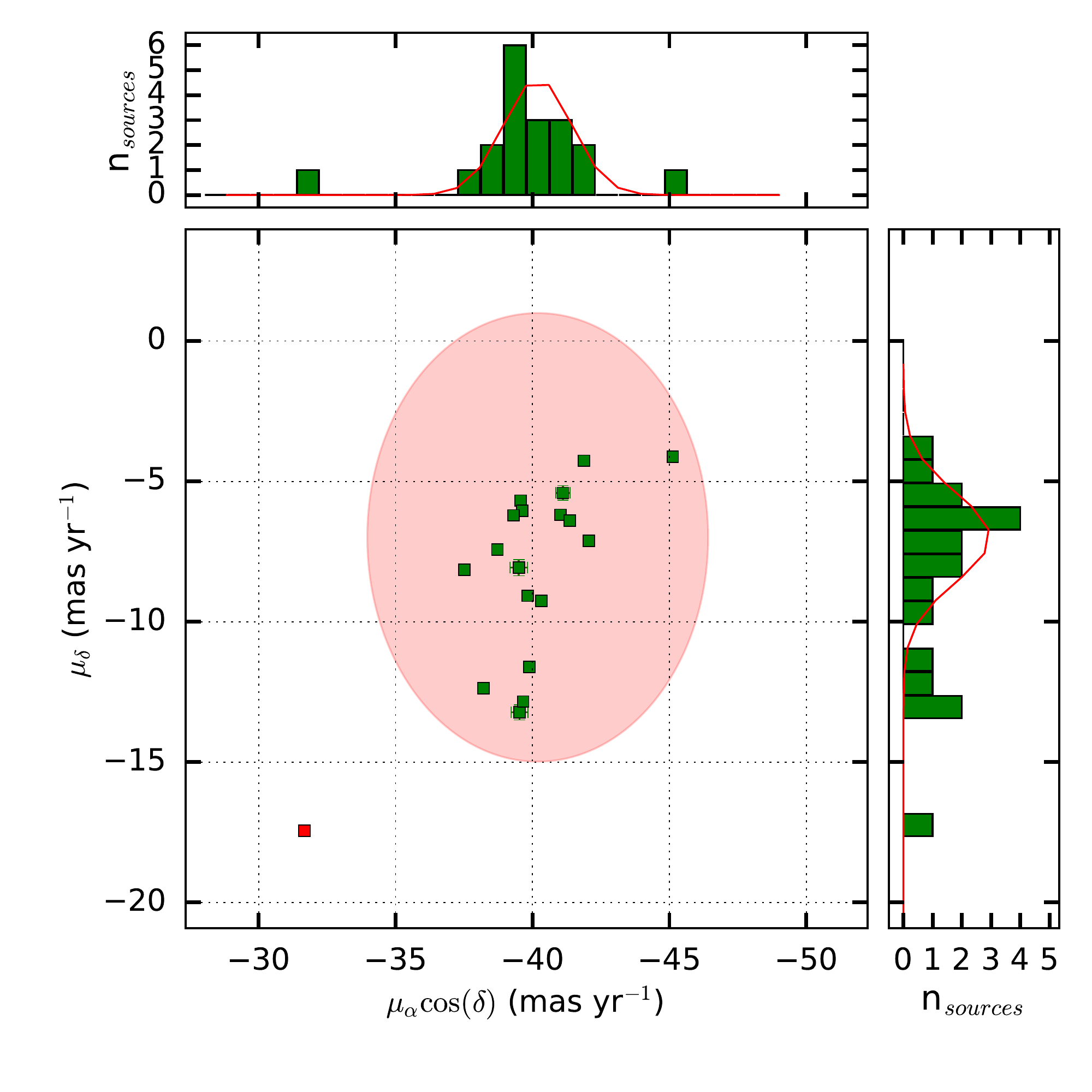}
  \end{tabular}
   \caption{Same as Figure~\ref{fig:b59}, but for $\epsilon$ Chamaeleontis. Red circle in the bottom right panel
   in this case is five times the proper motion dispersion measurement.}
   \label{fig:chan}
\end{figure*}

Most of the star formation in the Chamaeleon complex is occurring in the Cha~I cloud,
and we found that 71 of the YSOs in this cloud are in the Gaia DR2 catalog. We also found 13
YSOs in Cha~II and 19 dispersed YSOs that belong to $\epsilon$ Cha.
The trigonometric parallaxes and proper motions of the YSOs of each individual region have low dispersion
(see Tables~\ref{tab:DP} and~\ref{tab:kin} and Figures~\ref{fig:chai}, \ref{fig:chaii}, and~\ref{fig:chan}),
as expected if they are part of the same group of stars.

Our results also show that the two main clouds of the complex are at similar distances as it has been 
found previously. These distances agree within errors with those recently found by \cite{voirin2018},
but their errors are slightly larger. The distance to the group $\epsilon$ Cha is shorter than earlier estimations
\citep[i.e., 114~pc;][]{feigelson2003}.

The mean proper motions of Cha~I and Cha~II show that they are moving in slightly different directions.
The mean proper motions in right ascension are similar for objects in both clouds, but not in the 
declination direction. Also, while all the proper motions of YSOs in the Cha~II are inside the 3$\sigma_{\mu}$  from the mean,
in Cha~I three objects show peculiar motions: [CCE98]~1-74, Ass~Cha~T~1-23, and V*~CS~Cha.  Their peculiar motion may be originated from
interaction with other YSOs in the cloud. In $\epsilon$ Cha, on the other hand, the proper motion 
dispersion is larger, as expected
for a more dispersed population. To take this into account for the identification of objects with 
peculiar motions we relaxed our constrain from 3$\sigma_{\mu}$ to 5$\sigma_{\mu}$.  Only the YSO RX~J1123.2-7924
has a large peculiar motion.  The Gaia DR2 proper motions of this object are 
$\mu_\alpha=-31.7\pm0.1$~mas~yr$^{-1}$ and $\mu_\delta=-17.4\pm0.1$~mas~yr$^{-1}$, which corresponds to a 
relative proper motion with respect to the group of $\Delta\mu=13.3\pm0.7$~mas~yr$^{-1}$. At a distance of 100.7 pc,
the star is moving at a velocity of $6.4\pm0.3$~km~s$^{-1}$. In conclusion, even when  the star has a high peculiar 
motion with respect to the rest of the group it is not a fast-moving source.

\subsection{Corona Australis}

The mean trigonometric parallax of CrA suggests a distance of $154\pm4$~pc, which is
significantly larger than that presented by \citet[][and references therein]{neuhauser2008}.
The mean proper motions are in good agreement with the values of 
$(\mu_\alpha\cdot\cos{\delta},\mu_\delta)\simeq(5.5,-27.0)$~mas~yr$^{-1}$ estimated by \cite{neuhauser2000}
based on results of different catalogs. 

One of the stars in the CrA region is above the 3$\sigma_\mu$ criteria, it is named [MR81] H$\alpha$~17.
The peculiar 
proper motion of this star is $5.4\pm0.4$~mas~yr$^{-1}$ and corresponds to peculiar velocity of 
$4.0\pm0.3$~km~s$^{-1}$.

\begin{figure*}[!th]
   \centering
   \begin{tabular}{cc}
  \includegraphics[height=0.33\textwidth, trim=0 0 0 0, clip]{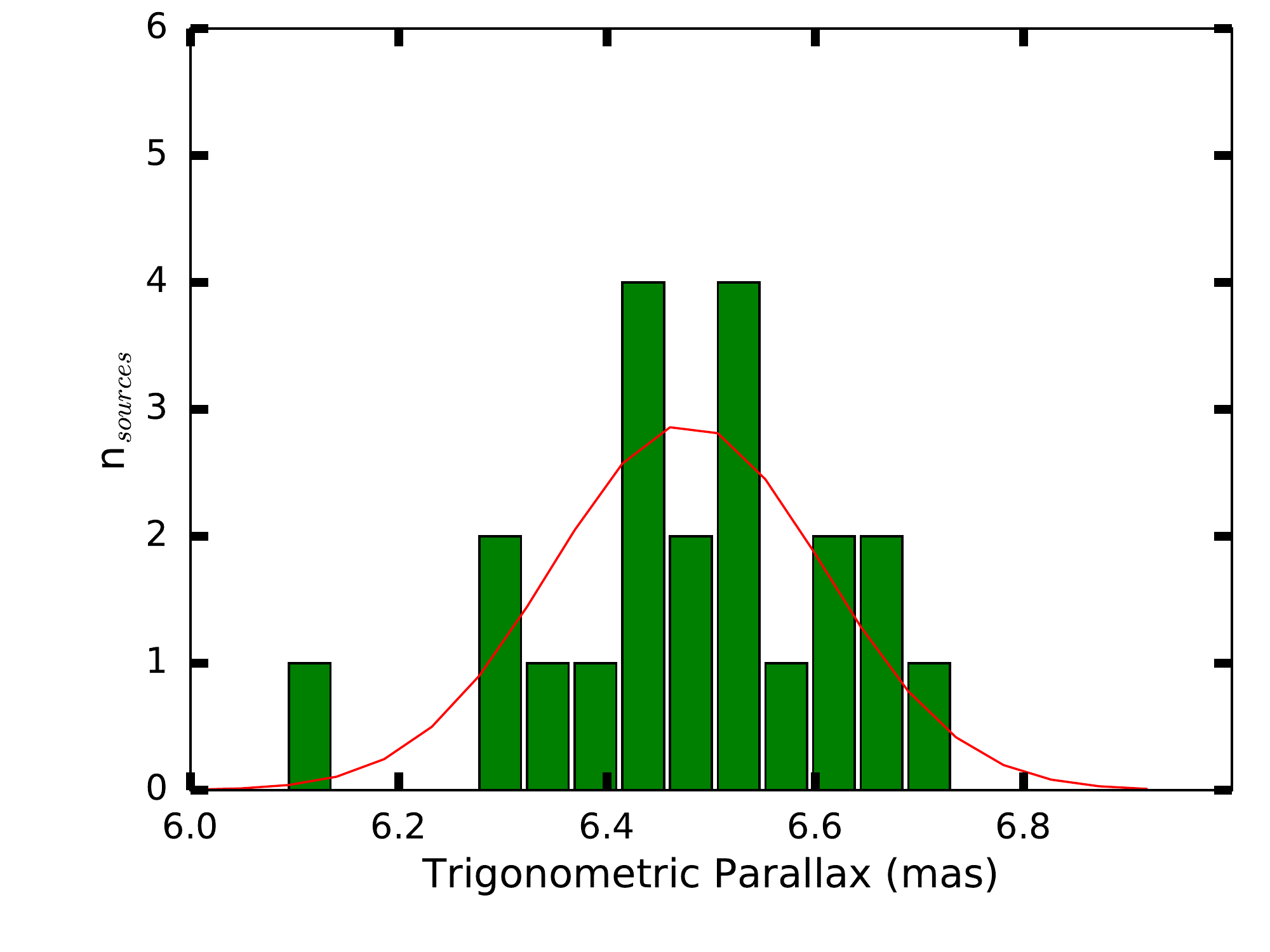} &
  \includegraphics[height=0.33\textwidth, trim=0 10 0 0, clip]{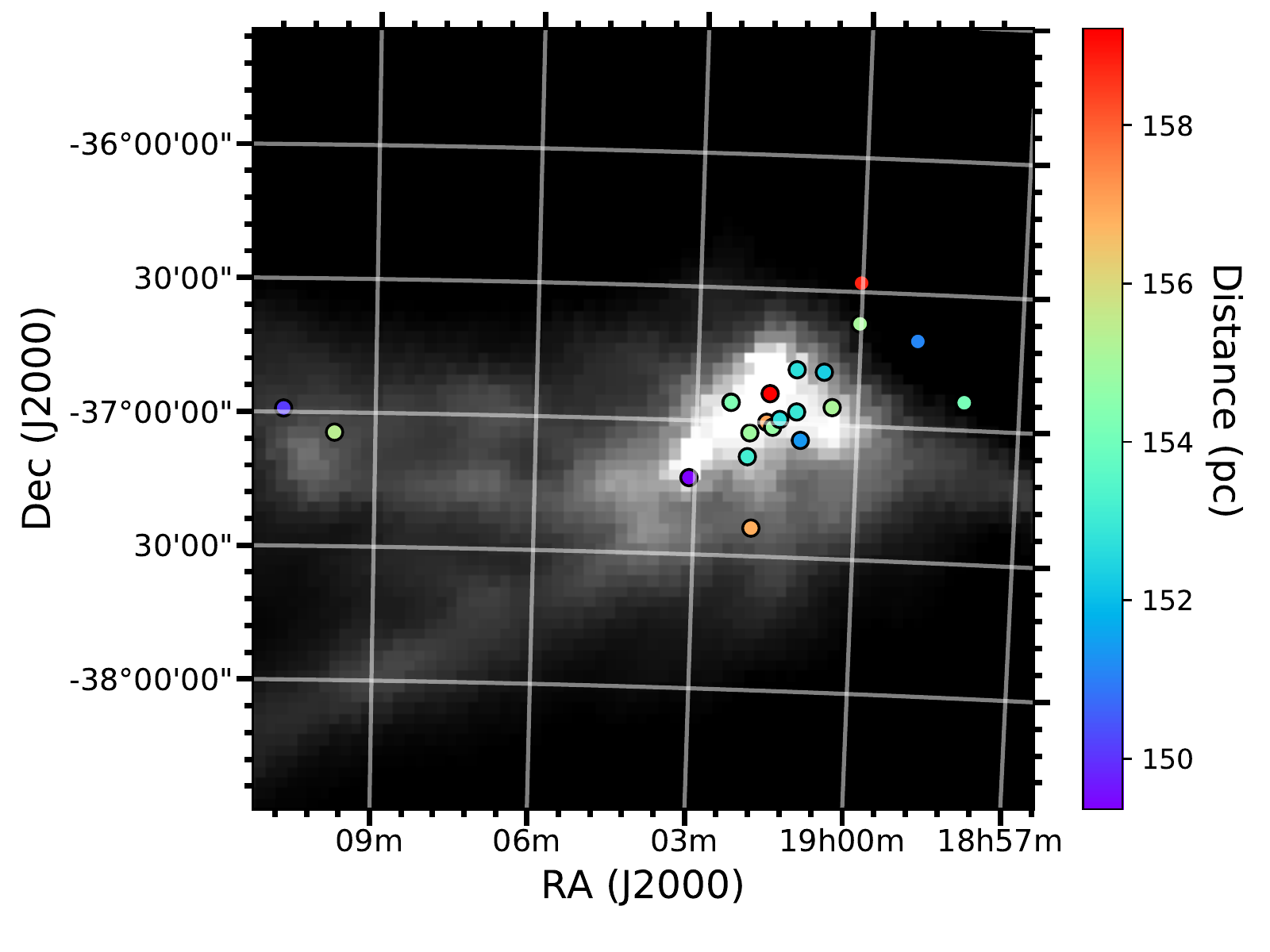}\\
  \includegraphics[height=0.33\textwidth, trim=0 0 40 0, clip]{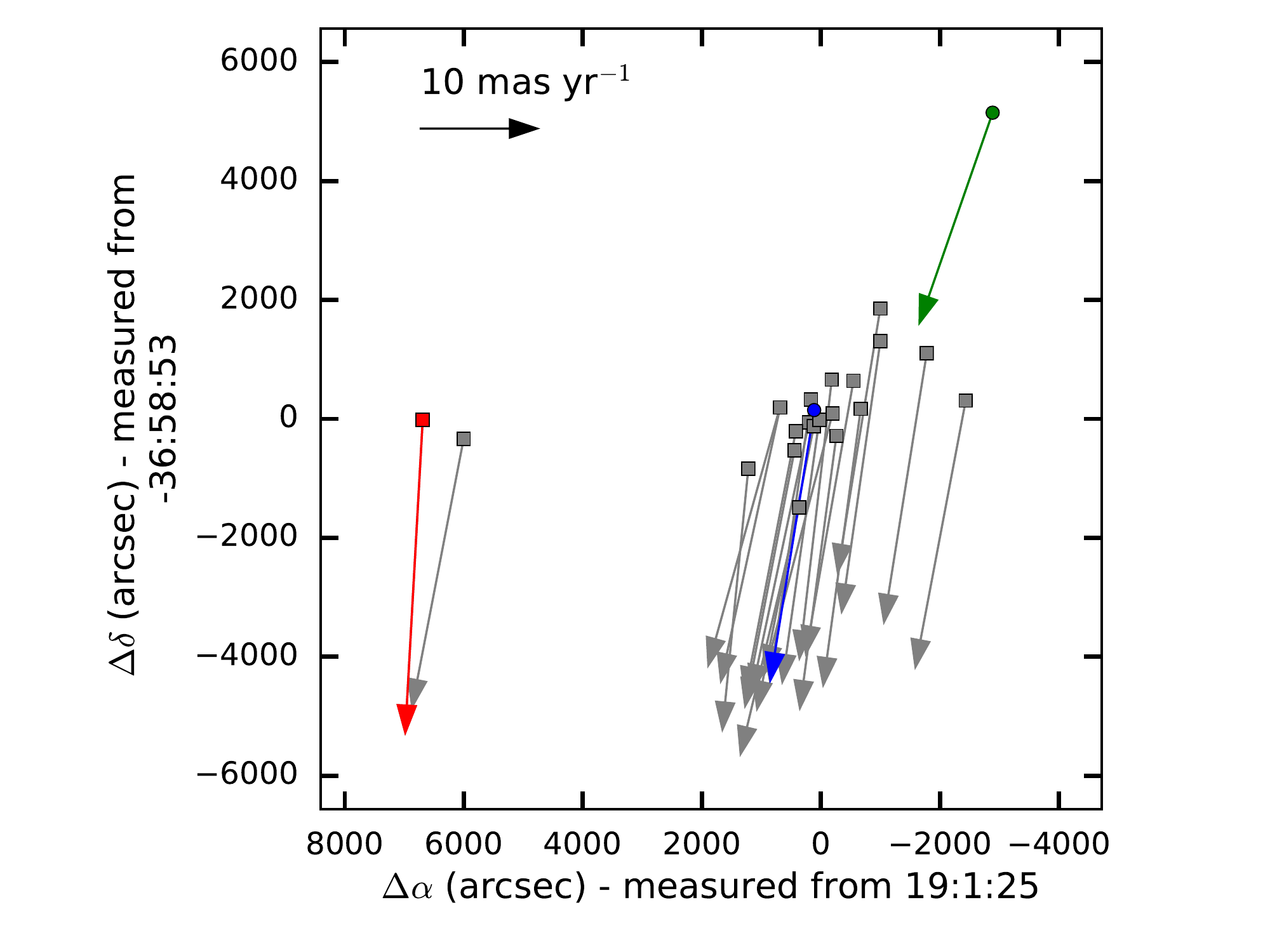} &
  \includegraphics[height=0.40\textwidth, trim=0 35 0 0, clip]{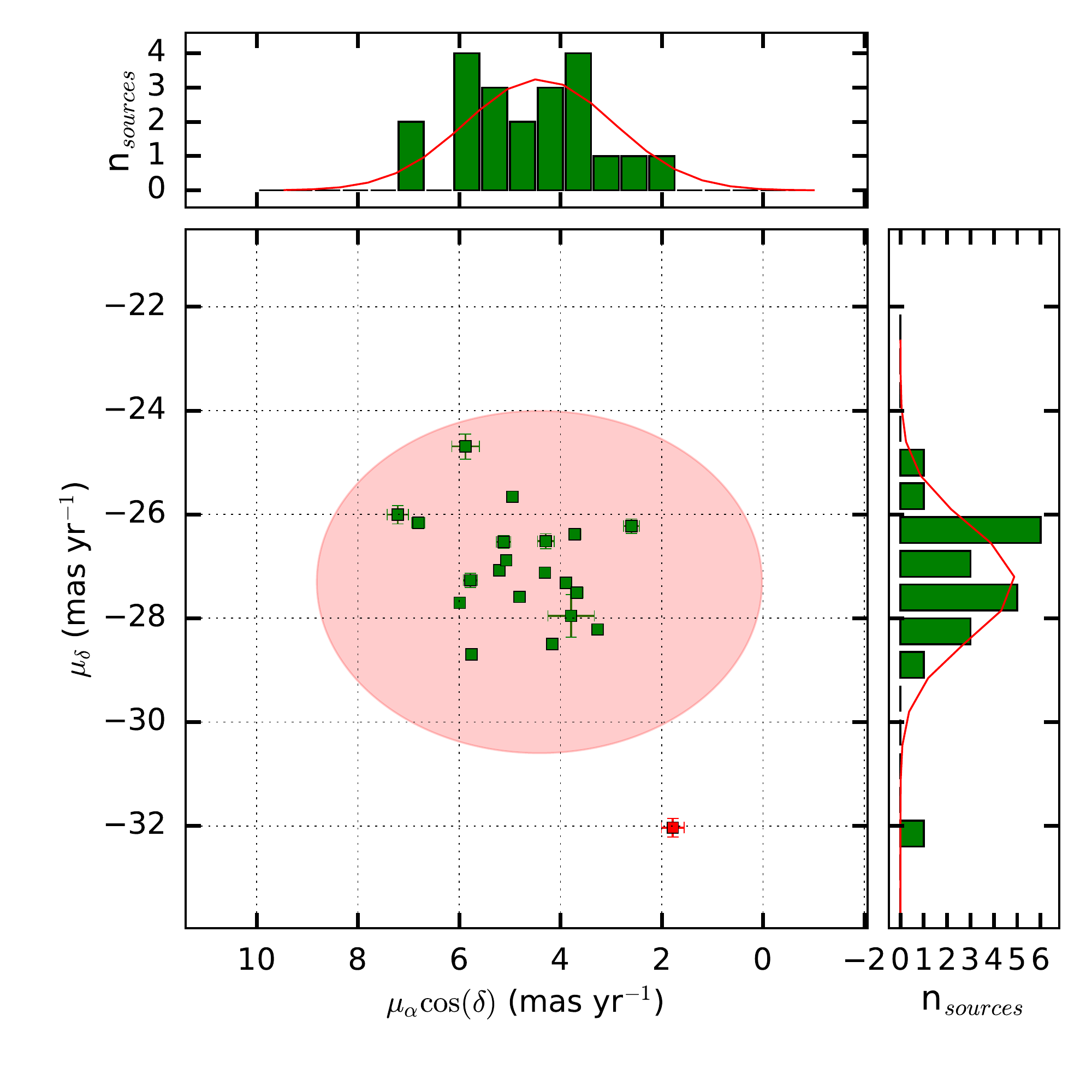}
  \end{tabular}
   \caption{Same as Figure~\ref{fig:b59}, but for Corona Australis}
   \label{fig:cra}
\end{figure*}

\subsection{IC~5146}

Our results show that the mean trigonometric parallax of YSOs in IC~5146
agrees with a distance of 813$\pm106$~pc. This value is consistent with
the more recent values found in this nebula and discards the short distance of 460~pc 
obtained by \cite{lada1999} and \cite{lada1994}. Thus, this region is not 
part of the Gould Belt.

\begin{figure*}[!th]
   \centering
   \begin{tabular}{cc}
  \includegraphics[height=0.33\textwidth, trim=0 0 0 0, clip]{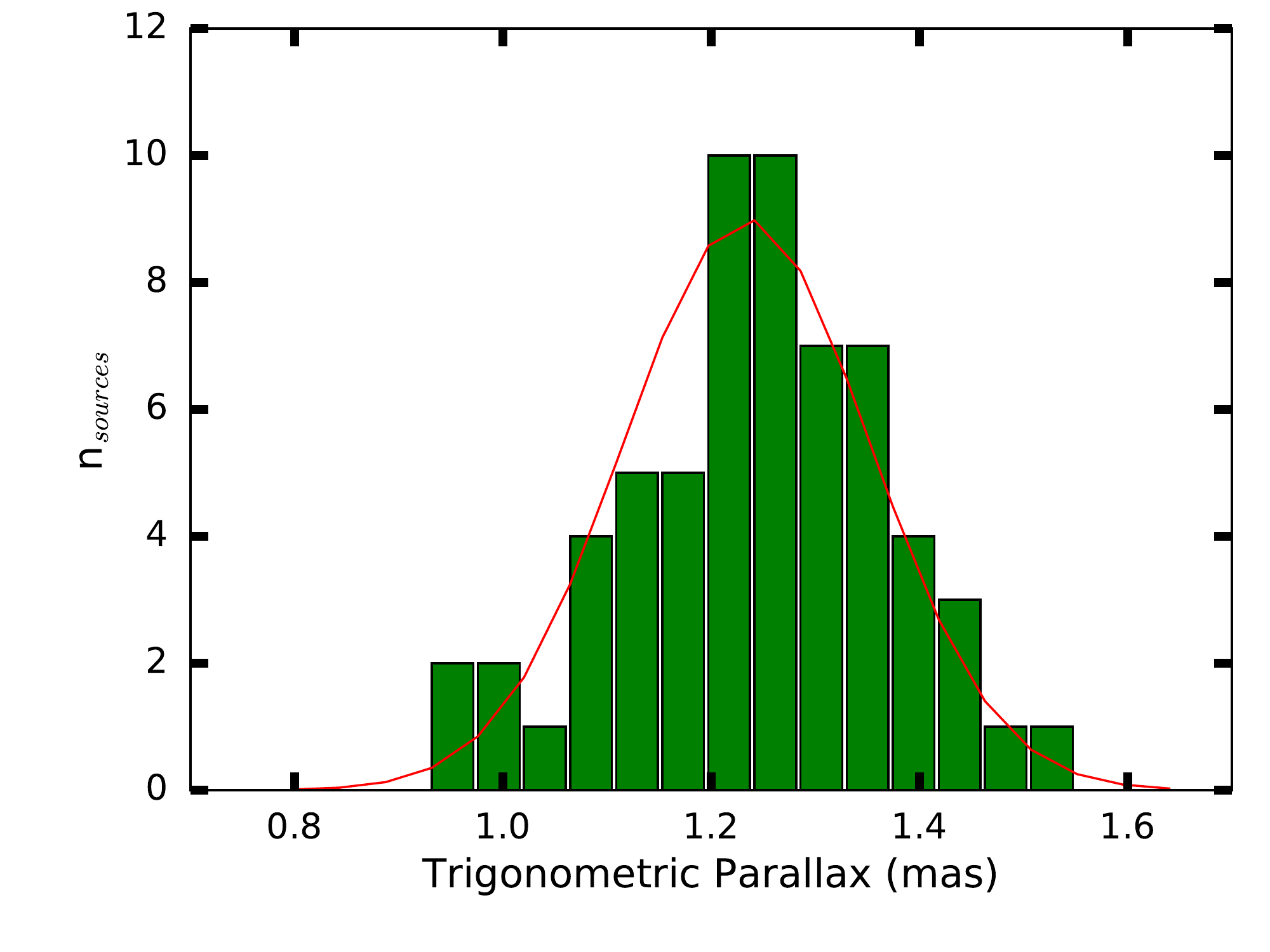} &
  \includegraphics[height=0.33\textwidth, trim=0 10 0 0, clip]{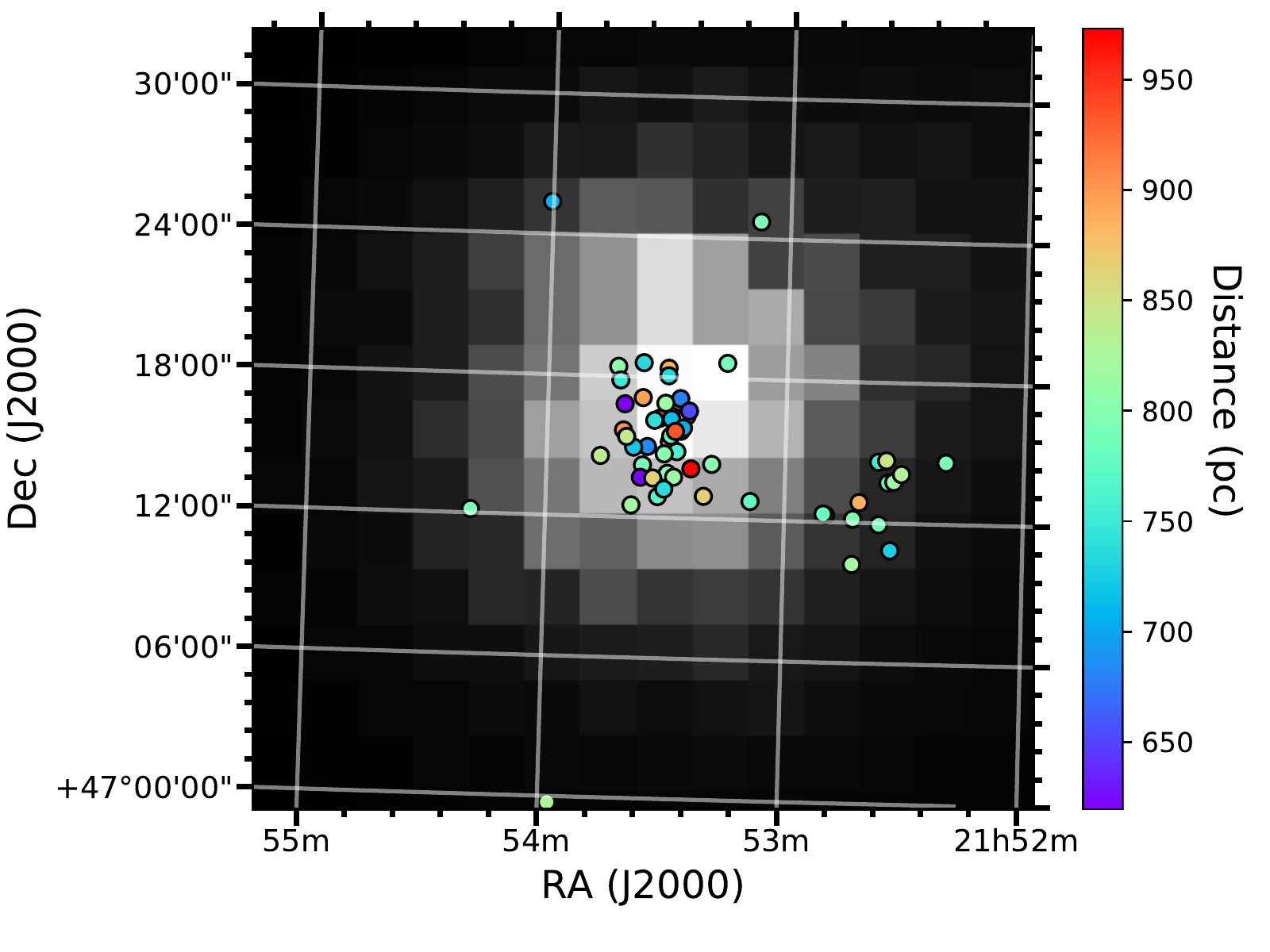}\\
  \includegraphics[height=0.33\textwidth, trim=0 0 40 0, clip]{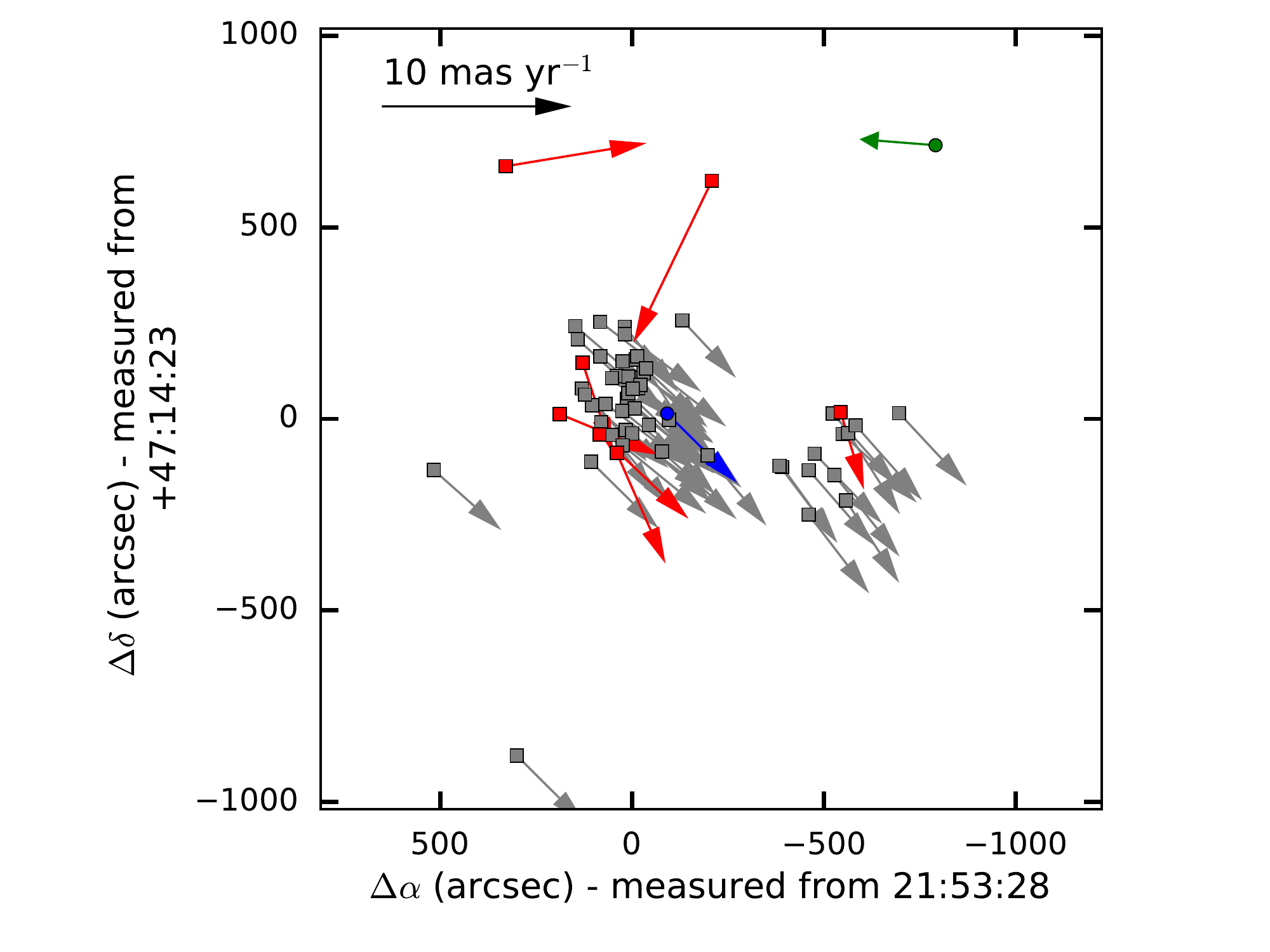} &
  \includegraphics[height=0.40\textwidth, trim=0 35 0 0, clip]{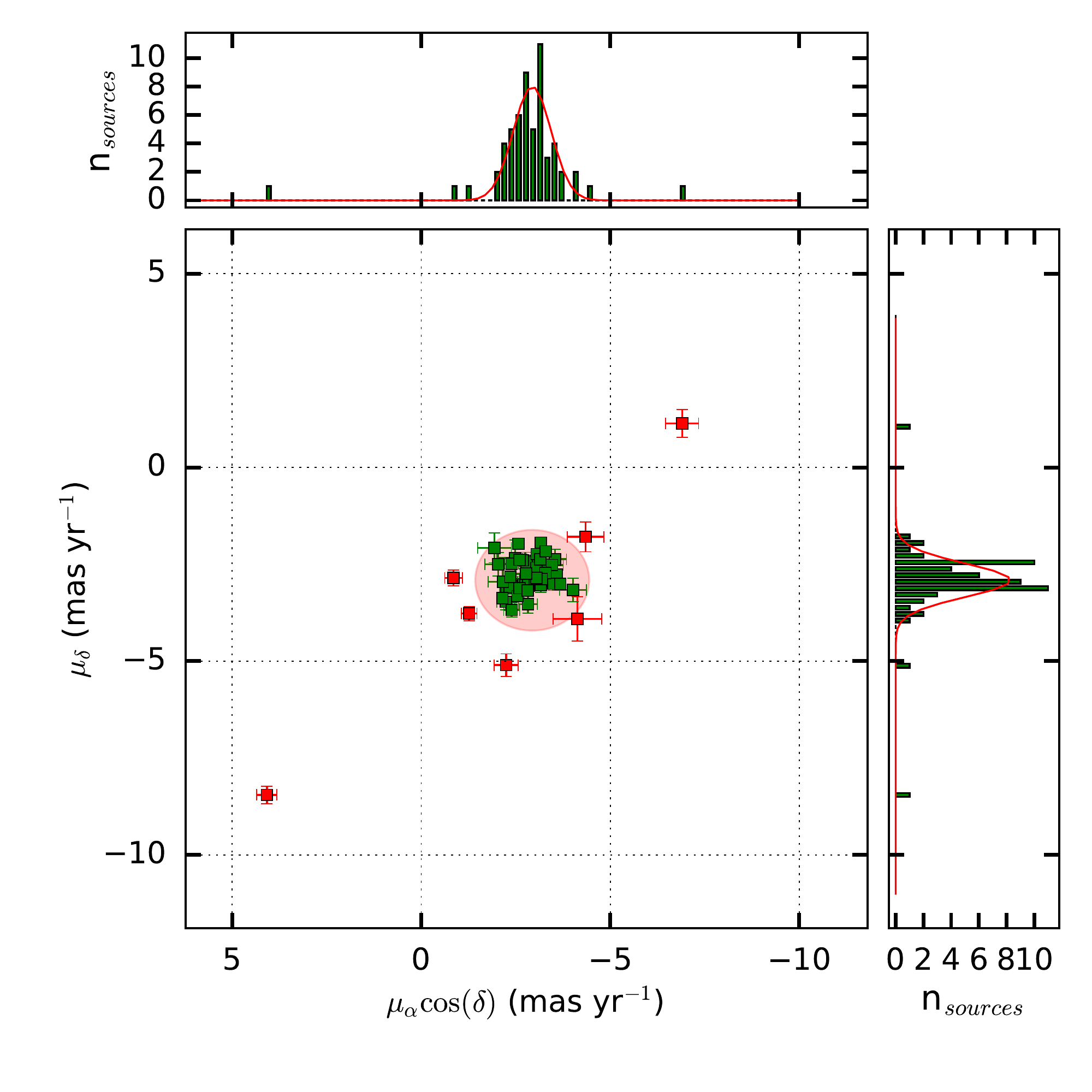}
  \end{tabular}
   \caption{Same as Figure~\ref{fig:b59}, but for IC 5146.}
   \label{fig:ic5}
\end{figure*}

Five of the stars in IC~5146 have proper motions that do not agree, within errors,
with our 3$\sigma_\mu$ criteria. 
Two of these stars, LkH$\alpha$~255 and LkH$\alpha$~241, have the largest peculiar motions
and are at the edges of the cloud (see Figure~\ref{fig:ic5}). The proper motions of LkH$\alpha$~241
are $(\mu_\alpha\cdot\cos{\delta},\mu_\delta)\simeq(4.1\pm0.3,-8.5\pm0.2)$~mas~yr$^{-1}$, indicating a relative 
proper motion with respect to the main group of $9.4\pm0.4$~mas~yr$^{-1}\simeq36.3\pm1.5$~km~s$^{-1}$,
with a direction towards the cluster. The motion of LkH$\alpha$~255 points away from the cluster.
Its proper motions are $(\mu_\alpha\cdot\cos{\delta},\mu_\delta)\simeq(-6.9\pm0.4,1.1\pm0.4)$~mas~yr$^{-1}$,
or a relative motion of $5.6\pm0.4$~mas~yr$^{-1}\simeq21.3\pm2.3$~km~s$^{-1}$. The velocities
of these two stars indicate that they are fast moving stars.

\vspace{1.5cm}
\subsection{Lupus}

Contrary to what was suggested by \cite{comeron2008} the distances to
the YSOs in the Lupus clouds indicate that they all are at a distance around 160~pc.
This distance is also significantly shorter than the kinematic distance found by
\cite{galli2013}, and in agreement with the earlier determination of the distance to this 
region \citep[e.g.,][]{franco1990,crawford2000M,lombardi2008}. 

The mean proper motions of the four clouds seem to be also similar. Only the Lupus 1 
cloud has a slightly different mean proper motion in the R.A. direction. This may indicate
that even when the four regions are well constrained, the clouds where they form were not
independent and probably were part of a large cloud. 

As in the other regions, we also identified YSOs with peculiar motions in the Lupus clouds.
The star with the most peculiar velocity in this region is the YSO THA~15~36.
It has  proper motions of 
$(\mu_\alpha\cdot\cos{\delta},\mu_\delta)\simeq(-17.5\pm0.1,-26.4\pm0.1)$~mas~yr$^{-1}$.
This motion will imply a relative motion of  $8.0\pm0.2$~mas~yr$^{-1}\simeq6.1\pm0.2$~km~s$^{-1}$
with respect to the other YSOs. The velocity of this star indicates that it is not a fast moving object.
As the same analysis may be applied to the other object, we conclude that  none of the YSOs
that we analyzed  in the Lupus clouds are fast moving sources.

\begin{figure*}[!th]
   \centering
   \begin{tabular}{cc}
  \includegraphics[height=0.33\textwidth, trim=0 0 0 0, clip]{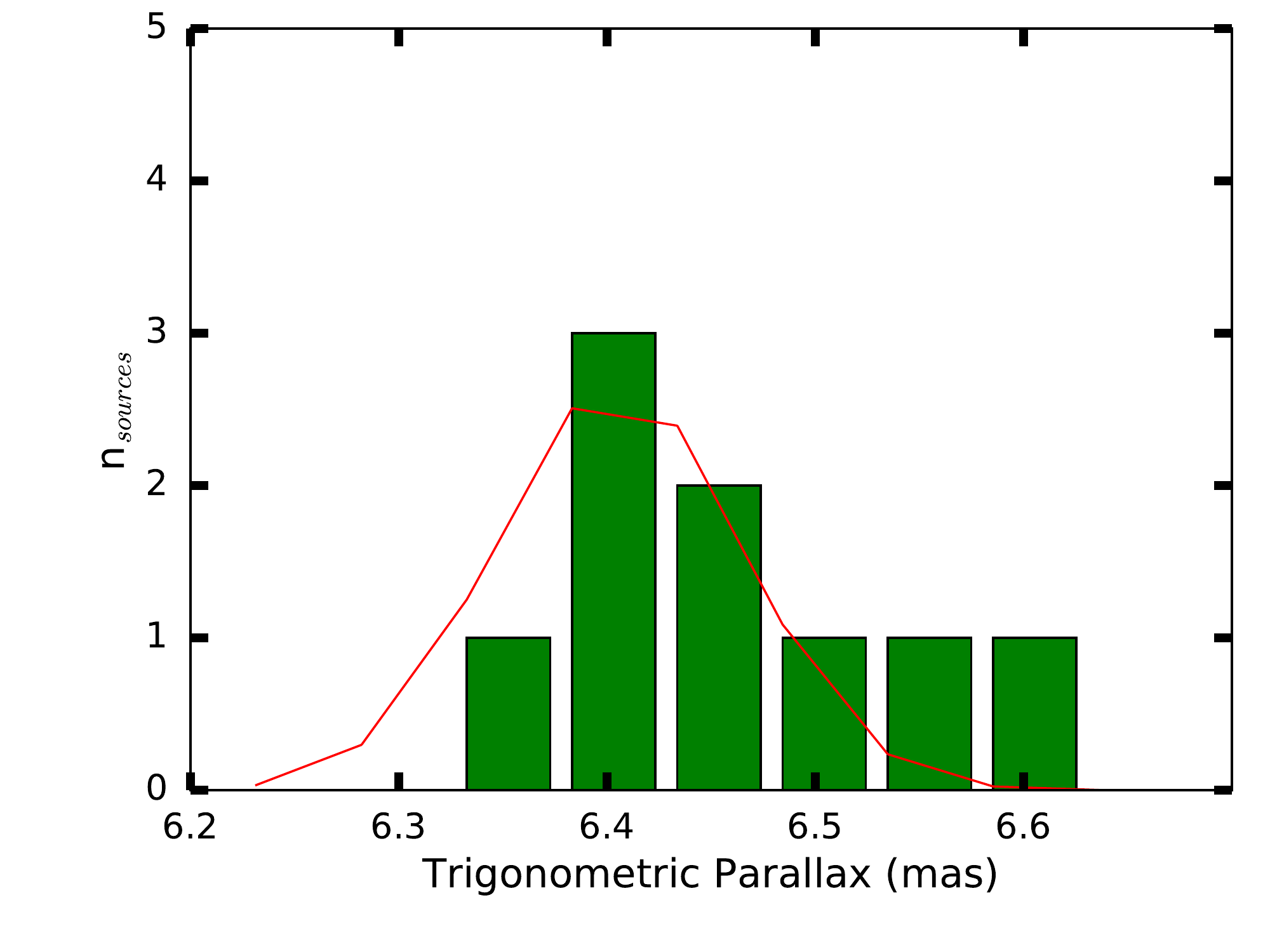} &
  \includegraphics[height=0.33\textwidth, trim=0 10 0 0, clip]{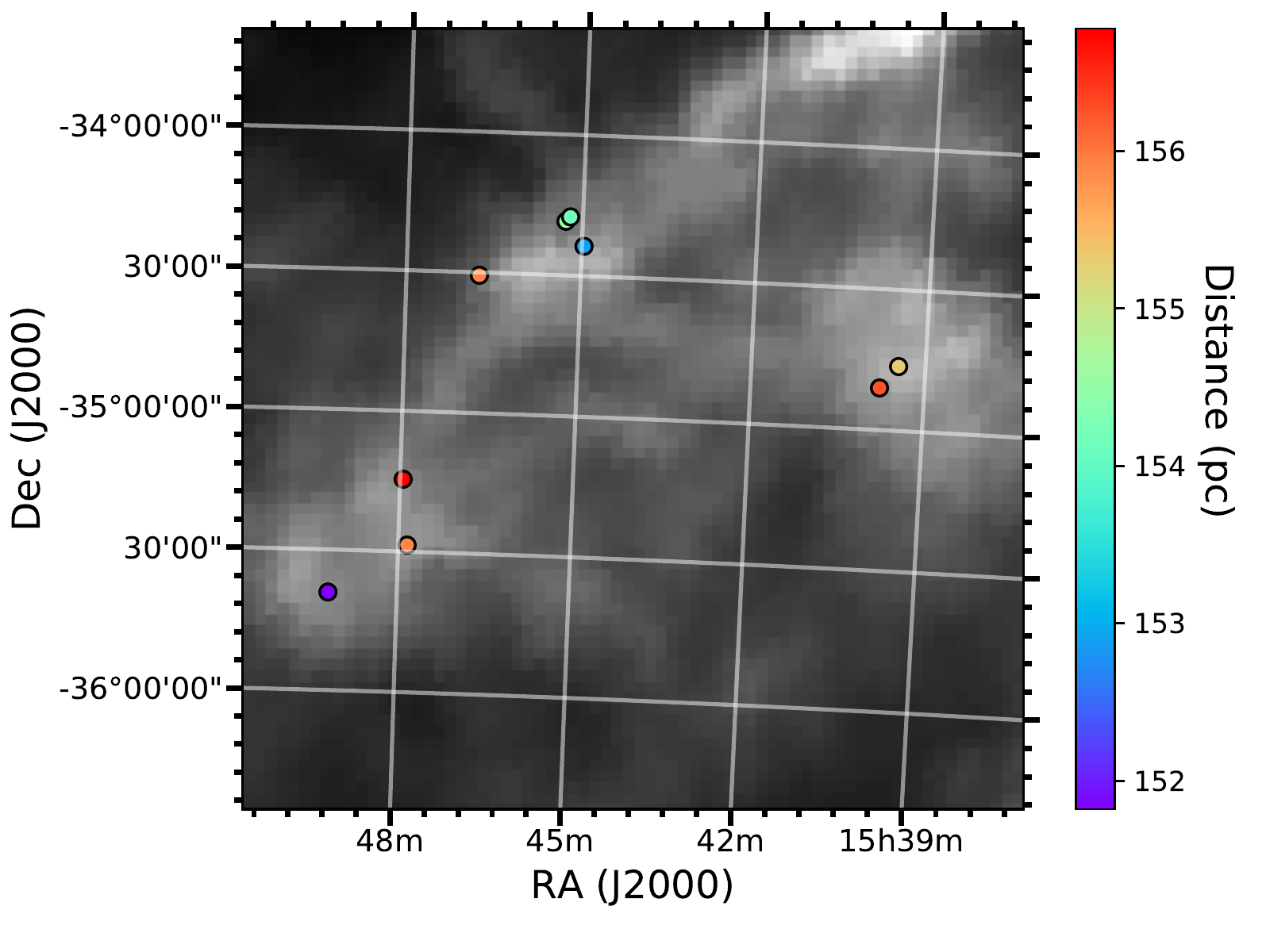}\\
  \includegraphics[height=0.33\textwidth, trim=0 0 40 0, clip]{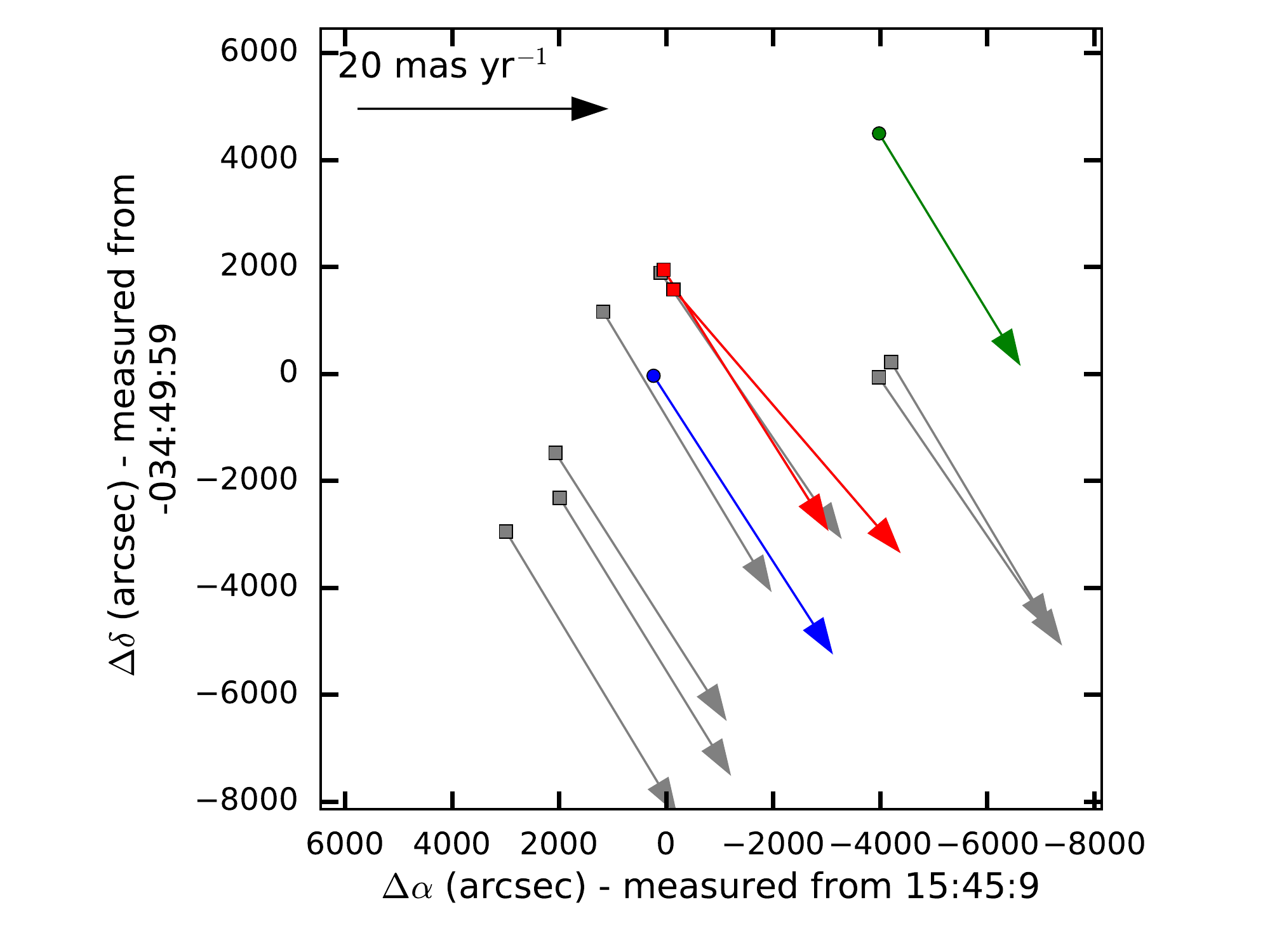} &
  \includegraphics[height=0.40\textwidth, trim=0 35 0 0, clip]{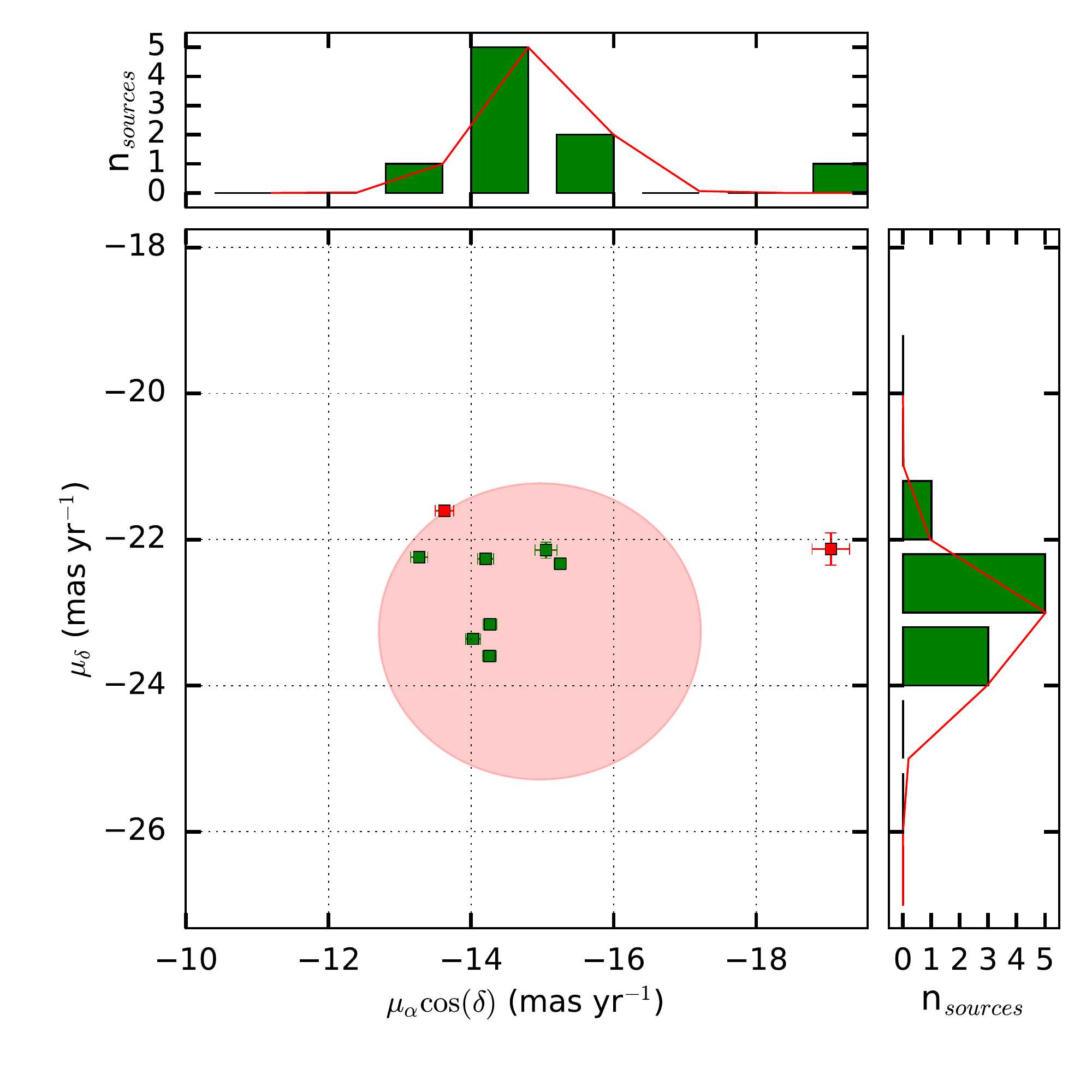}
  \end{tabular}
   \caption{Same as Figure~\ref{fig:b59}, but for Lupus 1.}
   \label{fig:li}
\end{figure*}

\begin{figure}[!h]
   \centering
  \includegraphics[height=0.38\textwidth, trim=0 10 0 0, clip]{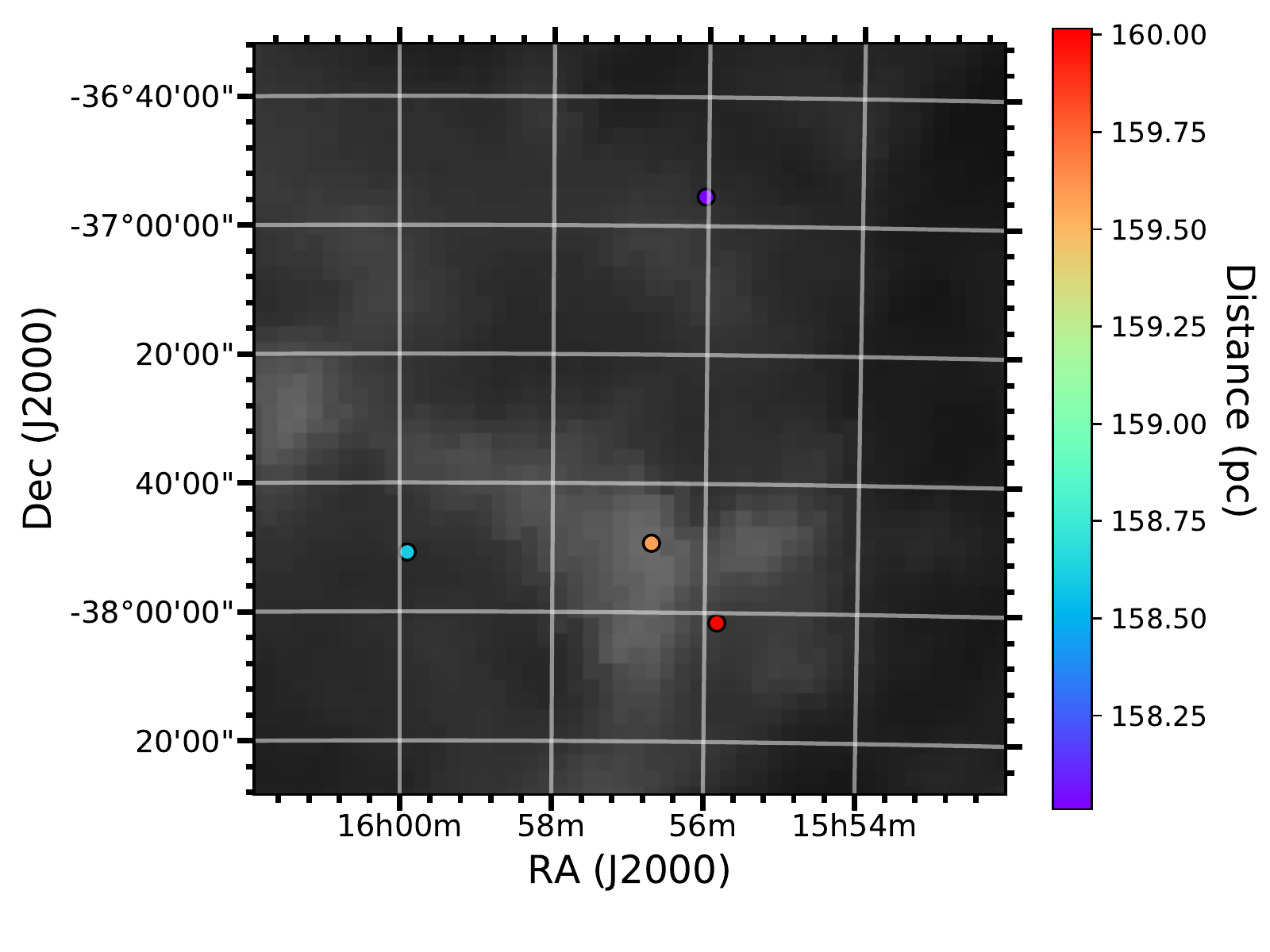}
   \caption{Spatial distribution of YSOs in Lupus 2 that appear in the Gaia DR2 catalog.}
   \label{fig:lii}
\end{figure}

\begin{figure*}[!th]
   \centering
   \begin{tabular}{cc}
  \includegraphics[height=0.33\textwidth, trim=0 0 0 0, clip]{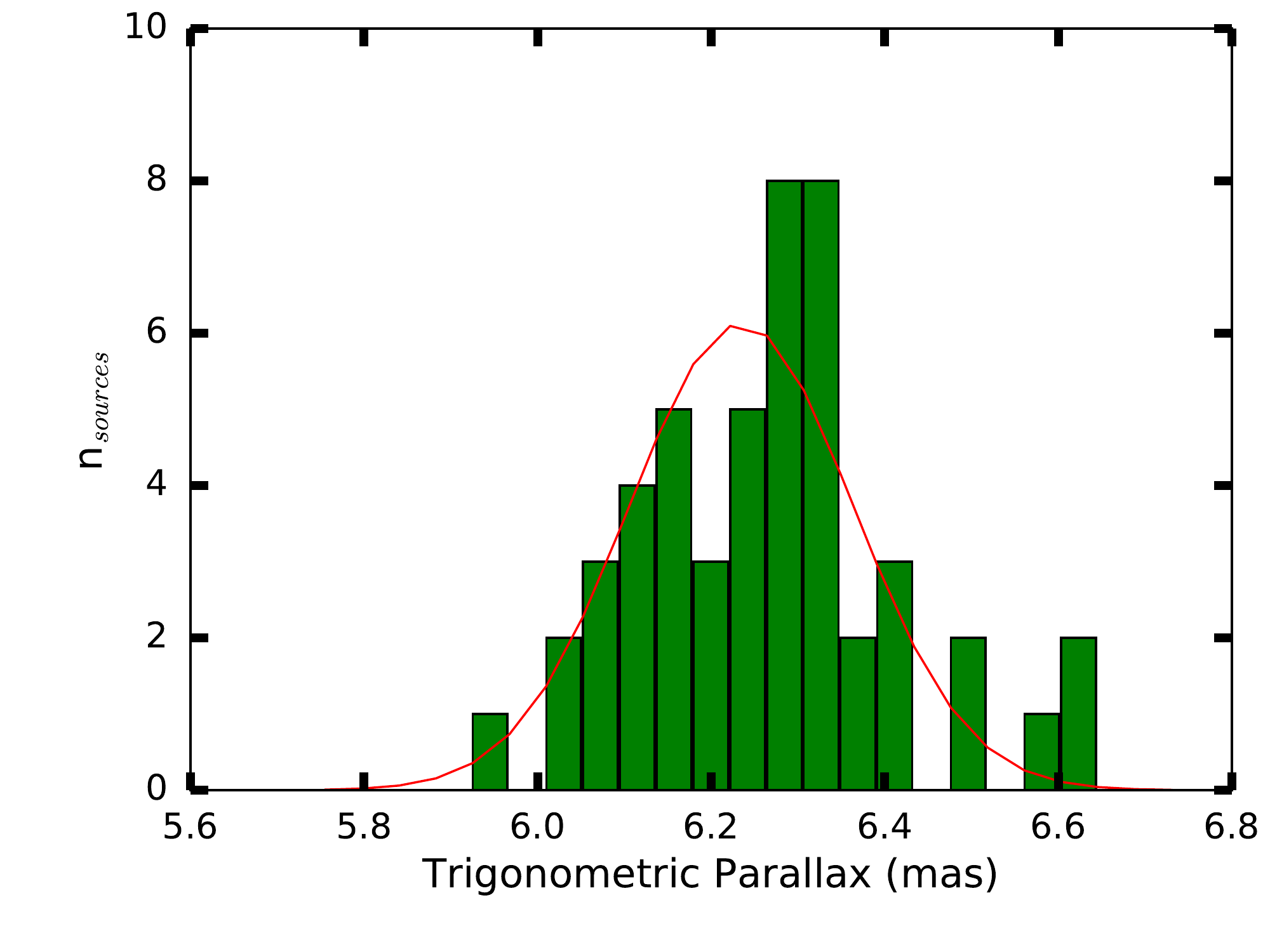} &
  \includegraphics[height=0.33\textwidth, trim=0 10 0 0, clip]{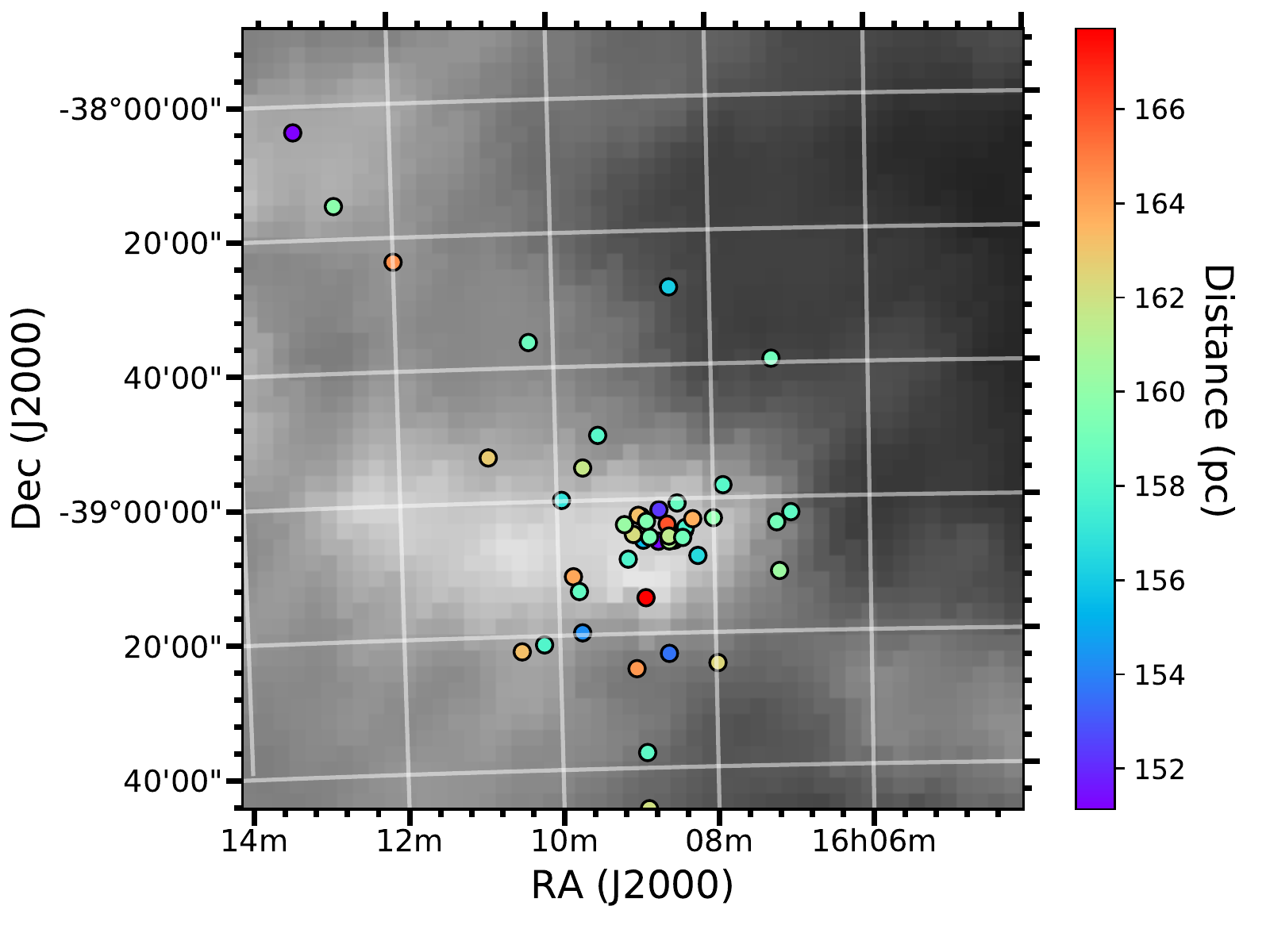}\\
  \includegraphics[height=0.33\textwidth, trim=0 0 40 0, clip]{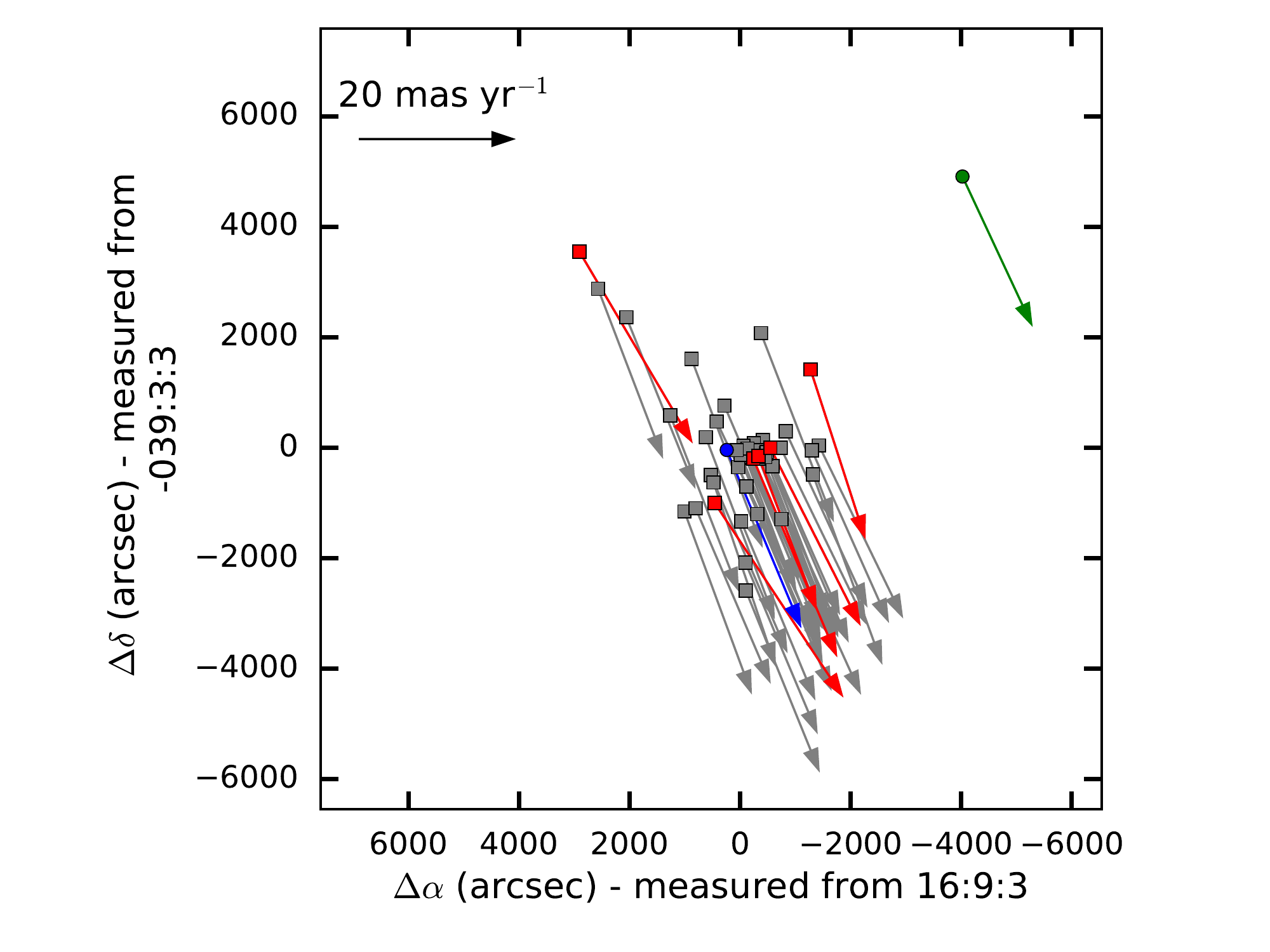} &
  \includegraphics[height=0.40\textwidth, trim=0 35 0 0, clip]{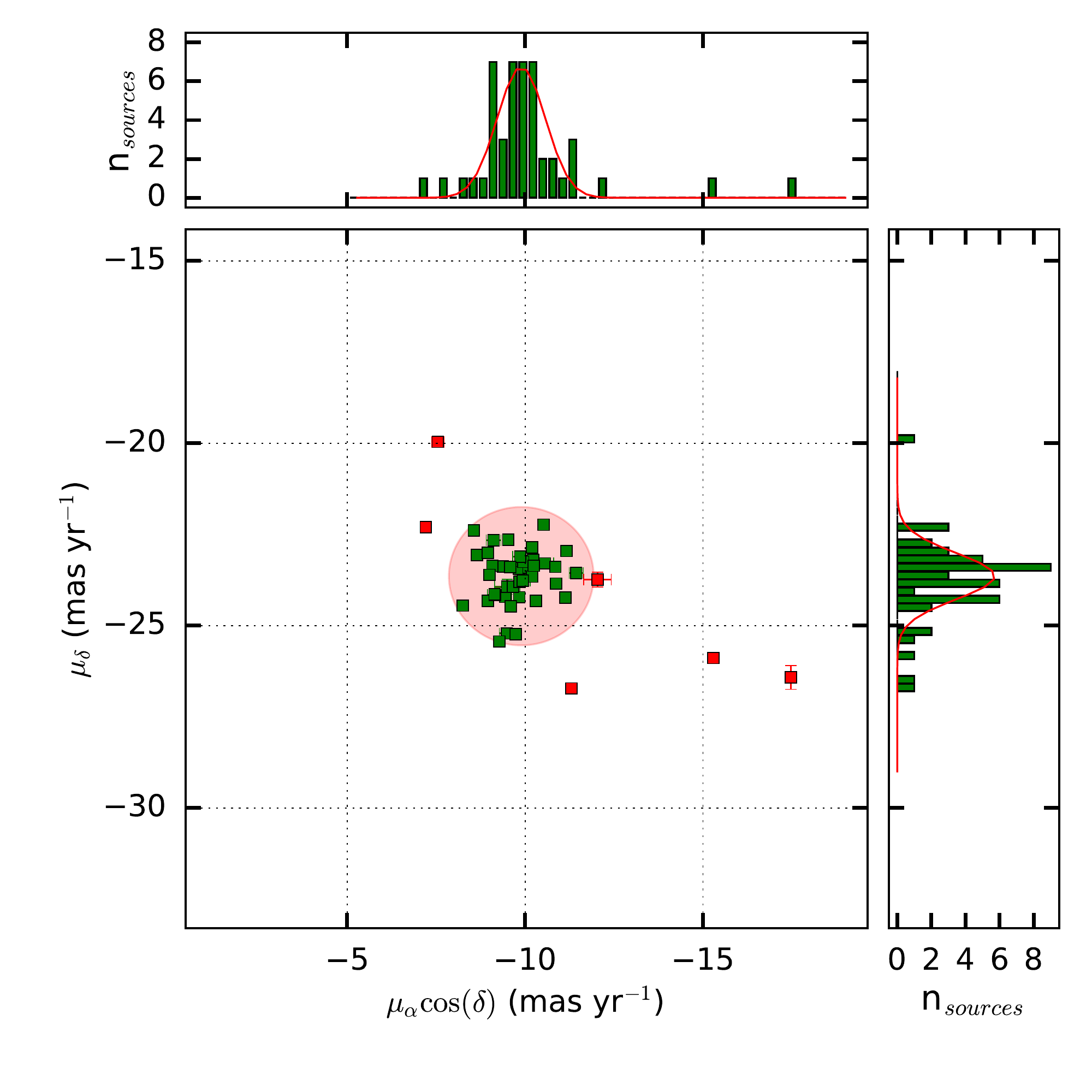}
  \end{tabular}
   \caption{Same as Figure~\ref{fig:b59}, but for Lupus 3.}
   \label{fig:liii}
\end{figure*}

\begin{figure*}[!th]
   \centering
   \begin{tabular}{cc}
  \includegraphics[height=0.33\textwidth, trim=0 0 0 0, clip]{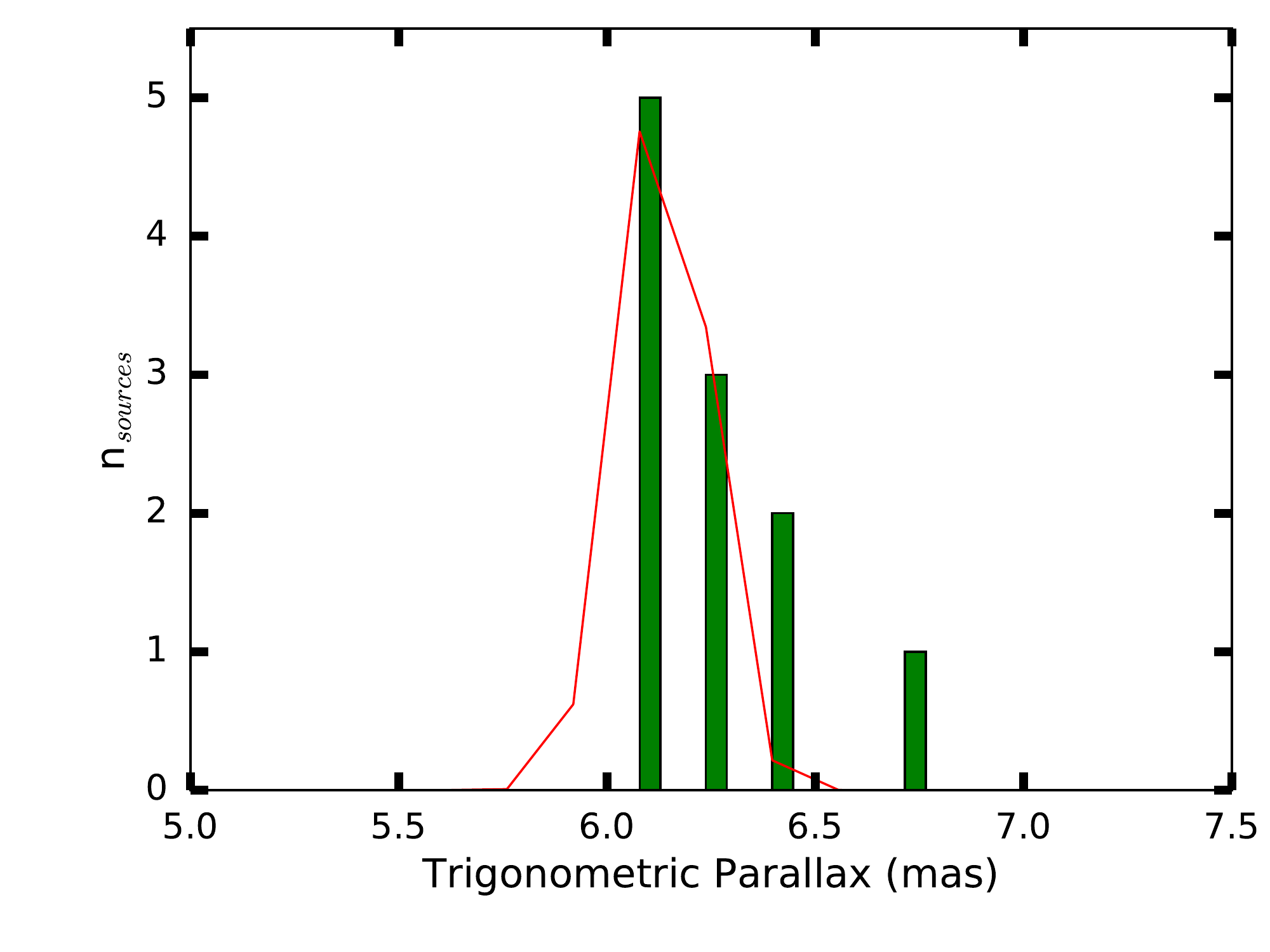} &
  \includegraphics[height=0.33\textwidth, trim=0 10 0 0, clip]{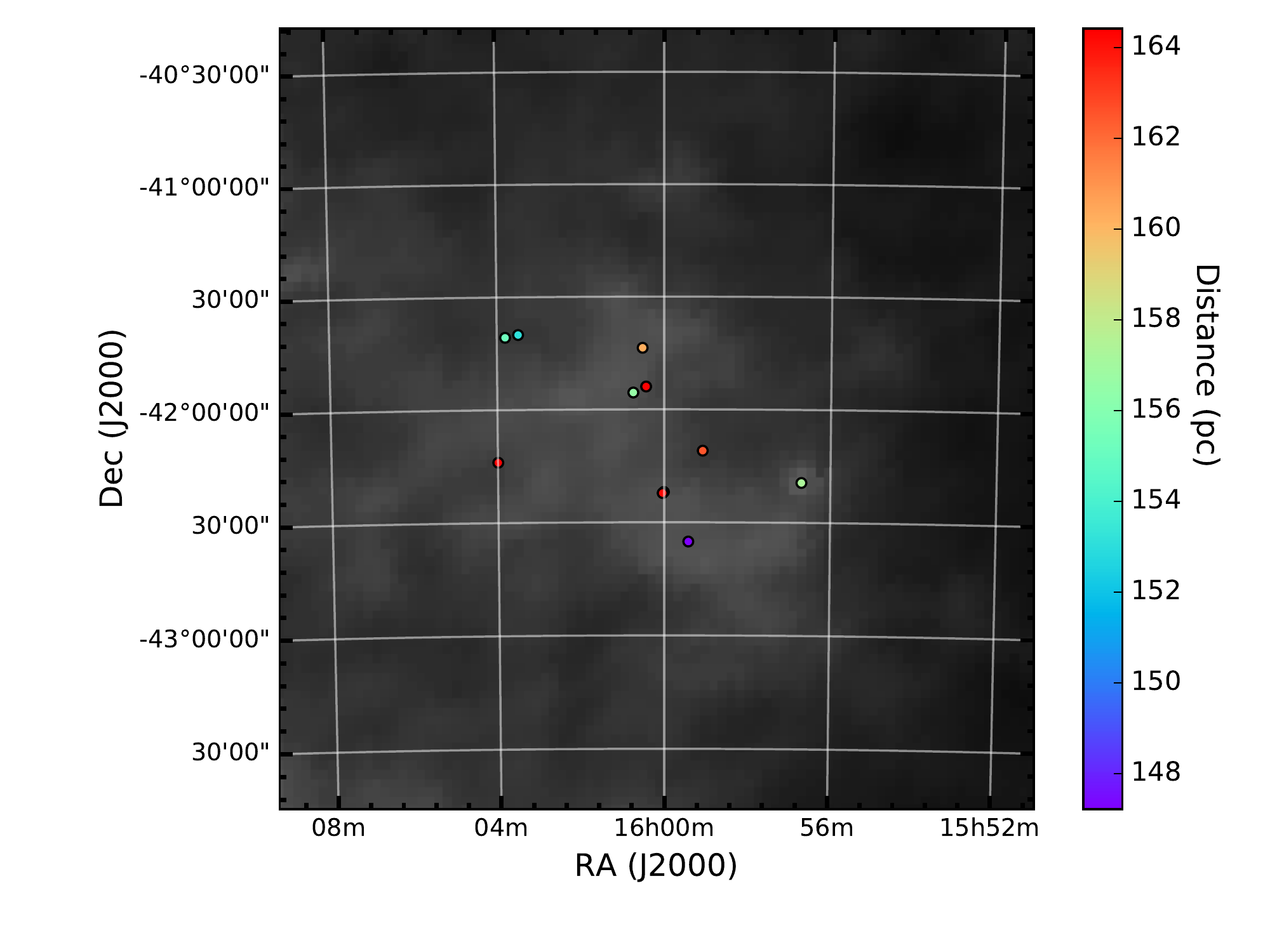}\\
  \includegraphics[height=0.33\textwidth, trim=0 0 40 0, clip]{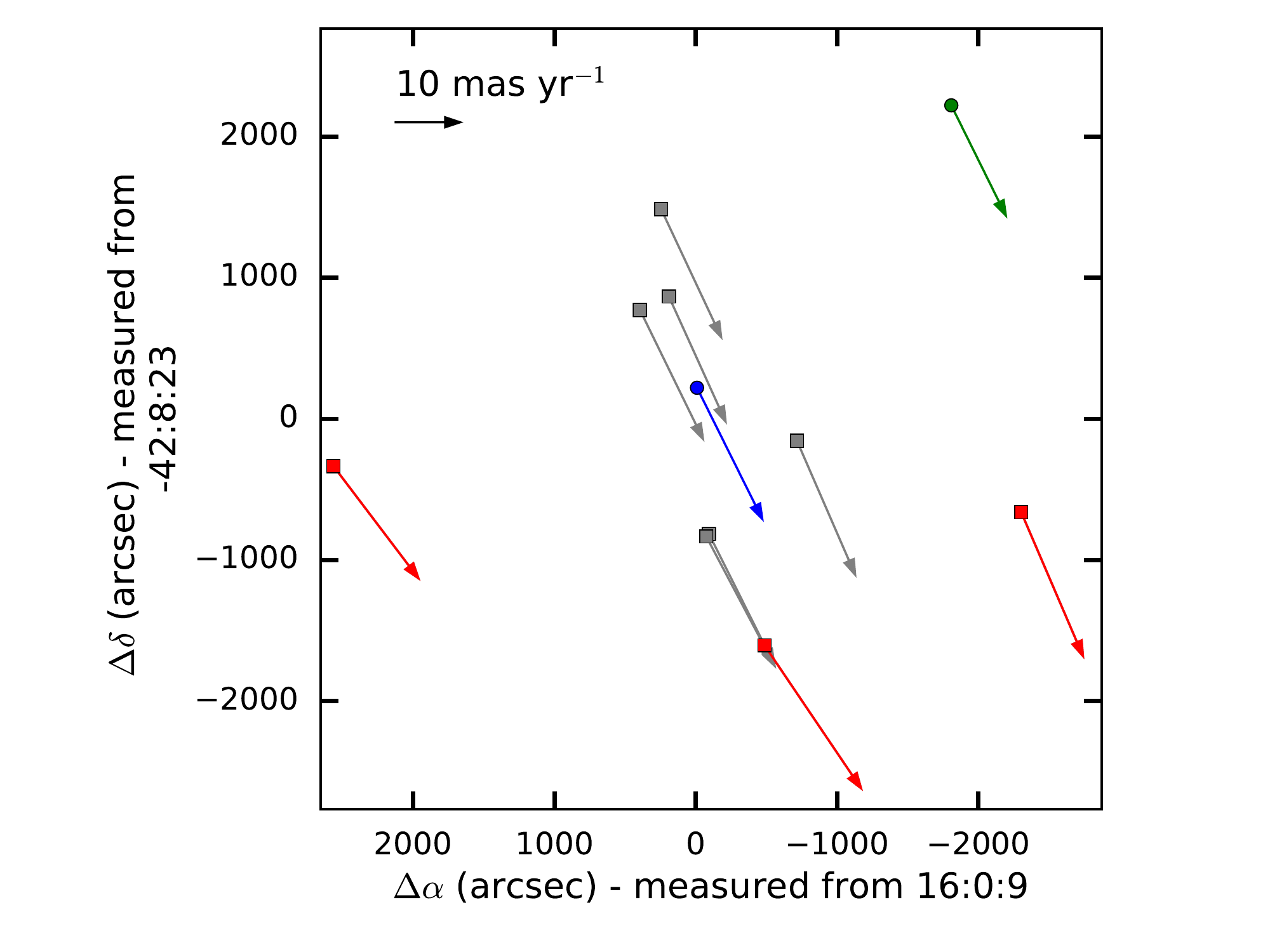} &
  \includegraphics[height=0.40\textwidth, trim=0 35 0 0, clip]{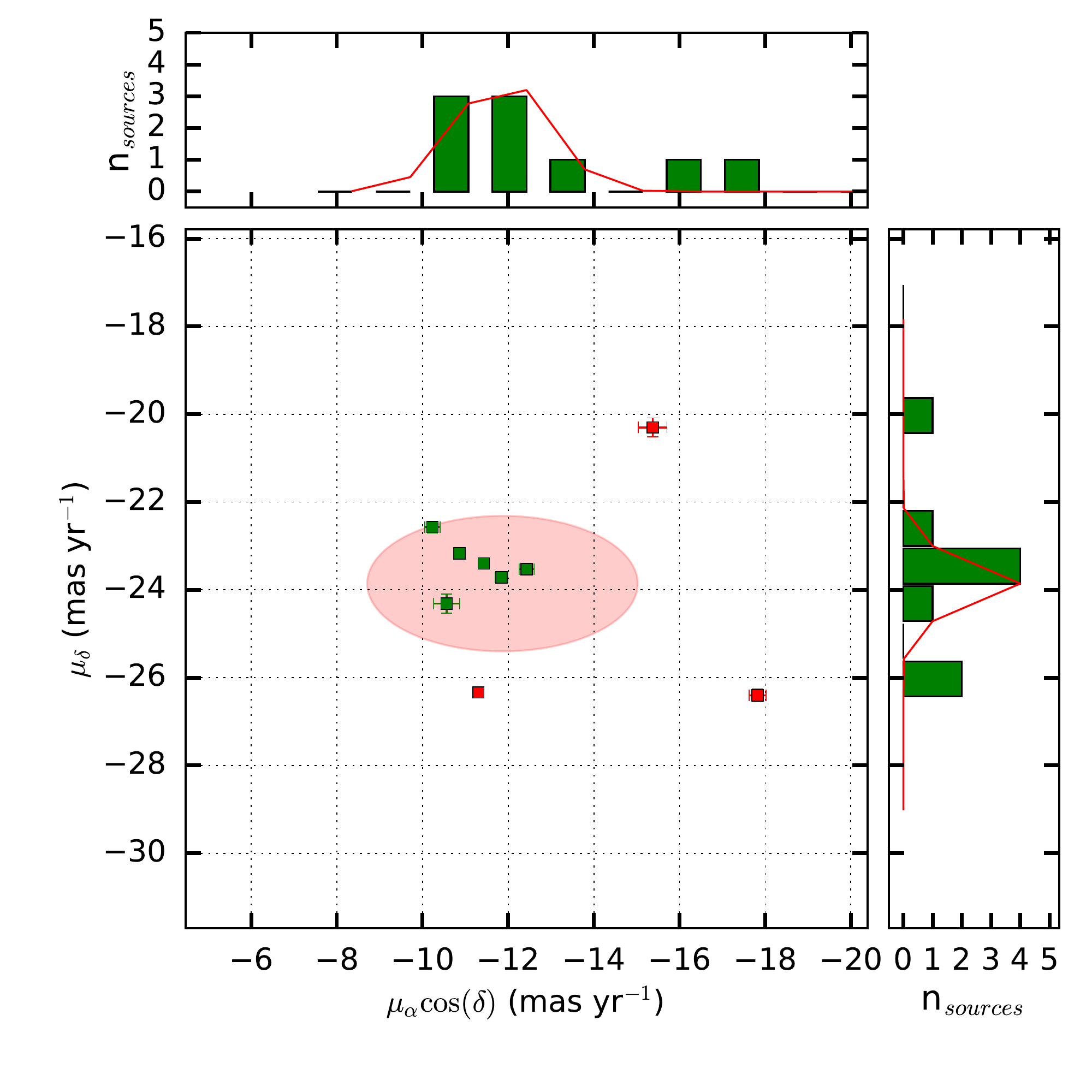}
  \end{tabular}
   \caption{Same as Figure~\ref{fig:b59}, but for Lupus 4.}
   \label{fig:liv}
\end{figure*}

\section{Gould Belt Kinematics}

In order to re-construct the three dimensional kinematics of star-forming regions in the Gould Belt, we will use
the mean proper motions of the YSOs of regions listed in Tab.~\ref{tab:kin} with distances $120<$d[pc]$<500$. 
Additionally to them, we will also use other star-forming regions whose distances and mean proper motions 
have been studied in other papers and are also suggested to belong to the GB. These regions are: the 
Orion Nebula Cluster (ONC), L1688 in Ophiuchus, IC 348 and NGC 1333 in Perseus, and L~1495 in Taurus.   The clouds in the Taurus complex are known to lie near the GB
center, which is inconsistent with the classical picture of the ring structure, 
however, since we are also considering the possibility of a filled disk, they will be used 
for the following analysis. 
All the regions used for this new analysis are listed in 
Table~\ref{tab:GBk}. Most of the YSOs in these regions have no radial velocities in the 
Gaia DR2 catalog, thus we will use literature values of the radial velocities of YSOs or the gas in the clouds.

\begin{figure*}[!htb]
   \centering
  \includegraphics[height=0.35\textwidth, trim=10 0 0 0, clip]{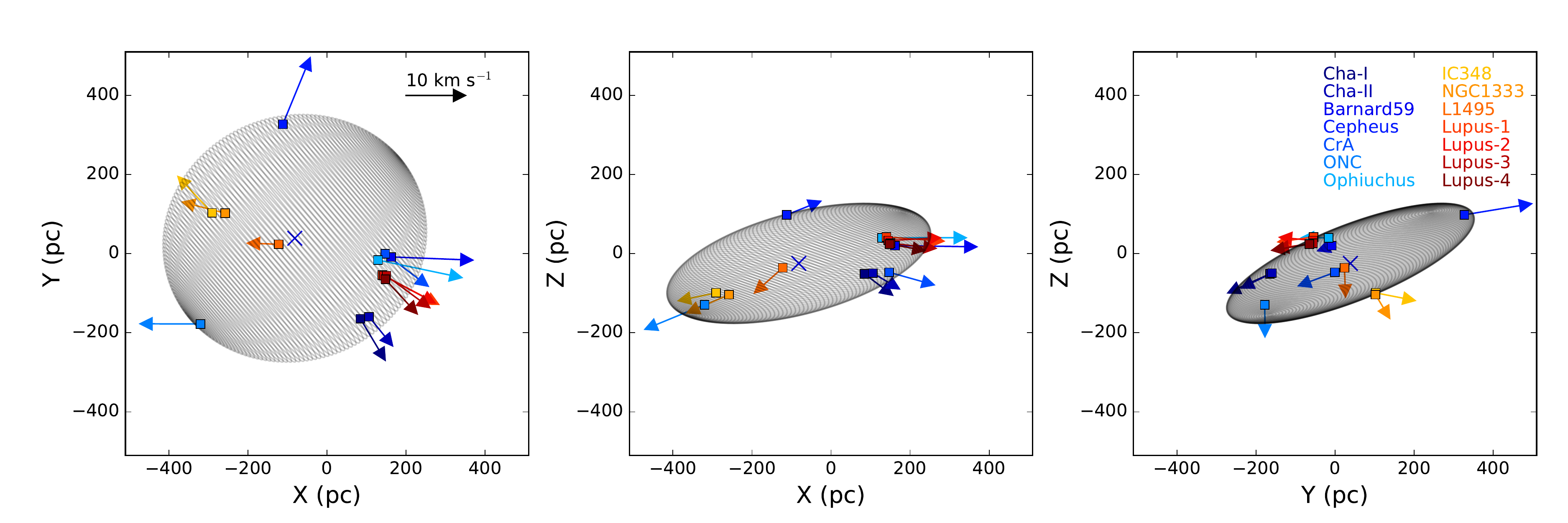}
  \caption{Motion of the star-forming  region in the Gould Belt in the three planes: X-Y (left),
  X-Z (center), and Y-Z (right). The projection of the fitted ellipsoid is also plotted.
  }
   \label{fig:kin2d}
\end{figure*}

\begin{figure*}[!htb]
   \centering
  \includegraphics[height=0.65\textwidth, trim=70 10 30 10, clip]{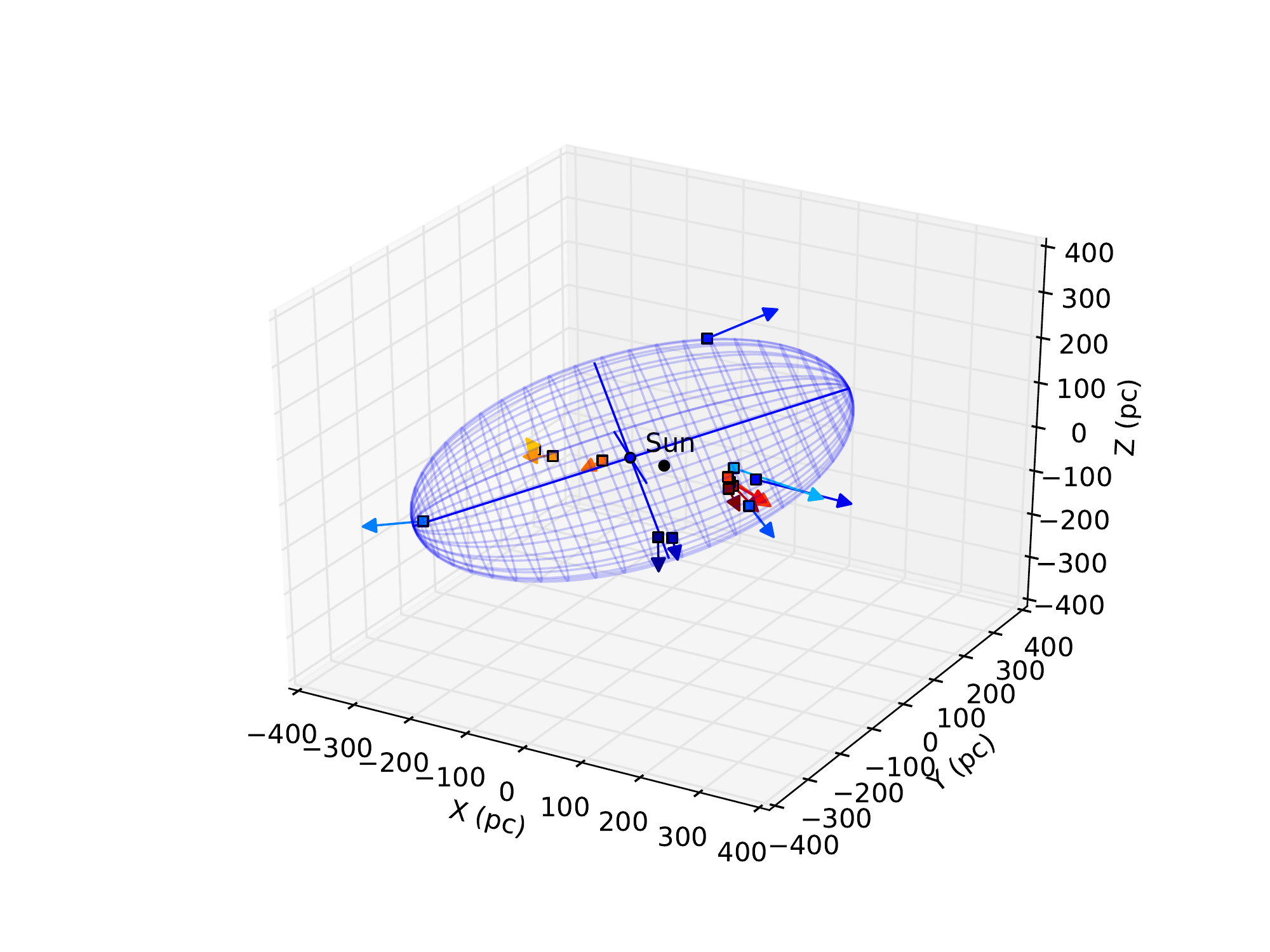}
  \caption{Three dimensional motion of star-forming regions in the Gould Belt (colored arrows same as in figure~\ref{fig:kin2d}). Position of the Sun and the Center of the Gould Belt (GBC) are indicated as  black and blue circles, respectively. Fitted ellipsoid is also plotted. 
  }
   \label{fig:kin3d}
\end{figure*}

The central position of the star-forming regions will be expressed in 
a rectangular coordinate system ($x,y,z$) with the Sun at the origin. The 
$(Ox)$-axis is defined in the line of the Sun-Galactic center direction
(the positive direction is towards the Galactic center). The $(Oy)$-axis is in 
the galactic plane and orthogonal to the $(Ox)$-axis (positive values in the 
direction of the galactic rotation). The $(Oz)$-axis is 
orthogonal to the Galactic plane with positive values towards the 
Galactic North pole.  The coordinate system $(Oxyz)$ is right handed.
The positions ($X,Y,Z$) of the star-forming regions 
are listed in Table~\ref{tab:uvw}. The three-dimensional heliocentric velocities 
(U,V,W) of the star-forming regions in this reference frame are obtained 
directly from the mean proper motions and radial velocities of Table~\ref{tab:GBk} 
and listed in Table~\ref{tab:uvw}. The ($U,V,W$) velocities are transformed to the Local Standard 
of Rest (LSR) velocities ($u,v,w$) by subtracting the motion of the Sun relative 
to the LSR. This motion is ($U_\odot,V_\odot,W_\odot$)= ($11.1\pm0.7,12.2\pm0.47,7.25\pm0.37$)~km~s$^{-1}$ \citep{schonrich2010}. We also assumed a distance to the Galactic center of 8.4 kpc \citep{reid2009}.
The ($u,v,w$) velocities of the stars forming regions are listed in Table~\ref{tab:uvw} and plotted in Figures~\ref{fig:kin2d} and~\ref{fig:kin3d}.

Now we use the $(X,Y,Z)$ positions listed in Table \ref{tab:uvw} to fit an ellipsoid that approximates the spatial distribution of the regions. To fit the ellipsoid, we use a singular value decomposition approach and the minimum volume enclosing ellipsoid method \citep{Moshtagh_minimumvolume}. The resulting best-fit parameters for the geometric center of the ellipsoid are $(X_0,Y_0,Z_0)=(-82\pm15, 39\pm7, -25\pm4)$~pc, while the resulting ellipsoid sizes are $(358\pm7)\times(316\pm13)\times(70\pm4)$~pc.  The inclination angle of the ellipsoid with respect to the Galactic plane is $21\rlap{.}^{\circ}4\pm0\rlap{.}^{\circ}9$, and the direction of the ascending node is at  $\ell=319\rlap{.}^{\circ}4\pm2\rlap{.}^{\circ}3$.
These parameters are in good agreement with recent estimates of the Gould Belt properties based on samples of massive O--B stars, which trace a fairly flat system with semiaxes $\sim350\times250\times50$~pc \citep[i.e.][]{Bobylev2014} and center in the second Galactic quadrant.  We note that the shape of our ellipsoid, however, is closer to an oblate spheroid than other previous results. Differences may come from the different studied objects and from the more accurate
distances used for our analysis.


The velocity vectors ($\mathbf{v}$) of the star forming regions can now be used to investigate ordered motions
with respect to the central position of our fitted ellipsoid. The position of each region is given by a vector 
${r_R}$ in this reference system, and it
has an associated unit vector ${\widehat{r}_R}={{r}_R}/{\mid{r}_R\mid}$.
\cite{rivera2015} argue that the mean value of the scalar products of the unit vector and the velocity vector 
may be used as a proxy to the values of expansion/contraction of a cluster. Following these authors, 
we now use the scalar products to measure the expansion velocity of the star-forming regions 
in the Gould Belt relative to the center of the fitted ellipsoid. Using the LSR velocities from Table~\ref{tab:uvw}
we estimated the mean of their scalar products to be
$\overline{{\widehat{r}_R \cdot v}}=2.5\pm0.1$~km~s$^{-1}$, and a dispersion of 
$\sigma_{\widehat{r}_R \cdot v}=0.8\,$km~s$^{-1}$. 
Our results indicate that the ordered motion of the analyzed star-forming regions agrees with an expansion motion.

 The initial expansion velocity of the GB is unknown.
By assuming a simple free expansion model (i.e. the interaction of the GB with the 
interstellar medium (ISM) is negligible), \cite{linblad1973} estimated 
an initial expansion velocity of $3.6$~km~s$^{-1}$.  By assuming that the GB-ISM 
interaction has decelerated the expansion, other authors have proposed even higher values, 
from 20 to 60~km~s$^{-1}$ \citep[e.g.,][]{olano1982, moreno1999}.
 Even when considering the former case, our result 
indicates that 
the GB expansion has   decreased.  This effect may be caused by the 
GB-ISM interaction, or be due to distortion of the expansion by the Galactic gravitational potential 
\citep[e.g.,][]{perrot2003,vasilkova2014}. The Galactic gravitational potential causes the 
elongation of the initial circular expansion, but it also affects the velocities of 
the components, with respect to the center of the expansion, as each component follows its own 
epicyclic orbit \citep[see for example the model in Figure 9 from][]{perrot2003}. From Figure~\ref{fig:kin2d}
we see that the velocity vectors of the studied regions are consistent with an average of 
$10$~km~s$^{-1}$ (see also information on 
Table~\ref{tab:kin}). The small dispersion value of $\sigma_{\widehat{r}_R \cdot v}=0.8\,$km~s$^{-1}$, points out to a small distortion of the expansion.  
Both the small distortion of the expansion and the near oblate spheroid shape obtained from our 
geometric fit, suggest that the age of the GB must be younger than previously thought or 
that the initial expansion velocity was high \citep[see discussion by][]{moreno1999}.

Recently, the existence of the GB has been proposed as a projection effect rather than a true 
ring of clouds \citep{buoy2015}. Our results, however, show that even when assuming that the 
star-forming regions in
the GB are not part of a continuous ring, they seem to share the same kinematic expansion.

\begin{table*}
\small
\centering
\renewcommand{\arraystretch}{1.1}
\caption{Distances, mean proper motions, and mean radial velocities of star-forming regions in the Gould Belt.}
\begin{tabular}{lcccccc}
\hline\hline
                & {D} & $\mu_{\alpha*}$  & $\mu_\delta$    & RV &  References\\
Region          & (pc)& \multicolumn{2}{c}{(mas yr$^{-1}$)}&(km s$^{-1}$)& \\
\hline
Barnard 59      & $163\pm5$ & $-1.2\pm0.2$&$-19.2\pm0.1$ & $3.4\pm0.11$  & 1,2\\ 
Cepheus Flare   & $360\pm32$ & $6.0\pm0.2$ &$0.4\pm0.5$   & 0.0 & 1,3\\
Chamaeleon I    & $192\pm6$ & $-22.8\pm0.1$&$0.3\pm0.1$  & $14.7\pm1.3$  & 1,4\\
Chamaeleon II   & $198\pm6$ & $-20.7\pm0.1$&$-8.0\pm0.1$ & $11.4\pm2.0$  & 1,5\\
Corona Australis& $154\pm4$ & $4.4\pm0.2$&$-27.3\pm0.2$  & $-1.0\pm1.0$ & 1,6\\
Lupus 1         & $156\pm3$ & $-15.0\pm0.1$&$-23.3\pm0.1$& $2.5\pm1.6$ & 1,7\\
Lupus 2         & $159\pm3$ & $-10.7\pm0.1$&$-22.0\pm0.1$& $2.2\pm0.9$ & 1,7\\
Lupus 3         & $162\pm3$ & $-10.0\pm0.1$&$-23.7\pm0.1$& $1.0\pm0.7$ & 1,7\\
Lupus 4         & $163\pm4$ & $-11.9\pm0.2$&$-23.9\pm0.2$& $0.3\pm3.8$ & 1,7 \\
ONC             & $388\pm5$ & $-1.1\pm0.1$&$-0.8\pm0.2$  & $27\pm2$ & 8,9,10 \\
L 1688 (Ophiuchus)& $137\pm1$ &$-5.28\pm0.02$&$-26.06\pm0.08$&3.6 & 11,12 \\
IC 348 (Perseus) & $321\pm10$&$4.34\pm0.03$&$-6.76\pm0.01$&  $15.9\pm1.8$ & 13\\
NGC 1333 (Perseus)&$293\pm22$&$7.34\pm0.05$&$-9.90\pm0.03$& $14.8\pm1.3$ & 13 \\
L 1495 (Taurus)  & $129.5\pm0.3$ &$8.56\pm0.10$&$-25.3\pm0.1$ &$15.1\pm0.6$& 14\\
\hline\hline
\label{tab:GBk}
\end{tabular}
\tablerefs{1=This work, 2=\cite{red2017}, 3=\cite{kun2008}, 4=\cite{joergens2006},
5=\cite{biazzo2012}, 6=\cite{neuhauser2000}, 7=\cite{galli2013}, 8=\cite{kounkel2017},
9=\cite{dzib2017}, 10=\cite{gd2008}, 11=\cite{ortiz2017a}, 12=\cite{friesen2017}, 
13=\cite{ortiz2018}, 
 and 14=\cite{galli2018}.
}
\end{table*}

\begin{table*}
\small
\scriptsize
\centering
\renewcommand{\arraystretch}{1.1}
\caption{Heliocentric and LSR velocities, and heliocentric positions of star-forming regions in the Gould Belt.}
\begin{tabular}{lcccccccccc}
\hline
                & $U$ & $V$  & $W$    & $u$ &  $v$  & $w$ & X & Y & Z\\
Region          & (km s$^{-1}$)& (km s$^{-1}$)& (km s$^{-1}$)& (km s$^{-1}$)& (km s$^{-1}$)& (km s$^{-1}$) & (pc)& (pc)& (pc)\\
\hline\hline
Barnard 59&$3.6\pm1.0$&$-12.8\pm0.1$&$-7.5\pm0.2$&$14.7\pm1.2$&$-0.6\pm0.5$&$-0.2\pm0.4$&$161.9\pm5.0$&$-8.3\pm0.3$&$20.3\pm0.6$\\
Cepheus&$-6.3\pm3.8$&$-0.6\pm2.4$&$-5.3\pm3.4$&$4.8\pm3.9$&$11.7\pm2.4$&$1.9\pm3.4$&$-111.9\pm10.0$&$327.1\pm29.3$&$98.2\pm8.8$\\
Cha I&$-7.2\pm0.9$&$-18.8\pm0.9$&$-10.2\pm1.1$&$3.9\pm1.1$&$-6.6\pm1.0$&$-3.0\pm1.1$&$84.4\pm2.6$&$-164.7\pm5.1$&$-51.0\pm1.6$\\
Cha II&$-7.7\pm1.2$&$-16.4\pm1.6$&$-9.3\pm0.6$&$3.4\pm1.4$&$-4.2\pm1.7$&$-2.0\pm0.7$&$106.2\pm2.7$&$-159.5\pm4.1$&$-49.4\pm1.3$\\
CrA&$-4.0\pm1.0$&$-17.6\pm0.8$&$-9.2\pm1.0$&$7.1\pm1.2$&$-5.3\pm0.9$&$-2.0\pm1.1$&$147.0\pm3.9$&$-0.3\pm0.1$&$-47.1\pm1.2$\\
Lupus 1&$-1.5\pm1.5$&$-17.2\pm0.8$&$-7.9\pm0.7$&$9.6\pm1.6$&$-5.0\pm0.9$&$-0.7\pm0.8$&$140.2\pm2.7$&$-54.3\pm1.0$&$42.2\pm0.8$\\
Lupus 2&$-2.6\pm0.9$&$-16.8\pm0.7$&$-6.8\pm0.6$&$8.5\pm1.1$&$-4.6\pm0.8$&$0.4\pm0.7$&$144.5\pm2.8$&$-55.7\pm1.1$&$33.2\pm0.6$\\
Lupus 3&$-4.0\pm0.7$&$-17.5\pm0.6$&$-8.0\pm0.5$&$7.1\pm1.0$&$-5.2\pm0.7$&$-0.8\pm0.6$&$150.2\pm3.9$&$-55.8\pm1.4$&$26.4\pm0.7$\\
Lupus 4&$-6.1\pm3.4$&$-17.7\pm1.5$&$-8.1\pm0.6$&$5.0\pm3.5$&$-5.4\pm1.6$&$-0.9\pm0.7$&$147.7\pm3.4$&$-65.0\pm1.5$&$24.0\pm0.5$\\
ONC&$-21.3\pm1.7$&$-12.2\pm1.0$&$-11.5\pm0.7$&$-10.2\pm1.8$&$0.0\pm1.1$&$-4.2\pm0.8$&$-319.9\pm4.4$&$-177.7\pm2.4$&$-129.1\pm1.8$\\
Ophiuchus&$4.1\pm0.9$&$-15.6\pm0.1$&$-7.3\pm0.3$&$15.2\pm1.2$&$-3.3\pm0.5$&$-0.1\pm0.5$&$130.2\pm1.1$&$-16.3\pm0.1$&$40.5\pm0.3$\\
IC348&$-17.0\pm1.6$&$-6.1\pm0.6$&$-8.5\pm0.6$&$-5.9\pm1.8$&$6.1\pm0.7$&$-1.3\pm0.7$&$-287.7\pm9.2$&$101.8\pm3.2$&$-98.0\pm3.1$\\
NGC1333&$-17.3\pm1.2$&$-10.7\pm0.5$&$-9.9\pm0.5$&$-6.2\pm1.3$&$1.6\pm0.7$&$-2.7\pm0.6$&$-257.8\pm19.7$&$102.3\pm7.8$&$-103.5\pm7.9$\\
L 1495&$-15.2\pm0.6$&$-12.1\pm0.1$&$-10.9\pm0.2$&$-4.1\pm0.9$&$0.1\pm0.5$&$-3.6\pm0.4$&$-122.2\pm0.3$&$23.8\pm0.1$&$-35.7\pm0.1$\\
\hline\hline
\label{tab:uvw}
\end{tabular}
\end{table*}

\newpage
\section{Conclusions}

We have presented an analysis of YSOs in star-forming regions related to the Gould Belt that have an astrometric solution in the Gaia DR2 catalog. The estimated mean distances to each star-forming region are consistent with previous estimations. However, these new estimations have significant improvements to the final errors.
The new distances were used to fit an ellipsoid to these regions, with the center located at $(X_0,Y_0,Z_0)=(-82\pm15, 39\pm7, -25\pm4)$~pc, and dimensions of  $(358\pm7)\times(316\pm13)\times(70\pm4)$~pc,
these values are consistent with those found for nearby O--B stars.
We also obtained the mean proper motions of these star-forming regions that were combined with radial velocities from the literature to obtain their three-dimensional motion. The motion of the star-forming regions is consistent with them moving away from the center of the Gould Belt at a velocity of 2.5~km~s$^{-1}$, in agreement with an expansion of the Gould Belt. This is the first time that YSOs motions are used to investigate the kinematics of the Gould Belt. 

As an interesting side result of our analysis, we also identified stars with motions significantly larger than
other stars in the same region. 
We speculated that some of them gained their peculiar motion via $n$-body interactions.

\acknowledgements 
{\small
G.-N.O.L acknowledges support from the Alexander von Humboldt 
Foundation in the form of a Humboldt Fellowship.
L.R. and L.L. acknowledges the financial support of DGAPA, UNAM (project IN112417), and CONACyT, M\'exico.
This work has made use of data from the European Space Agency (ESA) mission
{\it Gaia} (\url{https://www.cosmos.esa.int/gaia}), processed by the {\it Gaia}
Data Processing and Analysis Consortium (DPAC,
\url{https://www.cosmos.esa.int/web/gaia/dpac/consortium}). Funding for the DPAC
has been provided by national institutions, in particular the institutions
participating in the {\it Gaia} Multilateral Agreement.
This research has made use of the SIMBAD database,
operated at CDS, Strasbourg, France.
}

\facilities{Gaia} 

\bibliographystyle{aa}
\bibliography{references}

\end{document}